\documentclass{article}
\pdfoutput=1
\usepackage{fullpage}
\usepackage{graphicx}
\usepackage{color}
\usepackage{authblk}

\usepackage{slashed}

\usepackage{dcolumn}
\usepackage{tensor}
\usepackage{bm}
\usepackage{float}
\usepackage[usenames,dvipsnames]{xcolor}
\usepackage{listings}
\usepackage{fancyvrb}
\usepackage{slashed}
\usepackage{amsmath}
\usepackage{amssymb}
\usepackage{amsfonts}
\usepackage{enumitem}
\usepackage{listings}
\usepackage{mathtools}
\usepackage{slashed}
\usepackage{verbatim}
\usepackage{hyperref}
\usepackage[margin=1.0in]{geometry}

\usepackage{accents}
\newcommand{\myfigsep}{0.02 \textwidth}

\usepackage{titlesec}
\def\lsim{\mathrel{\rlap{\lower4pt\hbox{\hskip1pt$\sim$}}
    \raise1pt\hbox{$<$}}}         
\def\gsim{\mathrel{\rlap{\lower4pt\hbox{\hskip1pt$\sim$}}
    \raise1pt\hbox{$>$}}}         
    
\begin{document}

\begin{titlepage}
\vspace*{1cm}
\begin{center}
\Large{Neutrinos in Stellar Astrophysics}
\end{center}
\vspace{2cm}
\begin{center}
{G. M. Fuller}\\
\textit{Department of Physics,  University of California, San Diego, \\
 and the Center for Astrophysics and Space Sciences, La Jolla CA }\\
\vspace{0.7cm}
{W. C. Haxton}\\
\textit{Department of Physics, University of California, Berkeley, \\
and Lawrence Berkeley National Laboratory, Berkeley CA}\\
\end{center}

\vspace{1.5cm}
\begin{center}
Abstract
\end{center}
\vspace{0.3cm}
The physics of the mysterious and stealthy neutrino is at the heart of many phenomena in the cosmos. These particles interact with matter and with each other through the aptly named weak interaction. At typical astrophysical energies the weak interaction is some twenty orders of magnitude weaker than the electromagnetic interaction. However, in the early universe and in collapsing stars neutrinos can more than make up for their feeble interaction strength with huge numbers. Neutrinos can dominate the dynamics in these sites and set the conditions that govern the synthesis of the elements. Here we journey through the history of the discovery of these particles and describe their role in stellar evolution and collapse, the big bang, and multi-messenger astrophysics. Neutrino physics is at the frontier of elementary particle physics, nuclear physics, astrophysics and cosmology. All of these fields overlap in the neutrino story.

\end{titlepage}

\clearpage

\tableofcontents

~\\

\pagebreak

\pagebreak

\section{Introduction}
\noindent
The neutrino \cite{hh00} is an elementary particle that scatters only through the weak interaction, and consequently rarely interacts in matter.  Hints of the neutrino's existence emerged over a century ago, arguably first in an experiment performed by Chadwick in his student days on the spectrum of electrons emitted in beta decay.  Instead of detecting a line source of electrons from the process ${}^{214}$Bi $\rightarrow$ ${}^{214}$Po + e$^-$, Chadwick observed a continuous spectrum, and speculated that perhaps some unobserved radiation accompanied the decay (thereby carrying off the remainder of the decay energy) \cite{Chadwickbeta}.  Rutherford, Hahn, Meitner, and others offered a more prosaic explanation, that perhaps the spectrum reflected energy lost by the electron as it exited the target.  This alternative explanation was put to rest in 1927 by Ellis and Wooster, who observed the beta decay of $^{210}$Bi in a target sufficiently thick to stop most of the electrons, finding that the energy deposited per event in the target, 0.34 $\pm$ 0.04 MeV, differed significant from the energy (or Q value) for the decay of 1.05 MeV \cite{Wooster}.  In 1930, in a famous letter sent to his colleagues attending the Gauverein meeting, Pauli gave form to Chadwick's unobserved radiation, hypothesizing that an unobserved neutral spin-$\textstyle{\frac{1}{2}}$ particle with a mass less than 1\% that of the proton was emitted in the decay, with this particle and the outgoing electron sharing the decay energy \cite{Pauli}. This hypothesis preserved both energy conservation and the spin-statistics theorem that otherwise appeared to be violated in certain nuclear reactions.  Pauli called this new particle the ``neutron," later renamed the neutrino. Pauli did not published his ideas until three years later, when he gave the first public lecture on the neutrino at the 7th Solvay Conference of 1933.

Pauli viewed the neutrino as a component of the nucleus, thereby accounting for the integral spin of the Z=3 nucleus $^6$Li.   In 1932 Chadwick \cite{Chadwickneutron} discovered the particle we now know as the neutron, clarifying not only the ${}^6$Li statistics issue but resolving many issues connected with nuclear masses and charges.  In 1934 Fermi combined Pauli's neutrino and Chadwick's neutron in a quantum mechanical theory of beta decay in which the electron and neutrino are not pre-existing nuclear components,
but instead are produced spontaneously in the decay process.   In Fermi's remarkably insightful model, a neutron converts to a proton, while an electron and accompanying neutrino are produced.  Fermi assumed, in analogy with electromagnetism, that this semi-leptonic process was mediated by a vector charge operator, but replaced the long-range electromagnetic field with a contact interaction in which the four fermions interact at a point \cite{Fermi}.  Fermi recognized a bit later that an associated vector three-current interaction had to exist, due to the constraints of relativity.  Two years later, in another remarkable paper, Gamow and Teller deduced from the pattern of what we now call ``allowed" beta decays in heavy nuclei, that an axial coupling comparable in strength to Fermi's vector coupling had to be added \cite{GamowTeller}.  Arguably by 1936 a phenomenologically derived effective theory of beta decay had emerged that was consistent with today's standard model, apart from one ingredient -- twenty years later, the discovery of parity violation would lead to the conclusion that Pauli's vector interaction and the Gamow-Teller axial interaction interfere destructively in the beta decay amplitude.  

In 1956, approximately 25 years after Pauli's neutrino hypothesis, Cowan and Reines observed reactor neutrinos interacting in water via $\bar{\nu}_e + $p$ \rightarrow $n + e$^+$ \cite{CowanReines}.  The reaction was identified through the coincidence between the emitted positron and the delayed gamma-ray produced by the capture of the neutron.  The capture occurred on the absorber ${}^{108}$Cd that had been dissolved in the water.  Later it was established that, like its charged-lepton cousins, neutrinos come in three flavors, $\nu_e$, $\nu_\mu$, and $\nu_\tau$, which are produced when intense beams of electron, muons, and tauons interact in targets, respectively.  Nearly immediately after Pauli's hypothesis it was recognized that neutrinos have unusual properties.  Unlike the Standard Model's quarks, which participate in strong and electromagnetic interactions, or charged leptons, which interact electromagnetically, neutrinos interact only weakly.  Consequently they are extremely difficult to detect and can pass through astrophysical objects like stars without scattering.  Pauli argued that neutrinos are light, but are they massless, as postulated in the minimal Standard Model?   Not until 1999 was this question answered definitely and in the negative, when neutrino oscillations were found to be the source of the atmospheric and solar neutrino ``problems."  Today we know that the neutrino mass eigenstates, which govern neutrino propagation in free space, and the flavor eigenstates, the neutrinos accompanying their respective charged-lepton partners in reactions, do not coincide.  The entries in the unitary transformation connecting these alternative neutrino bases are the mixing angles that govern neutrino oscillations, a phenomenon we will discuss here in connection with several stellar environments.

And perhaps the most intriguing open question about neutrinos today is the anomalous scale of neutrino mass, which is now limited to $< 0.9$ eV/c$^2$ at 90\% CL by the KATRIN experiment, a value nearly six orders of magnitude lower than the mass of the lightest charged lepton.  Current cosmological limits, though somewhat model dependent, are significantly more stringent \cite{Hannestad}.   Why does the $\nu_e$ mass differ so radically from that of the electron or up and down quarks?  This question may be related to an observation made by Majorana in 1937, who pointed out that as neutrinos lack a charge or any other additively conserved quantum number, they could possibly be their own anti-particles.   More to the point, while the charged leptons and quarks must have distinct antiparticles and masses of the Dirac form, there is more freedom in describing neutrino masses, which can be Dirac, Majorana, or a combination of the two.   This flexibility  is exploited in the seesaw mechanism to account for the difference between the Dirac mass scale of the charged fermions, $m_D$, and the neutrino mass scale through the relation $m_\nu = m_D (m_D/m_R)$ where $m_R$ is a heavy right-handed Majorana mass.   This relationship identifies $m_D/m_R$ as the small parameter accounting for the difference between neutrino and charge lepton masses.   Indeed, part of the excitement engendered by the discovery of neutrino oscillations is that the atmospheric neutrino oscillation frequency suggests $m_R \sim 10^{16}$ GeV, near the Grand Unified scale.  Thus neutrinos may provide us a glimpse into new physics that resides far beyond the reach of any existing or anticipated accelerator.

The theme of this chapter is the very special role neutrinos play in astrophysics and cosmology, and specifically in various stages of stellar evolution \cite{bahcallbook}.  Astrophysical environments provide opportunities for probing unknown neutrino properties, often under conditions not possible to replicate in the laboratory.   There are several reasons that the ``inner space -- outer space" connection between neutrino properties and astrophysical environments has proven so rich.    First -- reminiscent of Willie Sutton's response when ask why he robbed banks, ``because that's where the money is" -- stellar objects are prodigious sources of neutrinos, producing some of our best understood distributions.
Neutrinos are the direct byproducts of the nuclear reaction chains by which stars generate energy: each solar conversion of four protons into helium produces two neutrinos, for a total of $\sim$ 2 $\times$ 10$^{38}$ neutrinos each second.  The resulting flux is observable on Earth.  These neutrinos carry information about conditions deep in the solar core, as they typically leave the Sun without further interacting.   They also provide experimentalists with opportunities for testing the properties of neutrinos propagating through matter over long distances.  Second, they are produced in nature's most violent astrophysical explosions, including the Big Bang, core-collapse supernovae (CCSN), neutron star (NS) mergers, and the accretion disks encircling supermassive black holes (BHs).  With the discovery of neutrino mass, the first component of dark matter was identified -- the primordial neutrinos.  Though these neutrinos are a minor component of the dark matter, they have an outsized effect on  the evolution of large-scale structure -- the pattern of voids and galaxies mapped by astronomers as a function of redshift -- because they evolve from being relativistic to nonrelativistic as the universe expands, thus producing effects with distinctive red-shift and scale dependence. Third, neutrinos dominate the cooling of many astrophysical objects, including young NSs and the degenerate helium cores of red giants.  Neutrinos can be radiated from deep within such bodies, in contrast to photons, which are trapped within stars, diffusing outward only slowly.  Finally, neutrinos are produced in our atmosphere and elsewhere as secondary byproducts of cosmic-ray collisions.  Detection of these neutrinos can help constrain properties of the primary cosmic ray spectrum.  Neutrinos produce
by reactions of ultra-high-energy cosmic rays can provide information on otherwise inaccessible
cosmic accelerators.

Neutrinos also mediate important astrophysical processes.  While the site of the r-process -- the
rapid-neutron-capture process thought to be responsible for the nucleosynthesis of
about half of the neutron-rich nuclear species heavier than iron -- 
remains uncertain, plausible sites include the debris expelled in the merger of two NSs
and the neutron-rich matter expelled from CCSNe.  The
energetics of these events are dominated by neutrinos, which play roles both in the ejection of
matter and in controlling its net isospin through reactions such as $\nu_e+n \leftrightarrow p +e^-$.
Observations of the kilonova produced in the binary NS merger GW170817
provided strong evidence that an associated r-process synthesized a sufficient number of heavy elements to
alter the opacity and thus the light curve of the radioactively powered debris. Other possible SN  r-process 
sites include the neutron-rich ``neutrino winds" that
blow off the proto-NS surface as well as the ${}^4$He zones of metal-poor
SNe, where neutrons can be produced by neutrino-induced spallation reactions.
SN neutrinos can also directly synthesize certain rare nuclei, 
such as ${}^{11}$B and ${}^{19}$F.

\begin{figure}
\begin{center}
\includegraphics[width=12cm]{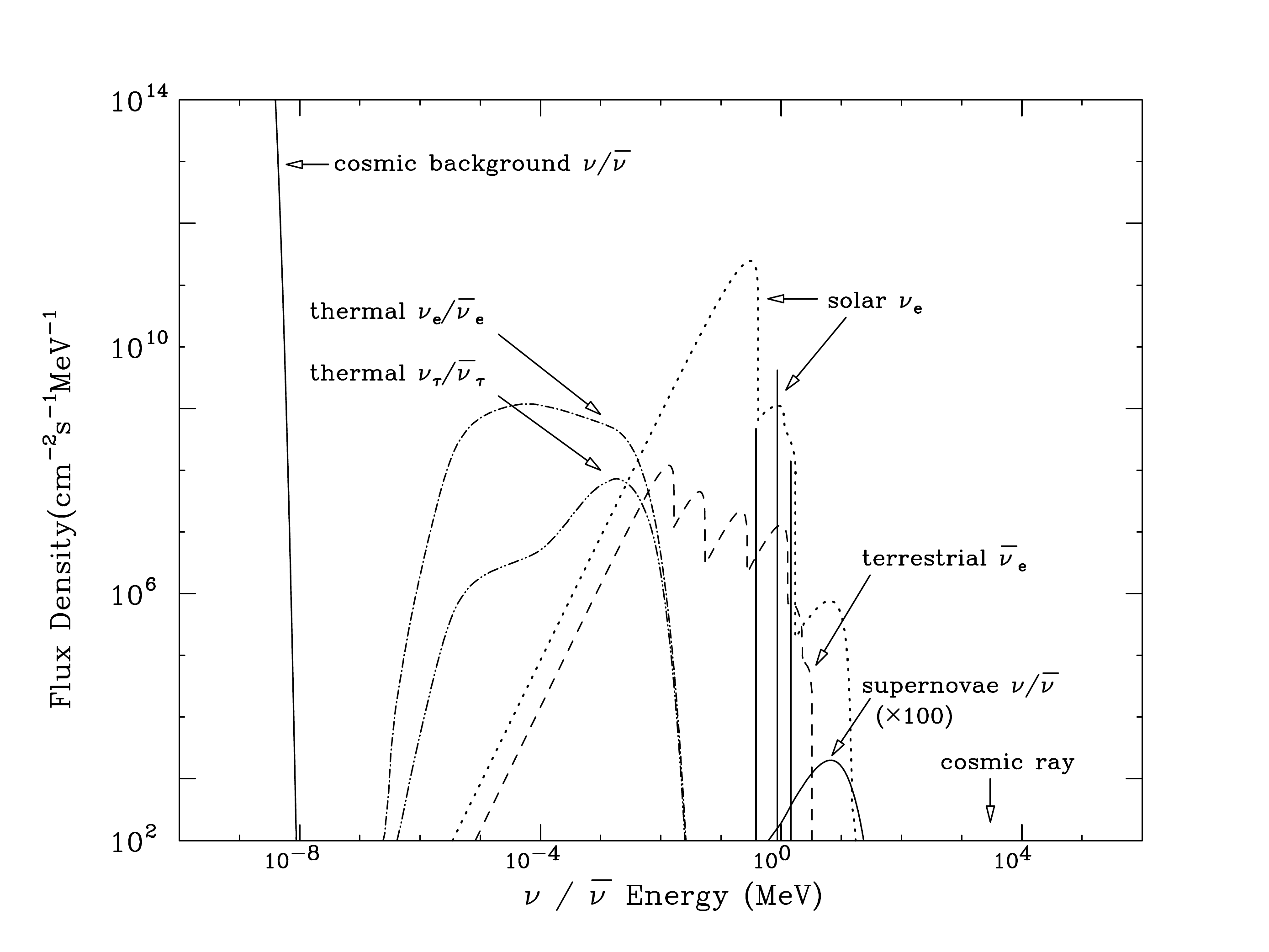}
\end{center}
\caption{The low-energy neutrino ``sky."  The principal sources
are: The relic neutrinos from the early universe, the cosmic neutrino background (C$\nu$B); The Sun; The integrated flux from past core collapse supernovae; The atmospheric neutrinos created by cosmic ray interactions; 
The  $\sim$ 1 keV thermal neutrinos of all flavors emitted
by the Sun; and the $\bar{\nu}_e$s generated by the Earth's natural radioactivity.
Reproduced with permission from Ref. \cite{lin}.  See \cite{edoardo1} for an updated estimate of the thermal flux and \cite{edoardo2} for a plot of astrophysical and terrestrial sources
ranging up to $10^{12}$ MeV.}
\label{fig:lowEnus}
\end{figure}

One way to characterize the natural neutrinos sources available for study on earth is the ``neutrino sky"
plot of the neutrino fluence vs. energy at low energies, defined here as $E_\nu <$ TeV.   Shown in Fig. \ref{fig:lowEnus} are the
significant steady-state sources.   Among these, three important stellar sources stand out, the solar electron neutrinos
produced in hydrogen fusion, the solar thermal neutrinos produced as  $\nu - \bar{\nu}$ pairs in the
$\sim$ 2 keV core of the Sun, and the continuous flux of SN relic neutrinos, generated by integrating
over all past SNe as a function of redshift.  The flavor labels are those at the source: oscillation
effects will alter propagating neutrinos.  These stellar sources appear in combination with other continuous sources,
including the primordial flux of neutrinos produced at the time of weak decoupling in the early universe,
at a temperature $T \sim$ 3 MeV.  Also shown is the flux of terrestrial antineutrinos produced 
primarily by the decays of uranium, thorium, and ${}^{40}$K in the earth's crust.

In this chapter we will also be interested in the extension of this plot along another axis -- transient 
neutrino sources where relevant parameters are the frequency, duration, and total flux contained within
the neutrino burst.  The most interesting of the transient sources are CCSNe within our galaxy, 
which occur at a frequency of one to a few in a century.  Such events could produce up to  $10^4-10^5$ events
in current and future massive detectors, such as Super- and Hyper-Kamiokande, given a SN
within 10 kpc \cite{suwa}.  There are other sources -- generally rarer and/or weaker -- associated with 
phenomena such as NS-NS or NS-BH mergers, BH accretion, and pre-SN evolution.

\section{Solar Neutrinos}
\subsection{Solar neutrinos and multi-messenger astrophysics}
\noindent
The quest to solve the solar neutrino puzzle not only led to a profound physics discovery, but provides an outstanding example of a theme now much in vogue in astrophysics, multi-messenger observations.
Measurements of the Sun's mass, radius, luminosity, photoabsorption lines, meteoritic abundances, and helioseismology of the convective and radiative zones were combined to tightly constrain a model
with a large number of free parameters, to produce neutrino flux predictions with well-defined errors.  The observation of fluxes in significant disagreement with standard solar model (SSM) predictions led to the
conclusion that neutrino oscillations associated with a well-determined mixing angle and neutrino mass difference were altering the neutrino event rates seen in terrestrial detectors.

As noted in the entertaining 1982 paper by Bahcall and Davis \cite{SNhistory}, ``An Account of the Development of the Solar Neutrino Problem," the classic papers of Bethe and Critchfield \cite{BetheC1938} and Bethe \cite{Bethe1939}
describing hydrogen fusion neither included neutrino production in the nuclear reactions nor discussed the possibility that neutrino detection might provide a test of the theory.  These papers were written five years
after Fermi developed his theory of low-energy weak interactions.  Bahcall and Davis cite a 1948 paper by H. R. Crane \cite{Crane48} for an early discussion of the flux: Crane used terrestrial heating by neutrinos to place an upper bound
on neutrino cross sections.

The first successful effort to detect neutrinos from the Sun got underway nearly six decades ago.  Ray Davis, Jr. and his collaborators constructed a 650-ton detector in the Homestake Gold Mine, one mile beneath
Lead, South Dakota \cite{davis}.  This radiochemical detector, based on the chlorine-bearing cleaning fluid C$_2$Cl$_4$, was designed to capture about one of the approximately 10$^{18}$  high-energy solar neutrinos that penetrated it each day -- the rest
passed through the detector, without interacting.  The neutrino-capture reaction was 
inverse electron capture
\[ {}^{37}\mathrm{Cl} + \nu_e^{solar} \rightarrow {}^{37}\mathrm{Ar} + e^-. \]
The product of this reaction, ${}^{37}$Ar, is a noble-gas isotope with a half life of about one month.  It can be efficiently removed from a large volume of organic fluid by a helium gas purge, then counted in miniature gas proportional counters as  ${}^{37}$Ar decays back to ${}^{37}$Cl.  Davis typically exposed his detector for
about two months, building up to nearly the saturation level of a few dozen argon
atoms, then purged the detector to determine the number of solar neutrinos captured during this period.

Within a few years it became apparent that the number of neutrinos detected was only about one-third that
predicted by the SSM \cite{bahcallbook}, that is, the model of the Sun based on the standard theory of main sequence stellar evolution. Some initially attributed this ``solar neutrino problem" to uncertainties in the SSM: As the flux of neutrinos most important to the Davis detector vary as $\sim$ $T^{22}_c$, where $T_c$ is the solar core temperature, a 5\% theory uncertainty in $T_c$ could explain the discrepancy.  In fact, the correct explanation for the discrepancy proved much more profound.  Davis was awarded the 2002 Nobel Prize in Physics for the chlorine experiment.

The solar neutrino problem stimulated a series of follow-up experiments to measure the different
components of the solar neutrino flux and to determine the source of the Cl experiment
discrepancy.  The SAGE and GALLEX/GNO  experiments, radiochemical detectors similar to Cl,  but using ${}^{71}$Ga as a target, were designed to measure the flux of neutrinos from
the dominant low-energy branch of solar neutrinos, the pp neutrinos.  The first detector to measure neutrinos event by event, recording neutrino interactions in real time, was the converted proton decay detector Kamiokande.   The detector contained three kilotons of very pure water, with solar neutrinos scattering off the electrons within the water.  Phototubes surrounding the tank recorded the ring of Cerenkov radiation produced by the recoiling relativistic electrons.  Kamiokande measured the high-energy neutrinos most important to the Davis detector, and thus confirmed the deficit that Davis had originally observed.  New and very massive water (Super-Kamiokande) and heavy-water (Sudbury Neutrino Observatory (SNO)) detectors were constructed.  Finally, Borexino, a detector using liquid scintillator, was constructed to measure 
low-energy solar neutrino branches in real time.  These experiments -- most particularly SNO, 
because of its multiple detection channels sensitive to different combinations of neutrino types -- showed that solar neutrinos were not missing, but rather hidden by a change of flavor occurring during their transit from the Sun to the Earth, as will be described later in this chapter.  For more detailed summaries of these experiments, see \cite{HRS}. 

Of these experiments Super-Kamiokande and Borexino remain in operation as solar neutrino detectors, while SAGE is currently being used as a target
for a neutrino-source experiment to probe possible oscillations at short distances.  Borexino
provided an exciting new result this past year, the first measurement of CN neutrinos, that we will discuss later.  Super-Kamiokande recently introduced
gadolinium to detector water to improve sensitivity to relic SN neutrinos.

\subsection{The Standard Solar Model}
Solar models trace the evolution of the Sun from its beginning -- when the collapse of the primordial gas cloud was halted by the turn-on of thermonuclear reactions -- to today, 4.6 Gyr later, thereby predicting contemporary
solar properties such as the composition, temperature, pressure, and sound-speed profiles and the neutrino fluxes.  

The Sun belongs to a class of stars that derive their energy from the conversion of hydrogen to He in their cores: approximately 80\% of stars
 lie along the ``main-sequence" path in the Hertzsprung-Russell color-magnitude diagram.  The SSM thus is an application of a more general theory of main-sequence stellar evolution, to a specific case
uniquely constrained by the many detailed measurements only possible for the Sun.  The SSM is based on four basic assumptions:
    
\begin{itemize}
\item The Sun evolves in hydrostatic equilibrium, maintaining a local balance between the gravitational force and the pressure gradient.  To describe this condition in detail, one must  specify the electron-gas equation of state as a function of temperature, density, and composition.  This requires attention to such issues
as the incomplete ionization of metals, the contribution of radiation to pressure, and the influence of
screening.  Because the ionization is nearly complete, the end result is not too different from an ideal gas.
\item The mechanisms for energy transport are radiation and convection.  The inner portion of today's Sun -- 98\% by mass or about 72\% by radius -- is radiative.
In the Sun's outer envelope, where the temperature gradient is larger, convection dominates the energy transport.
The region of the Sun where energy is generated through thermonuclear reactions is deep within the radiative zone.

\begin{figure}
\begin{center}
\includegraphics[width=12cm]{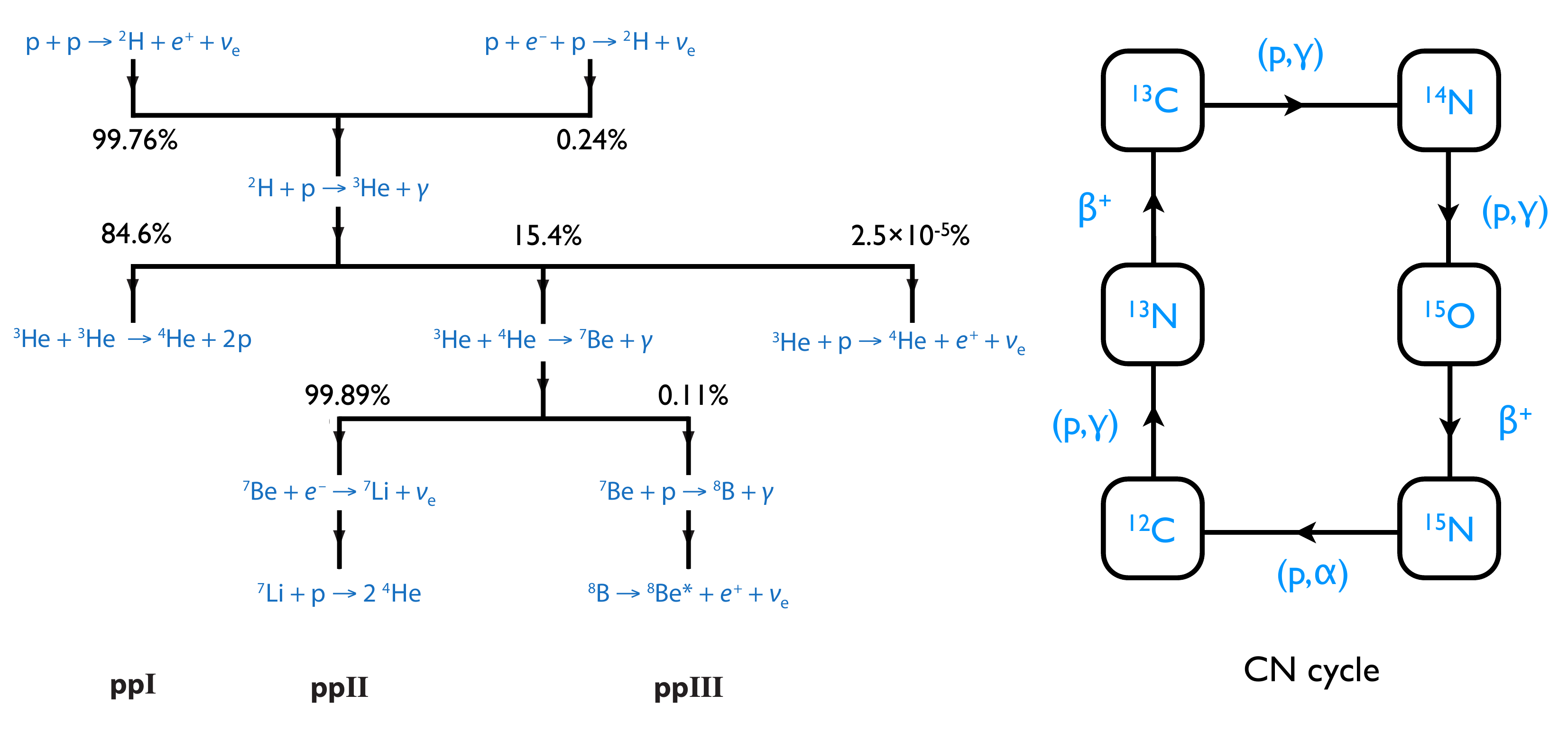}
\end{center}
\caption{ Left panel: The three principle cycles comprising the pp chain (denoted ppI, ppII, and ppIII).
Associated neutrinos ``tag" the three branches.  The SSM branching ratios come from the
GS98-SFII SSM \cite{SHP}.  Also shown is the minor branch $^3$He+p$\rightarrow {}^4$He+e$^+$+$\nu_e$,
which generates the Sun's most energetic neutrinos.  Right panel: The CN I cycle, which produces 
solar neutrinos from the $\beta$ decays of ${}^{13}$N and ${}^{15}$O. Reproduced from Ref. \cite{HRS}.}
\label{fig:ppCNO}
\end{figure}

The location of the boundary between the radiative interior and convective zone depends on the opacity, which also influences
other interior properties such as the sound-speed profile. In addition to
elementary processes such as the scattering of photons off electrons and fully ionized H and He, more complex
processes such as bound-free scattering off metals are important contributors to the opacity in the
Sun's interior regions: deeper in the Sun, energy-momentum conservation increasingly favors free-bound transitions on the more tightly bound electrons of  high-Z elements.  SSM convection is modeled through mixing length theory, in which volume elements are transported
radially over a characteristic distance determined empirically in the model, but typically on the order of the pressure scale height.
\item The Sun produces its energy by fusing protons into ${}^4$He,
\begin{equation}
2e^-+4\mathrm{p} \rightarrow {}^4\mathrm{He}  + 2 \nu_e +26.73 \mathrm{~MeV} 
\label{eq:Hburn}
\end{equation}
via the pp chain (99\%) and CN I cycle reactions of Fig. \ref{fig:ppCNO}.
The Sun is
a large but slow reactor.  As the Sun's core temperature is a modest  $T_c \sim 1.57 \cdot 10^7$K, the rate of energy 
production/volume is only $\sim$ 275 watts/m$^3$, comparable to that of a compost heap: the solar reactor produces its  $\sim 10^7$ terawatts of energy by
virtue of its enormous mass. Typical center-of-mass energies for reacting particles are $\sim$ 10 keV, much less than the Coulomb barriers inhibiting
charged-particle nuclear reactions.  Though enormous progress has been made in recent years in extending laboratory measurements to lower energies,
with the LUNA program \cite{LUNA} of particular note, several critical measurements still must be made at higher
energies and then extrapolated to the solar Gamow peak.  More will be said below about the reaction chains driving Eq. (\ref{eq:Hburn}).
\item The model is constrained to produce today's solar radius, mass, and luminosity.  An important assumption of the SSM is that the proto-Sun passed through a highly convective phase, rendering the Sun uniform in composition
until main-sequence burning began.  The initial composition by mass is conventionally divided into hydrogen (X$_\mathrm{ini}$),
helium (Y$_\mathrm{ini}$), and everything else (the metals, denoted Z$_\mathrm{ini}$),
with X$_\mathrm{ini}$+Y$_\mathrm{ini}$+Z$_\mathrm{ini}$=1.  The relative abundances of the metals
are determined from a combination of meteoritic and solar photospheric data.  The absolute 
abundance Z$_\mathrm{ini}$ can be taken from the modern Sun's surface abundance Z$_\mathrm{S}$,
after corrections for the effects of diffusion over 4.6 Gyr of solar evolution. 
Finally, Y$_\mathrm{ini}$/X$_\mathrm{ini}$ is adjusted along with $\alpha_\mathrm{MLT}$,
a parameter describing solar mixing, until the model reproduces the modern Sun's
luminosity and radius.  The resulting ${}^4$He/H mass fraction ratio is typically 0.27 $\pm$ 0.01, which can be compared to the Big-Bang value of 0.23 $\pm$ 0.01, showing that the
Sun was formed from previously processed material. 
\end{itemize}
  
The left panel of Fig.~\ref{fig:ppCNO} shows the three cycles, ppI, ppII, and ppIII, comprising the pp chain, along with a fourth termination through $^3$He+p that while rare, yet is of some interest because it produces the most energetic
solar neutrinos.  Note that each cycle within the pp chain is associated with a characteristic neutrino, and thus through solar neutrino measurements one can determine the current solar rate of energy generation through each cycle.
This provides one with a critical test of the SSM due to the distinct temperature dependences of the three main cycles: in SSMs constrained to produce the correct solar luminosity, the ppII/ppI and ppIII/ppI branching ratios vary approximately as 
$T_c^7$ and $T_c^{18}$, respectively, where $T_c$ is the core temperature.  The sharp temperature dependence, reflecting the relative ease or difficulty of Coulomb barrier penetration in the various reactions, make solar neutrino spectroscopy an inordinately
sensitive probe of the Sun's central temperature.  Given the precision of current determinations of the relevant nuclear cross sections, one can determine $T_c$ to an accuracy of $\lesssim$ 1\% through solar neutrino measurements.  

The right panel of Fig.~\ref{fig:ppCNO} shows the CNI cycle, a mechanism for hydrogen burning that depends on the pre-existing
abundances of the metals C, N, and O to catalyze the fusion.  While responsible for only $\sim$ 1\% of solar energy
production, this cycle becomes the dominant mechanism for energy generation in stars more massive than the Sun,
where higher central temperatures generate a rapid increase in cross sections for higher Z nuclei.  An important
recent achievement by Borexino was the first measurement of these neutrinos:  We discussed later in this section
the relevance of this measurement to the solar abundance problem mentioned below.

The neutrino-producing reactions of the pp chain and CN I cycle are summarized in Table \ref{tab:one}.  
The $\beta$ decay sources produce neutrinos with a continuous spectra, while the electron
capture reactions produce lines with widths $\sim$ 2 keV characteristic of the temperature of the solar core.
The Table shows two solar models, denoted GS98-SFII and AGSS09-SFII, which employ the same nuclear
physics (given by the Solar Fusion II evaluation \cite{SFII}) but differ in their
assumptions about solar surface metallicity due to the use of 1D or 3D models, respectively,
to interpret photospheric absorption lines.  The predictions of the  higher metallicity ($\sim$ +30\%) 
GS98-SFII SSM are generally in excellent agreement with solar helioseismic properties, including
interior sound speeds and the location of the base of the convective zone.  This is not the
case for the AGSS09-SFII SSM, which nevertheless uses a more sophisticated treatment of
the photosphere.   The unresolved conflict between SSMs that agree with our best description
of the solar interior and those that employ our best model of the solar surface is known as the
solar abundance problem. 

The SSM is dynamic.  As the Sun evolves, the SSM predicts a rather steep luminosity increase ($\sim$ 40\%
over 4.6 b.y.): as He is synthesized, the core's temperature must respond to the resulting
changes in the mean molecular weight and opacity.   The challenge of reconciling this increase with the Earth's
geo-history is sometimes described as the faint young Sun paradox \cite{faint}.  As the luminosity increases, the ppI/ppII/ppIII burning evolves,
with the fraction of energy produced through the more temperature-sensitive ppII and
ppIII branches increasing.  The ${}^8$B neutrino flux from the ppIII cycle 
has an exceeding sharp growth $\sim e^{t/\tau}$ where $\tau \sim$ 0.9 b.y.

The Sun's composition is impacted.  In the first 10$^8$ years of main-sequence burning
most of the carbon in the Sun's central core in converted to nitrogen, building up the core
abundance of $^{14}$N.   ${}^{14}$N(p,$\gamma$) is a CN-cycle ``bottleneck," regulating further burning.  
Over the lifetime of the Sun a rather steep gradient in $^3$He -- which is produced and consumed
in the pp chain -- is established, increasing with  $r$ and $\propto$ $T^{-6}$, where $T$ is local
temperature.  This gradient in $^3$He was
discussed as a potential trigger for periodic mixing of the core \cite{Dilke}.  There is
also a slow diffusion of metals toward the core, sufficient to impact helioseismology, 
because metals have a charge-to-mass ratio smaller than that of the gas on average.

The details of the evolution depend on a variety of model 
input parameters and their uncertainties, including the solar age, the mean
radiative opacity, the modern Sun's luminosity $L_\odot$ and radius $R_\odot$, the diffusion coefficient describing
the settling of He and metals, the abundances of key metals, and the nuclear cross
sections for the pp chain and CN cycle -- a total of about 20 parameters in total.  This is why
the Sun is such a good illustration of the power of multi-messenger astrophysics:  laboratory measurements of nuclear cross sections and opacities
were combined with a variety of astrophysics measurements of metallicities, photo-absorption lines,
and the Sun's physical attributes ($R_\odot$, $M_\odot$, $L_\odot$) to constrain
the parameters.  In the end modelers were able to propagate remaining uncertainties through
calculations, to produce predictions with defined error bars on our two most important
observables sensitive to conditions in the Sun's interior -- the solar neutrino fluxes and the sound-speed profile deduced 
from helioseismology.

\begin{table}
\caption{SSM neutrino fluxes from the GS98-SFII (high Z) and AGSS09-SFII (low Z) SSMs, which
differ in their assumptions about photospheric metallicity.  In cases where associated uncertainties
are asymmetric, an average has been used.  The solar values
come from a luminosity-constrained analysis of all available data by the Borexino Collaboration. For
$^7$Be electron capture, the branching ratios for the 860 and 380 keV lines are 90\% and 10\%, respectively.
From Ref. \cite{HRS}.}
\vspace{0.4cm}
\label{tab:one}
\begin{center}
\begin{tabular}{lcccc}
\hline \hline
 $\nu$ source & E$_\nu^\mathrm{max}$ & GS98-SFII & AGSS09-SFII & Solar Flux \\
   &  (MeV) & & & (cm$^{-2}$ s$^-1$) \\
\hline \\
p+p$\rightarrow^2$H+e$^+$+$\nu$ & 0.42 & $5.98(1 \pm 0.006)$ & $6.03(1 \pm 0.006)$ & $6.05 \cdot 10^{10}(1^{+0.003}_{-0.011})$  \\
p+e$^-$+p$\rightarrow^2$H+$\nu$ & 1.44 & $1.44(1 \pm 0.012)$ & $1.47(1 \pm 0.012)$ & $1.46 \cdot 10^8(1^{+0.010}_{-0.014})$ \\
$^7$Be+e$^-$$\rightarrow^7$Li+$\nu$ & 0.86  & $5.00(1 \pm 0.07)$ & $4.56(1 \pm 0.07)$ & $4.82 \cdot 10^9(1^{+0.05}_{-0.04})$ \\
 & 0.38 & & &  \\
$^8$B$\rightarrow^8$Be+e$^+$+$\nu$ & $\sim$ 15 & $5.58(1 \pm 0.14)$ & $4.59(1 \pm 0.14)$ & $5.00 \cdot 10^6(1\pm 0.03)$ \\ 
${}^3$He+p$\rightarrow^4$He+e$^+$+$\nu$  & 18.8 & $8.04(1 \pm 0.30)$ & $8.31(1 \pm 0.30)$ & --- \\
$^{13}$N$\rightarrow^{13}$C+e$^+$+$\nu$  & 1.20 & $2.96(1 \pm 0.14)$ & $2.17(1 \pm 0.14)$ &$\leq 6.7 \cdot 10^8$ \\ 
$^{15}$O$\rightarrow^{15}$N+e$^+$+$\nu$  & 1.73 & $2.23(1 \pm 0.15)$ & $1.56(1 \pm 0.15) $ &$\leq 3.2 \cdot 10^8$ \\ 
$\chi^2/P^\mathrm{agr}$ & & 3.5/90\% & 3.4/90\% &  \\ \\
\hline \hline
\end{tabular}
\end{center}
\end{table}

\subsection{SNO, Super-Kamiokande, and Borexino}
\begin{figure}
\begin{center}
\includegraphics[width=9cm]{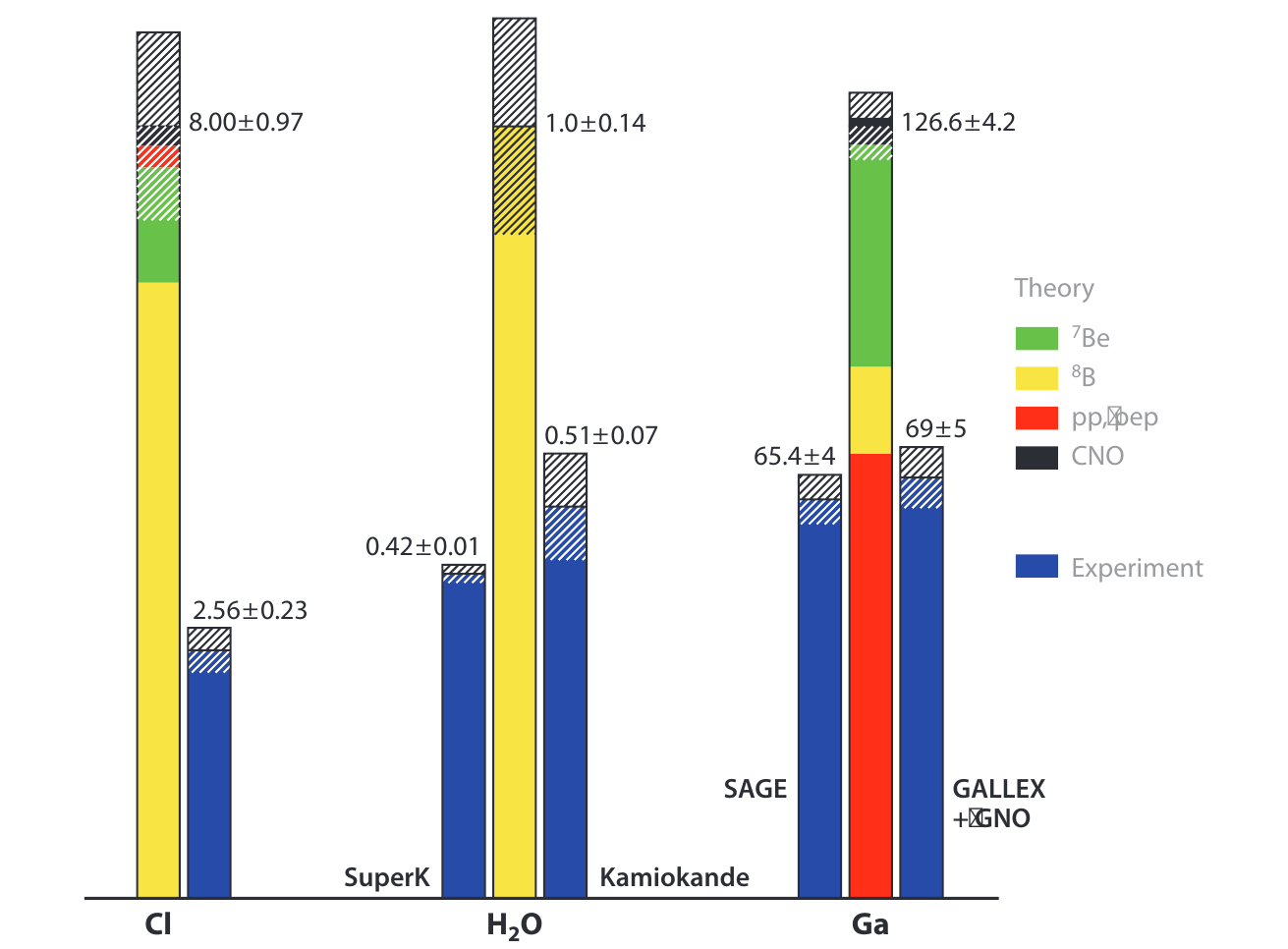}
\caption{Comparison of the measured neutrino rates for the a) chlorine,
b) Kamioka II/III, and c) SAGE and GALLEX/GNO experiments with the
SSM, computed with GS98 abundances.  From Ref. \cite{HRS}.}
\label{Fig:Old}
\end{center}
\end{figure}

After a wait of over two decades following the initial results of Davis's experiment, results from two new sets of experiments 
became available.  The Kamiokande experiment measured the high-energy $^8$B neutrino
flux by observing the recoiling electrons produced by neutrino-electron scattering in a water Cherenkov detector.  The SAGE/GALLEX/GNO 
experiments utilized the reaction  $^{71}$Ga($\nu_e,e^-)^{71}$Ge, exploiting its low threshold to capture pp neutrinos,
thus probing the dominant ppI cycle.  The results shown in Fig. \ref{Fig:Old} not only confirmed the deficit that Davis had found, but when combined produced a 
pattern of fluxes that was very hard to reconcile with the expected temperature dependence of neutrino fluxes described above.  This
suggested that some new physics was at play, and provided the impetus for new and more expensive experiments
with the sensitivity needed to resolve the solar neutrino puzzle.

Solar neutrino detection requires the combination of a large detector
volume (to provide the necessary rate of events), very low backgrounds (so
that neutrino events can be distinguished from backgrounds due to cosmic
rays and natural radioactivity), and a distinctive signal.  The first requirement
favors detectors constructed from inexpensive materials and/or materials having
large cross sections for neutrino capture.  The second generally requires
a deep-underground location for the detector, with sufficient rock overburden to
attenuate the flux of penetrating muons
produced by cosmic ray interactions in the atmosphere.  It also requires very careful attention to detector cleanliness,
including tight limits on dust or other contaminants that might introduce
radioactivity, use of low-background construction materials, control of radon, and
often the use of fiducial volume cuts so that the outer portions of a detector
become a shield against activities produced in the surrounding rock walls.

There are several possible detection modes for solar neutrinos,
interesting because of their different sensitivities to flavor.
The early radiochemical experiments using ${}^{37}$Cl and ${}^{71}$Ga
targets were based  on the charged current weak reaction 
\[ \nu_e + (N,Z) \rightarrow e^- + (N-1,Z+1) \]
where the signal for neutrino absorption is the growth over time of 
very small concentrations of the
daughter nucleus $(N-1,Z+1)$ in the detector.  As the spectrum of solar
neutrinos extends only to about 15 MeV, well below the threshold for
producing muons, this reaction is sensitive only to electron neutrinos.


A second possible nuclear detection channel is neutral-current
scattering
\[ \nu_x + (N,Z) \rightarrow \nu_x^\prime + (N,Z)^*, \]
a process independent of the neutrino flavor.  If this scattering
leaves the nucleus in an excited state, the observable would be the
de-excitation of the final-state nucleus, e.g., by $\gamma$ decay or by
particle breakup.  Alternatively, if the scattering is elastic, 
a coherent process at low energies with a cross section approximately proportion to N$^2$ where N is
the neutron number, the signal would be the small recoil energy of the nucleus after
scattering.  

A third possibility, utilized by the Kamioka collaboration, is the scattering of neutrinos off electrons,
\[ \nu_x + e^- \rightarrow \nu_x^\prime + e^-.\]
Both electron- and heavy-flavor ($\nu_\mu$, $\nu_\tau$) solar neutrinos
can scatter off electrons, the former by charge and neutral currents,
and the latter by neutral currents only.  Consequently the cross
section for scattering heavy-flavor neutrinos is only about 0.15 that
for electron neutrinos.    An important aspect of electron-neutrino
scattering is its directionality:  for solar neutrino energies much above
the electron mass of 0.511 MeV, the electron scatters into a narrow
cone along the incident neutrino's direction.   This directionality 
provides a powerful tool for extracting solar neutrino events from 
background:  neutrino events correlate with the direction of the Sun, while
background events should be isotropic.  Thus neutrino events can be
identified as the excess seen at forward electron angles.

The puzzling flux pattern of Fig. \ref{Fig:Old} helped motivated a new generation of experiments,
Super-Kamiokande, the Sudbury Neutrino Observatory (SNO), and Borexino.  These
detectors record (or recorded, in the case of SNO) events in real time, and both separately
and together have yielded a great deal of information on the flavor content of the
solar neutrinos reaching earth.

\begin{figure}
\begin{center}
\includegraphics[width=12cm]{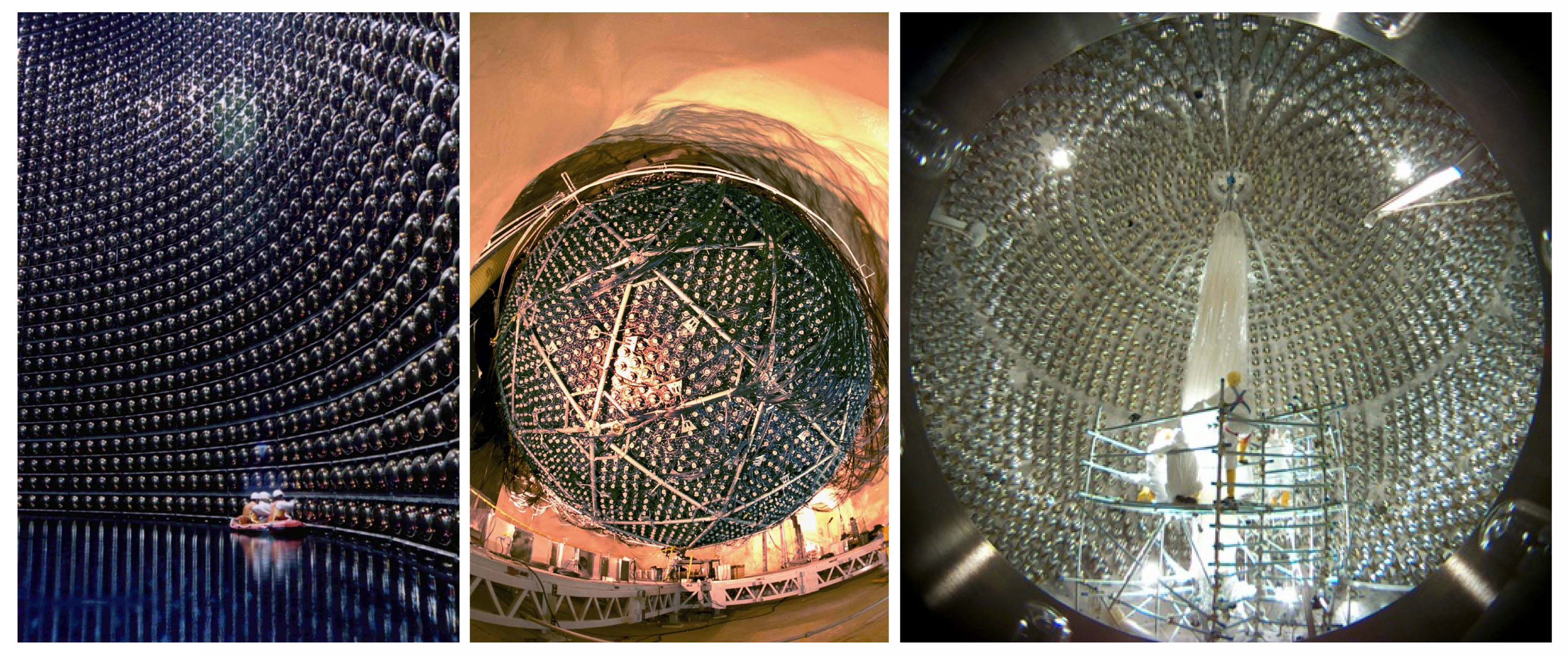}
\caption{The Super-Kamiokande (left), SNO (center) and Borexino (right) detectors.}
\label{fig:sksnob}
\end{center}
\end{figure}

The Super-Kamiokande detector (Super-K) \cite{SKgen} (left panel, Fig.~\ref{fig:sksnob}) consists of 50 kilotons of ultra-pure water
held within a cylindrical stainless steel tank, 39m in diameter and 42m tall.  
Two meters inside the walls a scaffold supports a dense array of 50-cm-diameter hemispherical 
photomultiplier tubes (PMTs), which face inward and view the inner 32 kilotons of
water.  Additional 20-cm tubes face outward, viewing the outer portion
of the detector that serves as a shield and as a veto.   A solar neutrino 
can interact in the inner detector, scattering off an electron.  The recoiling,
relativistic electron then produces a cone of Cherenkov radiation, a pattern
that can be reconstructed from the triggering of the phototubes that
surround the inner detector.  The detector is housed deep within Japan's
Kamioka Mine, approximately one kilometer underground.

Super-K began operations in 1996, progressing from Phase I
to the current Phase V.  The collaboration reported solar neutrino data recorded through Phase IV (completed in 2018) at Neutrino 2020, a total
of 5805 live days of solar neutrino data, corresponding to over 100K $^8$B neutrino events \cite{SKgen}.
The counting rate corresponds to a $\nu_e$ flux of $(2.35 \pm 0.02 \pm 0.04) \times 10^{-6}$/cm$^2$ s, which can be 
compared to the SSM results in Table 1.  The experiment has provided tentative evidence for both day-night differences
in the counting rate (significance $> 2\sigma$) and energy-dependent distortion of the spectrum ($\gtrsim 1 \sigma$),
both signatures of the matter effects on oscillations discussed in the next section \cite{SKgen}.  In summer 2020  SK
began SK-Gd operations, dissolving 14 tons of a gadolinium-bearing compound into the otherwise ultra-pure water in order
to enhance neutron detection efficiency.  One particular motivation is to improve the detection of the diffuse background
of relic SN antineutrinos by tagging the neutron produced in
\[ \bar{\nu}_e + p \rightarrow e^+ + n \]
$^{155}$Gd and $^{157}$Gd  have huge neutron absorption cross sections.   First results from the SK-Gd phase have
not yet been announced.

SNO \cite{SNOgen} (center panel, Fig.~\ref{fig:sksnob}) was constructed two kilometers 
underground in the INCO Creighton nickel mine in Ontario, Canada.
The detector took data from May 1999 through November 2006, operating
in three different modes over its 7.5-year lifetime.   SNO employed a
one-kiloton target of heavy water, contained within a spherical acrylic vessel
six meters in radius.  This sphere was surrounded by an additional five meters
(seven kilotons) of very pure ordinary water, filling the rock cavity 
that housed the entire detector.   
SNO operated with a threshold of  5 MeV through much of its lifetime, detecting the
portion of the ${}^8$B solar neutrino spectrum from 5-15 MeV, though a low-energy
reanalysis was later completed in which a
electron kinetic energy threshold of 3.5 MeV was employed.

The choice of a heavy-water target allowed SNO experimentalists to 
exploit all three of the reaction channels described above, with their
varying flavor sensitivities
\begin{eqnarray}
\nu _x + e^- \rightarrow \nu_x^\prime +
e^{-}~~~&&\mathrm{ES:~elastic~scattering} \nonumber \\
\nu _e + d \rightarrow  p + p + e^{-}~~~&&\mathrm{CC:~charged~current} \nonumber \\
\nu _x + d \rightarrow  \nu_x^\prime + n +
p~~~&&\mathrm{NC:~neutral~current} \nonumber 
\end{eqnarray}
The elastic scattering (ES) reaction is the same as that employed by Super-Kamiokande, 
with its differing sensitivities to electron- and heavy-flavor neutrinos.  The charged-current (CC)
reaction on deuterium is sensitive only to electron-flavor neutrinos,
producing electrons that carry off most of 
the incident neutrino's energy (apart from the 1.44 MeV needed to
break a deuterium nucleus into p+p).   
Thus, from the energy distribution of the electrons, one can 
reconstruct the incident $\nu_e$ spectrum (and
possible distortions discussed below) more accurately than in the
case of ES.

The NC reaction, which is observed through the produced neutron,
provides no spectral information, but does measure the total solar
neutrino flux, independent of flavor. The SNO experiment used
three techniques for measuring the neutrons during the course of the experiment:
1) neutron captured on deuterium,
producing 6.25 MeV $\gamma$s; 2) addition of salt to the detector to allow capture on Cl,
 producing 8.6 MeV $\gamma$s; and 3)
direct neutron detection using an array of ${}^3$He--filled proportional counters.

\begin{figure}
\begin{center}
\includegraphics[width=10cm]{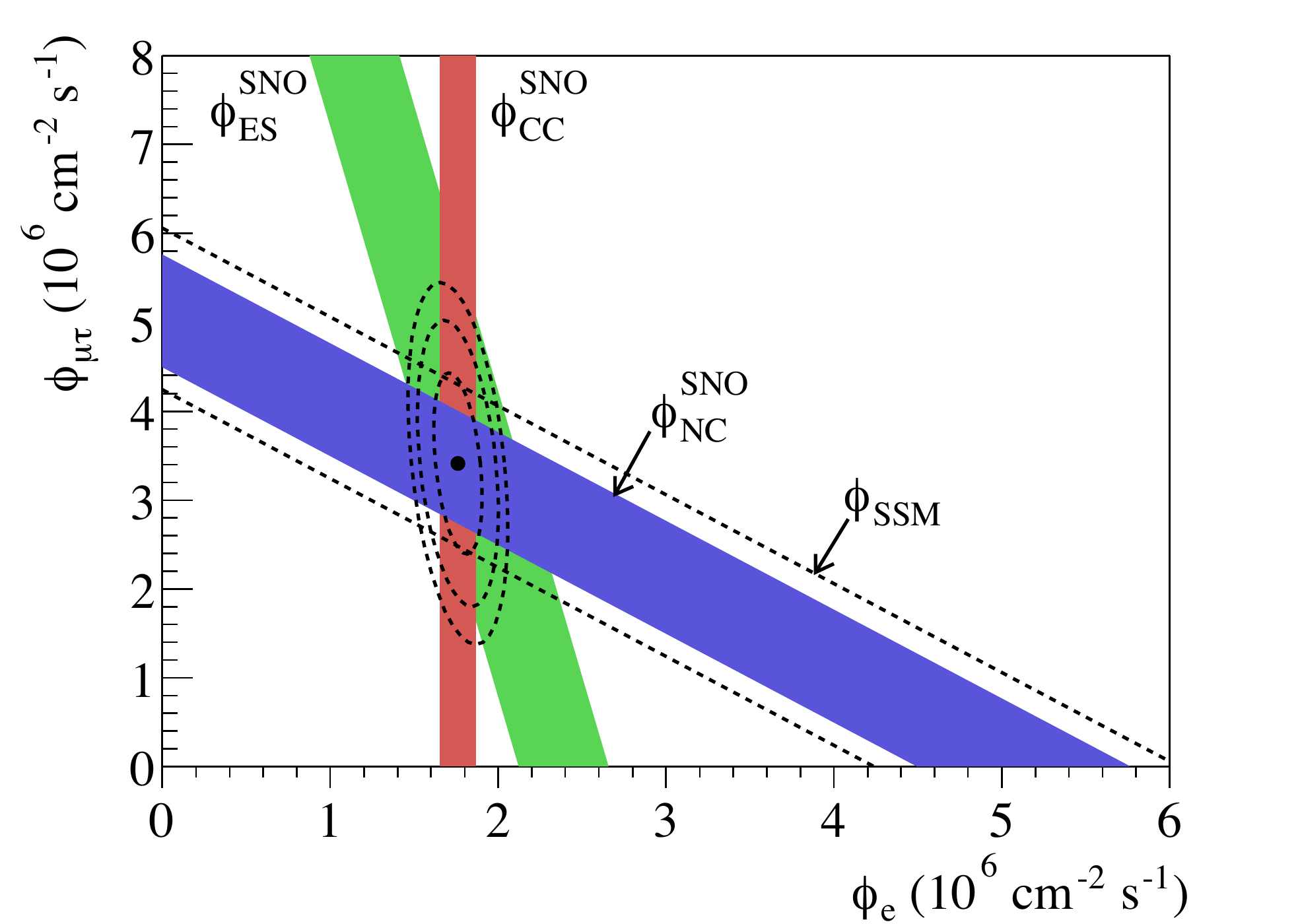}
\end{center}
\caption{ Results from the D$_2$O phase of the SNO experiment.  The allowed bands for
CC, NC, and ES reactions of solar neutrinos intersect to show a flux that is
one-third $\nu_e$s and two-thirds heavy-flavor
neutrinos.  There is agreement between the NC total-flux measurement and
the predictions of the SSM (band indicated by the dashed lines).  Figure from Ref. \cite{SNOgen}, with permission of the
Sudbury Neutrino Observatory collaboration.}
\label{fig:SNOresults}
\end{figure}

The results from the three phases of SNO are in generally good agreement,
separately and in combination establishing a total flux of active neutrinos from
${}^8$B decay of $\phi_\mathrm{NC}(\nu_\mathrm{active}) = (5.25 
\pm 0.16 (\mathrm{stat})^{+0.11}_{-0.13} (\mathrm{syst})) \times 10^6$/cm$^2$/s, in good agreement
with SSM predictions.  SNO also established $\phi_\mathrm{CC}(\nu_e) \sim 0.34
\phi_\mathrm{NC}(\nu_\mathrm{active})$.  Thus, as
Fig.~\ref{fig:SNOresults} illustrates,
about two-thirds of the electron
neutrinos produced in the Sun arrive on Earth as heavy-flavor
(muon or tauon) neutrinos.   The Davis detector and the
CC channel in SNO are blind to these heavy flavors, seeing only
the portion with electron flavor.  Thus the solar neutrino problem
was not a matter of missing neutrinos, but rather one of neutrinos
in hiding.  The implications of this discovery -- that neutrinos are
massive and violate flavor -- are profound, indicating that our
Standard Model of particle physics is incomplete.

Other potential signals of neutrino oscillations in matter, such as an energy-dependent
distortion in the $\nu_e$ survival probability or day-night differences due to neutrino
passage through the earth, were not seen in SNO.

The Borexino experiment \cite{Borgen}  (right panel, Fig.~\ref{fig:sksnob}), located in Italy's Gran Sasso Laboratory, is the
first to measure low-energy ($\lesssim$ 1 MeV) solar neutrinos in real time.   The
detector is housed within a 16.9m domed tank containing an outer layer of ultrapure water that provides
shielding against external neutrons and gamma rays.  At the inner edge of the water a stainless steel 
sphere serves as a support structure for an array of photomultiplier tubes that view both the inner
detector and the outer water shield, so that the Cherenkov light emitted by muons passing through
the water can be used to veto those events.  Within the steel sphere there are two regions, separated
by thin nylon vessels, containing high-purity buffer liquid, within which is sequestered a central volume
of 278 tons of organic scintillator.  The fiducial volume consists of $\sim$ 100 tons of the
liquid scintillator at the very center of the detector.  Scintillation light produced by recoil
electrons after ES events is the solar neutrino signal.  The 862 keV $^7$Be 
neutrinos produce a recoil electron spectrum with a distinctive cut-off edge 
at 665 keV.

Results reported by the  Borexino Collaboration from 2008 through 2020 \cite{Borgen}
constrain four low-energy
solar neutrino branches.  The Collaboration's achievements include
\begin{enumerate}
\item the first direct detection (detection in real time) of the low-energy pp neutrinos, the dominant
solar neutrino branch;
\item the first direct measurement of the line neutrino source from electron capture on $^7$Be,
at a rate that corresponds to an un-oscillated flux of
$(3.10 \pm 0.15) \times 10^9$/cm$^2$s, or about 62\% of the GS98-SFII SSM central value;
\item the first direct measurement of the pep neutrinos; and
\item the first measurement of neutrinos from the CNO cycle, determining the CNO contribution to the Sun's energy budget and
probing the hydrogen-burning mechanism that dominates energy production in stars above $\sim$ 1.3 $M_\odot$.
\end{enumerate}
In addition, Borexino found a $^8$B flux consistent with the more precise measurements by
Super-Kamiokande and SNO.  Borexino's comprehensive pp-cycle measurements provide an important test of
matter-enhanced solar neutrino oscillations, described in Sec. \ref{sec:MSW}.   The CNO neutrino
measurement opens up the possibility that a future, more precise flux determination can help
resolve the solar metallicity problem, discussed in Sec. \ref{sec:future}.


\subsection{Neutrino Mass and Oscillations \cite{Maltoni}}
\label{sec:MSW}
The discovery that the solar (and atmospheric) neutrino problems were a consequence
of neutrino oscillations arguably provides our most direct evidence of physics beyond the minimal
standard model: the phenomenon requires both massive neutrinos and flavor mixing.
Neutrino oscillations arise if the basis of neutrino eigenstates of definite flavor
does not coincide with the basis of neutrino eigenstates of definite mass, e.g., in the 
simplified case of two flavors, if the
mass eigenstates $|\nu_1 \rangle$ and $|\nu_2 \rangle$ (with masses
$m_1$ and $m_2$) are related to the weak interaction eigenstates by
\begin{eqnarray}
|\nu_e\rangle  &=& \cos \theta_v |\nu_1\rangle  + \sin \theta_v|\nu_2 \rangle  \nonumber \\
|\nu_\mu\rangle &=& - \sin \theta_v |\nu_1 \rangle + \cos \theta_v |\nu_2 
\rangle \nonumber
\end{eqnarray}
where $\theta_v$, the (vacuum) mixing angle, is nonzero.   
 
\begin{figure}
\begin{center}
\includegraphics[width=12cm]{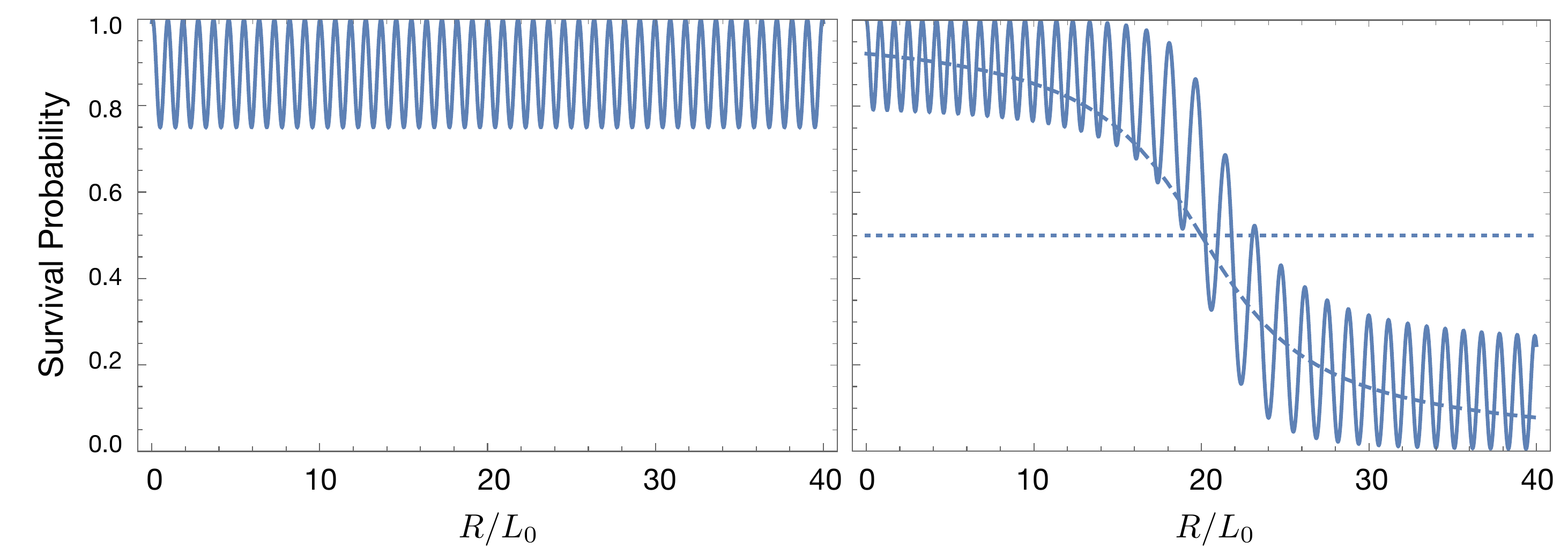}
\caption{Illustration of the MSW mechanism.  The left panel
shows vacuum oscillations for a $\nu_e$ propagating
to the right for 40 oscillation lengths, with $\theta_v$=15$^\circ$.  The average $\nu_e$ survival
probability is large, 87.5\%.  On the right, the same propagation in matter.  The dotted horizontal line
represents $\delta m^2/E$ (arbitrary units).  The dashed line is the corresponding MSW potential.  The
crossing at $R/L_0=20$ is the point where the matter potential exactly cancels the vacuum mass difference --
the level crossing point (where the local oscillation length is longest).  As the matter potential changes slowly over each oscillation,
the propagation is adiabatic.  Matter dramatically reduces the survival probability.}
\label{fig:osc2}
\end{center}
\end{figure}

In this case a state produced as a $|\nu_e\rangle$
or a $|\nu_\mu\rangle$ at some time $t$ --- for example, a neutrino
produced by $\beta$ decay in the Sun's core --- does not remain a pure flavor eigenstate
as it propagates away from the source.   If at time $t$=0 the neutrino is a flavor eigenstate,
$ |\nu(t=0)\rangle  = |\nu_e \rangle$, then on propagating in free space each mass eigenstate
component of the $|\nu_e \rangle$ accumulates a distinct phase,
\[e^{i(\vec{k} \cdot \vec{x} - \omega t)} =
e^{i [ \vec{k} \cdot \vec{x} - \sqrt{m_i^2 + k^2}~t ]} . \]
If the neutrino mass is small compared to the neutrino 
momentum/energy, one finds
\[ |\nu(t) \rangle = e^{i(\vec{k} \cdot \vec{x} - kt
-(m_1^2+m_2^2)t/4k)} \left( \cos \theta_v |\nu_1 \rangle e^{i \delta m^2 t/4k}
+ \sin \theta_v |\nu_2 \rangle e^{-i \delta m^2 t/4k} \right) \] 
There is a common average phase (which has no physical
consequence) as well as a beat phase that depends on
\[ \delta m^2 \equiv \delta m_{21}^2 = m_2^2 - m_1^2.  \]
From this one can find the probability that 
the neutrino state remains a $|\nu_e\rangle$ at time t
\[ P_{\nu_e} (t) = | \langle \nu_e | \nu(t) \rangle |^2  =
1 - \sin^2 2 \theta_v \sin^2 \left({{\delta m^2c^4 x}\over{4 \hbar c E}} \right)  \]
The probability oscillates from 1 to  $1-\sin^2 2 \theta_v$ and back to 1
over an oscillation length scale
\[ L_\mathrm{o} = {4 \pi \hbar c E \over \delta m^2 c^4} ,\]
as depicted in the left panel of Fig.~\ref{fig:osc2}.
In the case of solar neutrinos, if $L_\mathrm{o}$ were comparable to or shorter than one astronomical unit, a 
reduction in the solar $\nu_e$ flux would be expected in terrestrial detectors.
  
The suggestion that neutrinos could oscillate was first made by
Pontecorvo in 1958, who pointed out the analogy with $K_0 \leftrightarrow \bar 
K_0$     
oscillations.  If the Earth-Sun separation is much larger than $L_\mathrm{o}$, one expects 
an average flux reduction due to oscillations of
\[ 1 - {1 \over 2} \sin^2 2 \theta_v . \]
For a 1 MeV neutrino, this requires $\delta m^2c^4 \gg 10^{-12}$ eV$^2$.  
But such a reduction $-$ particularly given the initial theory prejudice that neutrino mixing
angles might be small $-$ did not seem sufficient
to account for the factor-of-three discrepancy that emerged from Davis's early
measurements.

The view of neutrino oscillations changed 
when Mikheyev and Smirnov \cite{MS} showed in 1985 that neutrino
oscillations occurring in matter -- rather than in vacuum -- could
produce greatly enhanced oscillation probabilities.   This
enhancement comes about because neutrinos propagating through matter 
acquire an additional mass due to their interactions with the
matter.  In particular, because the Sun contains many electrons, 
the electron neutrino becomes heavier in proportion to the
local density of electrons.   An enhanced probability for oscillations can
result when an electron neutrino passes from a high-density region
(such as the solar core) to a low-density one (such as the surface
of the Earth).  This matter enhancement is
called the MSW mechanism after
Mikheyev, Smirnov, and Wolfenstein \cite{Wolf} (who first described the 
phenomenon of neutrino effective masses).

To explain this enhancement, consider the case where the vacuum mixing
angle $\theta_v$ is small  and $m_2 > m_1$.
Then in vacuum $|\nu_e\rangle  \sim  |\nu_1\rangle \equiv |\nu_L(\rho=0) \rangle$
where $\rho=0$ is the electron density in vacuum, that is,
the $\nu_e$ and the light vacuum eigenstate $|\nu_L(\rho=0) \rangle$ are almost identical.
(Correspondingly, the heavy eigenstate $|\nu_2 \rangle \equiv |\nu_H(\rho=0) \rangle
 \sim |\nu_\mu \rangle$ in vacuum.)
Now what happens in matter?  As matter makes the $\nu_e$
heavier in proportion to the electron density, if that density
is sufficiently high, clearly the electron neutrino
must become the (local) heavy mass eigenstate.  That is, 
$|\nu_e\rangle  \sim |\nu_H(\rho \rightarrow \infty) \rangle$ (and
consequently $|\nu_\mu\rangle  \sim  |\nu_L(\rho \rightarrow \infty) \rangle$).  
That is, we conclude that there must be a local mixing angle $\theta(\rho)$
that rotates from $\theta_v \sim 0$
in vacuum to $\theta(\rho) \sim \pi/2$ as $\rho \rightarrow \infty$.  As the neutrino
transits the Sun, the local mass eigenstates cross at a critical density, with
the vacuum mass difference compensated by the effective mass coming from
interactions with the surrounding matter.  (More correctly, this is an avoided crossing.)
The heavy mass eigenstate in the center of the Sun may be approximately coincident
with the flavor eigenstate $\nu_e$, but the rotation described above renders the heavy
eigenstate at the Sun's surface approximately coincident with the $\nu_\mu$.

MSW enhancement of oscillations is thus a consequence of neutrinos transiting a region
of changing density.    Electron neutrinos
produced in the high-density solar core are created as heavy mass
eigenstates.  If these neutrinos now propagate to the solar surface
adiabatically  -- this means that they stay on the heavy-mass trajectory, which requires that changes in the solar density scale height
$\rho^{-1}~ d \rho/dx$ are 
small over an oscillation length, at all points along
the neutrino trajectory -- they will emerge 
from the Sun as $|\nu_H(\rho=0)\rangle =
|\nu_2 \rangle \sim |\nu_\mu \rangle$.  There
may be an almost complete conversion of the $\nu_e$s produced
in the solar core to $\nu_\mu$s.  The MSW mechanism is thus a consequence of adiabatically transiting a level
crossing,  a familiar phenomenon in quantum
mechanics.

The right panel of Fig. \ref{fig:osc2} illustrate the effects of matter on oscillations.  The matter transition between electron
and muon flavors is centered around a density where the vacuum mass 
difference is compensated by the matter contributions.

The discussion above was presented for two neutrino flavors, and thus a single
vacuum mixing angle $\theta_v$ and mass difference $\delta m^2$.  But three neutrino
flavors appear in the Standard Model of particle physics.  In this case the
relationship between flavor \{$\nu_e,\nu_\mu,\nu_\tau$ \} and mass \{ $\nu_1,\nu_2,\nu_3$ \} eigenstates
is described by the  PMNS matrix \cite{MNS,Pontecorvo67}
\begin{eqnarray}
\left( \begin{array}{c} | \nu_e \rangle \\ | \nu_\mu \rangle \\ | \nu_\tau \rangle \end{array} \right) =
\left( \begin{array}{ccc} c_{12} c_{13} & s_{12} c_{13} & s_{13} e^{-i \delta} \\
-s_{12}c_{23}-c_{12}s_{23} s_{13} e^{i \delta} & c_{12}c_{23}-s_{12}s_{23}s_{13} e^{i \delta} & s_{23} c_{13} \\
s_{12}s_{23}-c_{12}c_{23}s_{13} e^{i \delta} & -c_{12}s_{23}-s_{12}c_{23}s_{13}e^{i \delta} & c_{23} c_{13} \end{array}
\right) \left( \begin{array}{c} e^{i \alpha_1 /2} | \nu_1\rangle \\ e^{i \alpha_2 /2} | \nu_2 \rangle \\ | \nu_3 \rangle \end{array} \right)
\label{eq:three}
\end{eqnarray}
where $c_{ij} \equiv \cos{\theta_{ij}}$ and $s_{ij} \equiv \sin{\theta_{ij}}$.  This matrix depends on
three mixing angles $\theta_{12}$, $\theta_{13}$, and $\theta_{23}$, of which the first and last are the
dominant angles for solar and atmospheric oscillations, respectively; a Dirac phase $\delta$ that
can induce CP-violating differences in the oscillation probabilities for
conjugate channels such as $\nu_\mu \rightarrow \nu_e$ versus
$\bar{\nu}_\mu \rightarrow \bar{\nu}_e$; and two Majorana phases $\alpha_1$ and $\alpha_2$
that will affect the interference among mass eigenstates in the effective neutrino mass probed
in the lepton-number-violating process of neutrinoless double $\beta$ decay.  There are
also two independent mass$^2$ differences, $\delta m_{21}^2 \equiv m_2^2-m_1^2$ and
$\delta m_{32}^2 \equiv m_3^2-m_2^2$.

In this framework, the dominant oscillation affecting solar neutrinos is that described
by $\delta m_{21}^2$ and $\theta_{12}$.
The results from SNO, Super-Kamiokande, Borexino, and earlier solar
neutrino experiments, and from the
reactor experiment KamLAND \cite{KamLAND}, have determined these parameters
quite precisely, as discussed below.   Unlike the MSW
example given above, the dominant mixing is characterized by a  large mixing angle,
$\theta_{12} \sim 34^\circ$.  Thus the vacuum oscillation probability is significant.
The mass$^2$ difference,
$\delta m_{21}^2 \sim 7.4 \times 10^{-5}$ eV$^2$, leads to important
matter effects in the higher energy portion of the solar neutrino
spectrum, thus influencing the rates found in the SNO, Super-Kamiokande,
and Borexino experiments.  

\begin{figure}
\begin{center}
\includegraphics[width=11cm]{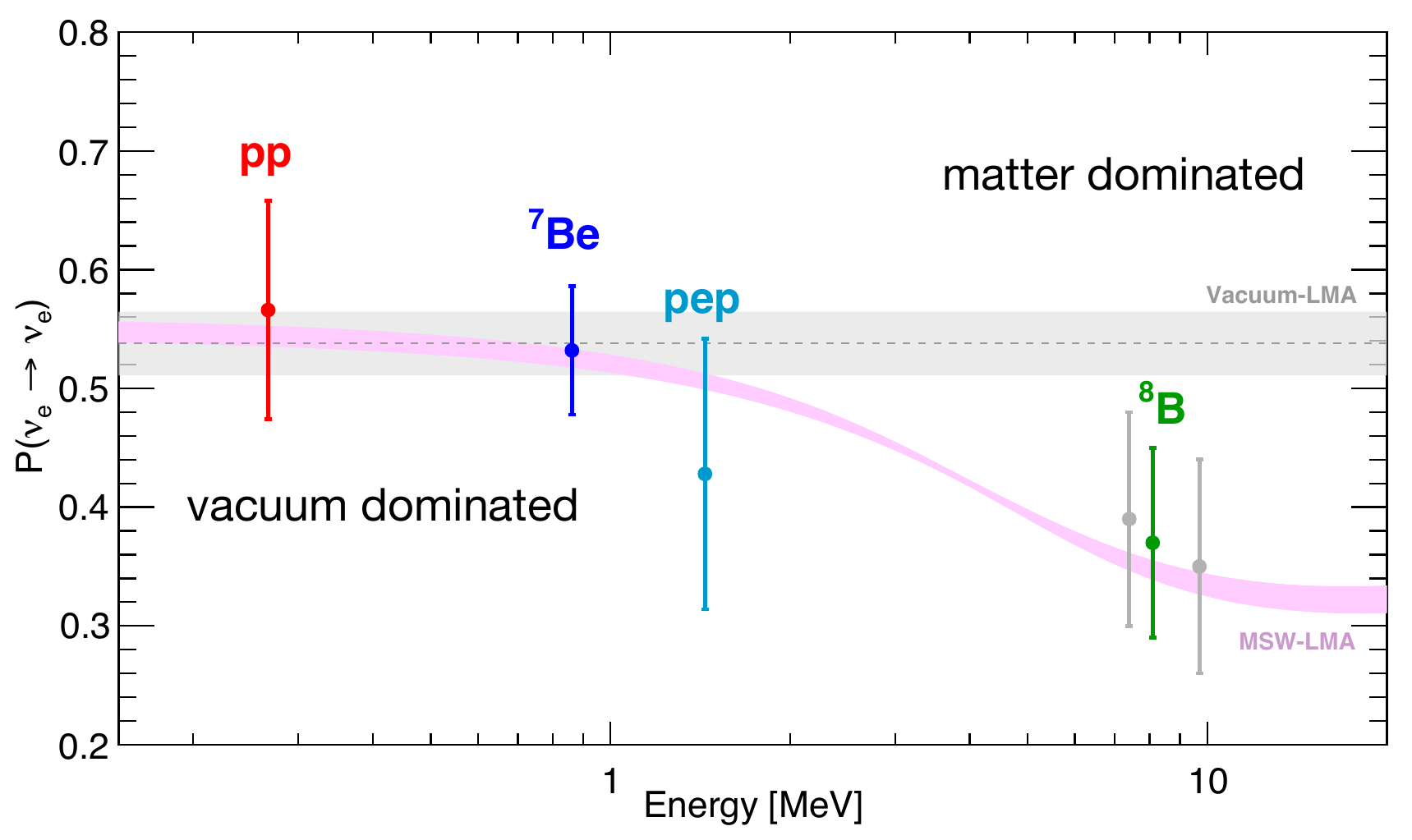}
\caption{The $\nu_e$ survival probability $P_{\nu_e}$ deduced from Borexino measurements.
The energy dependence is consistent with the MSW mechanism and exhibits the transition
from $\sim$ vacuum=dominated to matter-dominated oscillations as $E_\nu$ increases. Reprinted with
permission from Ref. \cite{Borgen}.}
\label{fig:Borexino}
\end{center}
\end{figure}

Figure \ref{fig:Borexino} displays the oscillation probabilities deduced from the Borexino experiment.  Remarkably 
there is a coincidence between the value of $\delta m_{21}^2$ and the solar neutrino spectrum.  The conditions for
the MSW enhancement of solar neutrino oscillations are 1) a level crossing that is traversed 2) adiabatically in the Sun.
The level crossing occurs at an electron density
\[ \sqrt{2} G_F \rho_e(r) = {\delta m_{21}^2 \cos{2 \theta_{12}} \over 2 E_\nu} \]
where $\rho_e(r)$ is the electron density at a radius $r$ in the Sun, $G_F$ is Fermi's weak coupling constant, and $E_\nu$ is the neutrino energy.
This relationship implies that as $E_\nu$ decreases, the level crossing radius moves toward the core of the Sun -- to higher densities.
Consequently the limiting density in the Sun, $\rho(0)$, determines a critical $E_\nu^0$.  Neutrinos with energies $\gtrsim E_\nu^0$ 
experience a level crossing at some radius $r$, and thus oscillations are enhanced by the MSW effect.  Neutrinos with energies $\lesssim E_\nu^0$
do not, and thus their oscillations are vacuum-dominated .  A more detailed discussion would show that $E_\nu^0$ is not a
sharp energy, but represents a region of energies in which oscillations transition from vacuum-dominated to matter-dominated.

To our great good fortune, $\delta m_{21}^2$ corresponds to an $E_\nu^0$ right in the middle of the solar neutrino
spectrum.  Thus Borexino's program to map the entire spectrum of solar neutrinos can alternatively be viewed as a
quest to map out the transition from vacuum-dominated to matter-dominated oscillations -- a remarkably successful quest. 
Figure \ref{fig:Borexino} shows the transition we have just described.  The fact that we can see the imprint of matter oscillations
on solar neutrino oscillations allows us to determine the sign of $\delta m_{21}^2$:  were $\delta m_{21}^2$ negative,
matter would increase the splitting of the eigenstates, and consequent there would be no level crossing and the 
oscillation effects would not resemble those seen in Fig. \ref{fig:Borexino}.   While we have seen the effects of matter on
solar neutrino oscillations, we have not yet isolated the effects of matter on atmospheric neutrino oscillations \cite{atmos}, which are
governed by the mass splitting $\delta m_{32}^2$.  Thus we know only $|\delta m_{32}^2|$ -- and consequently the pattern
of the mass eigenstates is not fully determined, with two hierarchies (normal and inverted) compatible with the data.

\subsection{Solar Neutrinos: Open Questions}
\label{sec:future}
Neutrino oscillations proved to be responsible for both the solar
neutrino problem and the atmospheric neutrino problem -- where the ratio $\nu_\mu/\nu_e$ was found to vary
with azimuthal angle and thus with distance, for neutrinos produced by cosmic rays impinging on the earth's atmosphere \cite{atmos}.
These discoveries demonstrated that neutrinos are massive and that the mass
eigenstates are not coincident with the states of definite flavor.  

Solar neutrino experiments, supplemented by reactor $\bar{\nu}_e$ oscillation results from 
KamLAND and other such experiments \cite{KamLAND}, provide our best constraints on 
$\theta_{12}$ and $\delta m_{21}^2$.  The constraints are typically folded into global neutrino analyses
that seek to determine all of the mixing angles appearing in Eq. (\ref{eq:three}) as well as the two mass
differences.  Results from various groups can be found 
in the Particle Data Group summary \cite{PDG}.  Adopted values \cite{Abe16c,Gando13}  are
\[ \sin^2{\theta_{12}} = 0.307^{+ 0.013}_{-0.012}~~~~~~~\delta m_{21}^2 =\left(7.53 \pm  \right) \times 10^{-5}
\mathrm{eV^2}~ \]
where the errors are $1\sigma$.  The determination  of $\theta_{12}$ to $\sim$ 4\% accuracy from solar
neutrino flux measurements stands as a remarkable illustration of the potential of multi-messenger astrophysics
 to probe fundamental physics with precision.  
 
Despite this success, relatively little new investment is being made in the field, particularly when
measured with respect to  investment in long-baseline neutrinos from accelerator sources.
We find this surprising, given both the intrinsic potential of solar neutrino physics and the intriguing
questions that remain open.   For example, the pp flux is arguably our best known neutrino source,
with theory determining its spectral shape and flux to sub-1\% accuracy.  Furthermore, when these neutrinos
interact in targets, the response can be predicted with similar precision: one typically exploits
either the elementary scattering off the electron, or nuclear transitions that can be calibrated with other
low-energy weak interactions, such as $\beta$ decay.  Yet, despite the fact that we are entering an era
of precision neutrino measurements, little use is being made of pp or other solar neutrino sources.

We will provide two examples here of opportunities, one connected with testing the equivalence 
of the Sun's electromagnetic and weak luminosities, and other dealing with the Sun's metallicity and 
its relevance to questions about the origin of our solar system and possibly others in which massive
planets form.

Many current analyses of solar neutrino data implicitly assume that neutrino fluxes are directly 
constrained by the Sun's luminosity:  some analyses utilize this constraint directly in extracting
fluxes from experiment, while in other cases this assumption is fed in through the standard solar model,
which uses the rather precisely known luminosity ($\sim$ 0.4\%) as a model constraint.  In either case,
assumptions are being made:  the solar neutrino events detected by Super-Kamiokande
and Borexino come from neutrinos produced in the solar core eight minutes earlier, while the luminosity
is associated with fusion in the core from an earlier time.  These measurements are related only if
our Sun burns in steady state.   Specifically, the Kelvin-Helmholtz time for the Sun is approximately 15 My.  
Alternatively, in models where the
Sun's core is perturbed in a specific way -- the Solar Spoon \cite{Dilke}  provides an example  -- the impact on luminosity
is seen $\sim$ a million years later.  A steady-state Sun burning in hydrostatic equilibrium is an
assumption, not established fact.   Our $\sim$ 60 years of experience with solar neutrinos provides no baseline 
for constraining luminosity variations on such timescales, and consequently
no meaningful experimental test of the SSM assumption of a steady-state Sun.

Alternatively, the luminosity constraints can be broken by new physics.  If in the high-temperature core
light particles are produced that escape the Sun without further scattering, they would carry
off energy (just as neutrinos do).  Consequently the core would have to burn at a somewhat higher 
temperature to maintain hydrostatic equilibrium, compensating for the lost energy.

Various indirect arguments can be used to check the luminosity constraint: for example, if the core
temperature is elevated by 1\%, the ratio of $^8$B/pp neutrino fluxes will as well, and the change
with be $\sim$ 20 times larger.  To the extent that the change exceeds SSM uncertainties -- this flux
ratio also depends on nuclear cross sections, the heavy element abundance in the core, and other 
parameters not precisely known -- this flux comparison could be useful as an indirect test of the luminosity constraint.

However, such a test would not be fully equivalent to a direct test of the equivalence of the Sun's
weak and electromagnetic luminosities \cite{vissani,vescovi}, namely
\begin{equation}
\displaystyle{ 4 \pi R_\mathrm{E-S}^2 ~\sum_i \Phi_i^\nu \left[ {\epsilon_{4 \mathrm{p} \rightarrow ^4\mathrm{He}} \over 2} - \langle E_i^\nu \rangle   \right] = L_\odot }~.
\end{equation}
Here $\epsilon_{4 \mathrm{p} \rightarrow ^4\mathrm{He}}  $ is the energy released when four protons fused to form $^4$He, $R_{E-S}$ is the Earth-Sun separation,
and $L_\odot$ (the right-hand side) is the rate of energy lost by the Sun due to photon emission.  The left-hand side calculates the rate of energy
deposition in the Sun:  the total solar neutrino rates are calculated from the terrestrial fluxes, then the energy deposition determined,
noting that two neutrinos are produce for every 4p $\rightarrow ^4$He conversion, and subtracting the energy $\langle E_i^\nu \rangle$ carried off by the
neutrinos, as this energy, about 1\% of the total, is not deposited.

Unfortunately, while the right-hand side is determined to $\sim$ 0.4\%, the left-hand side is only known to $\sim$ 10\%,
owing principally to the lack of a high precision measurement of the pp neutrino flux.   The lack of such a measurement 
has so far prevented a meaningful test of a key assumption of the SSM, a steady-state Sun burning in hydrostatic equilibrium
with no energy losses other than neutrinos.

Another key assumption of the SSM may be connected to a controversy we have already mentioned,
the solar metallicity problem.  In the SSM, at the onset of hydrogen burning 4.6 Gyr ago, the
Sun is taken to be homogeneous.  This assumption is based on the observation that the contracting gas cloud that formed
the Sun passed through the convective Hayashi phase.  Consequently inhomogeneities that might have existed in the gas cloud
would have been destroyed by the mixing.
As the SSM computes changes in composition generated during the Sun's subsequent evolution -- this includes
$^4$He synthesis in the core and the slow diffusion of metals with lower charge-to-mass ratios toward the core --
one can fix the SSM's initial metallicity by matching contemporary observables.

There are two opportunities to do so.  In helioseismology one measures solar surface oscillations, as deduced from the Doppler
shifts of spectral lines, some of which are associated with acoustic modes confined to the convective zone, and others reflect modes
that propagate deep into the Sun's interior.  Fourier inversions of the data have determined the sound speed
as a function of the solar radius to sub-1\% precision throughout most of the interior, as well the depth of the convective zone and related properties.
The sound speed is sensitive to the interior metallicity.

One can determine the metallicity of the Sun's photosphere by analyzing the Sun's photo-absorption lines.  Models are needed to interpret the data,
as line shapes and widths are distorted by the solar medium.  For many years medium effects were taken into account by modeling the photosphere as a 1D slab.  The resulting
photospheric metallicity was in good accord with that deduced from helioseismology, seemingly confirming the procedures followed.  However, more
than a decade ago 3D models of the photosphere were developed that
yielded significantly improved line widths and greater consistency among abundances
determined from different C and O line sources \cite{Asplund0}.  However, the improved analysis 
lead to a $\sim$ 25\% downward revision in convective-zone metal abundances.   The new metallicity yields an interior
sound speed profile in significant disagreement with that determined from helioseismology.

Thus the solar metallicity problem can be summarized:  There is a significant disagreement between SSMs that are optimized to agree with 
solar interior properties, determined through helioseismology, and those optimized to agree with solar surface photo-absorption lines.  
The former -- GS98-SFII an example -- predict  high-Z solar cores,
while the latter -- AGSS09-SFII is an example -- predict  low-Z cores.  

In the last decade solutions to this problem have been sought.  There have been small changes in 3D abundances determinations \cite{Asplund},
and new derivations of opacities \cite{Opacity}, which could in principle compensate for the impact of a low-Z Sun on helioseismology.
Some improvements have occurred 
in the atomic physics used in the photo-absorption line analysis \cite{Bergemann}.   Yet in large part, the controversy remains.

An alternative possibility has been suggested: that both measurements are correct, the Sun's convective zone metallicity is
indeed lower than that of its interior, and consequently that the SSM's assumption of a homogeneous zero-age Sun is not valid.  The authors of
\cite{HaxtonSerenelli} pointed to a possible mechanism for creating an inhomogeneous Sun subsequent to the Hayashi phase.  Others have also
considered related possibilities \cite{Guzik}.  When the Sun is
95\% formed, the remaining 5\% of gas has formed into a thin protoplanetary disk, carrying much of the solar system's angular moment.  
Most of the ice and dust within the disk are concentrated in the disk's midplane.  As the planets form they form sweep out material from the midplane,
becoming significant reservoirs of metal.  This excess, carried largely by the gaseous giants, corresponds to between 50-90 earth masses.
The disk gas not incorporated in the planets would thus be depleted of metals, comprised of H and $^4$He.  Consequently, as gas within 30 AU of the Sun is expected to
accrete onto the Sun, this H- and He-enriched gas could potentially dilute the convective zone.  The early Sun grows its radiative core over a period of about
30 My, expanding from the center outward, as the Sun's convective envelope -- which does not mix with the radiative core --  shrinks to its modern configuration, where it comprises about 2\% of the
Sun by mass.  If the depleted gas is deposited toward the end of this period, the resulting dilution of the convection zone would be sufficient
to account for the solar metallicity problem.

This suggestion is supported by some intriguing recent results.  Using data from Gaia, an analysis of the inner disks of 26 T Tauri stars -- systems similar to the early
Sun -- very large depletions in the carbon content of the accreting gas were observed, ranging up to a factor of $\sim$ 40 \cite{McClure}.  This result is consistent 
with recent theory:  Modeling of the disappearance of volatile species from the gas phase of the protoplanetary disk has been performed with both accretion
and radial drift included.  The results show that carbon-bearing molecular species like CO and CH$_4$ are always depleted \cite{Booth}.

If a host star's metallicity is altered by the formation of planets, there would be implications beyond our solar system.  Is there some way to test the
hypothesis that the Sun's convective zone and interior metallicity's differ?   In \cite{HaxtonSerenelli} it was pointed out that a direct measurement 
of core metallicity could be made, virtually free of solar model uncertainties, including issues like opacities.  The idea exploits the additional 
linear metallicity of CN neutrino production, as the cycle is catalyzed by metals.  A ratio can be formed between the CN neutrinos and the $^8$B neutrinos,
with the later serving as a ``thermometer" for the core, that isolates this linear dependence.  The Borexino Collaboration's remarkable recent measurement, which is
expected to improve further, has shown the feasibility of this measurement.  In order to measure core metallicity to $\lesssim$ 10\%, a new experiment
with a fiducial volume 10 times larger, $\sim$ kiloton, might be needed.  Such an experiment would need to replicate Borexino's exceptional radio-purity and thermal
stability.  There have been discussions about mounting a large scintillator-based experiment at the China Jinping Underground Laboratory.

\section{Supernova Neutrinos}
A cabal of the weak interaction and gravitation dooms massive stars.
Where neutrinos are passive messengers of the physics in the interior of the Sun, they
are stealthy accomplices to murder when it comes to stars with masses in excess of $\sim 10\,{\rm M}_\odot$.
These more massive stars require more pressure support in hydrostatic equilibrium. In turn, that implies a higher central temperature.
Coulomb barriers make thermonuclear reaction rates extremely sensitive to temperature and, consequently, massive stars burn 
through their initial hydrogen stock quickly compared to the main sequence lifetime of the Sun \--- $10$ billion years for the Sun, but only some
millions of years for a $25\,{\rm M}_\odot$ star. When the hydrogen is exhausted, the core contracts
until the density and temperature are reached where 3$\alpha \rightarrow
^{12}$C can take place.  The helium is then burned to exhaustion.
This pattern (fuel exhaustion, contraction, heating, and ignition of the 
ashes of the previous burning cycle) repeats several times,
leading finally to the burning of Si to Fe, if not interrupted first by instability and collapse.

In a sense the higher temperatures at each successive nuclear burning stage belie what is really going on.
Higher temperatures also mean more neutrino energy emission from deep in the star. While nuclear reactions proceed at a furious pace, liberating nuclear energy, 
at the same time neutrinos take away energy and entropy, in essence \lq\lq refrigerating\rq\rq\ the star. It is true that at the endpoint of nuclear burning the iron ashes left over are at a temperature nearly a thousand times
that of the solar core. However, the entropy there is also nearly ten times lower than that of the Sun. The core of this massive star is thermodynamically well-ordered and \lq\lq cold,\rq\rq\ with the nucleons locked up in iron-peak nuclei in nuclear statistical equilibrium (NSE). (Later, during the collapse, those nuclei may be arranged in a frozen Coulomb crystalline solid!) The pressure support for the star at this point is coming almost entirely from degenerate electrons,
with average electron energies ($3/4$ of the electron Fermi energy) nearly an order of magnitude larger than the temperature. The electrons are moving at almost light speed. The mass of this core is essentially the Chandrasekhar mass $\sim 1.4\,{\rm M}_\odot$.

A self-gravitating equilibrium configuration is trembling on the verge of instability whenever the pressure support is coming from particles with relativistic kinematics.
The nonlinear nature of gravitation, plus the capture of electrons by protons and attendant NSE changes, join forces to guarantee instability, shoving the core over the edge.
Collapse is inevitable. 

The collapse of this earth-sized core with a mass that detailed numerical simulations show is $\sim 1.2\,{\rm M}_\odot$ to $1.6\,{\rm M}_\odot$, proceeds apace, falling at an appreciable fraction of the free-fall rate.  The collapse may be halted when the nucleons touch, at nuclear density, where the nonrelativistic nucleons suddenly dominate the pressure. This \lq\lq bounce\rq\rq\ at super-nuclear densities launches a shock wave and creates a hot proto-neutron star. 
Or the collapse may never be halted, creating a black hole. In the former case, the gravitational binding energy ($\sim {10}^{53}\,{\rm ergs}$) of the final configuration, some ten percent of the {\it rest mass} of the core, is 
converted by the weak interaction during the $\sim 1\,{\rm s}$ duration of the collapse into seas of neutrinos of all types. This scenario for stellar death, a Core Collapse Supernova (CCSN), is the perfect engine for creating titanic bursts of neutrinos. 

Ultimately, the energy of a core collapse supernova explosion comes from gravitation. The details of how this scenario creates a supernova {\it explosion} with a total optical and kinetic energy of $\sim {10}^{51}\,{\rm ergs}$ remain a subject of intense research \cite{VartBur20}. In essence, we need to figure out how nature couples one percent of the gravitational binding energy released in the collapse into the explosion. Since $99\%$ of that gravitational binding energy appears as neutrino radiation, neutrinos are the prime suspects for how that energy gets transferred into the mantle and envelope of the star.
Short of understanding the explosion mechanism, the neutrino bursts associated with a CCSN are among
the most interesting sources of neutrinos in astrophysics \cite{models}.  


\subsection{Core Collapse and the Neutrino Burst}

It was Hans Bethe and his collaborators \cite{BBAL} who first recognized the particular importance of entropy as the key characteristic that dictates how the interplay of the weak interaction and nuclear physics proceeds in gravitational collapse. The conditions in the iron core at the onset of collapse are well ordered and, thermodynamically speaking, \lq\lq cold.\rq\rq\ The central density in units of ${10}^{10}\,{\rm g}\,{\rm cm}^{-3}$ is $\rho_{10} \approx 0.4$, while the temperature is $T \sim 0.7\,{\rm MeV}$, and the entropy-per-baryon in units of Boltzmann's constant $k_{\rm b}$ is $s \sim 1$.
The net number of electrons per baryon in terms of the number densities of electrons, positrons and baryons is $Y_e = (n_{e^-}-n_{e^+})/n_{\rm b}$. At this initial stage of collapse $Y_e \sim 0.42$, implying an electron chemical potential (Fermi energy) $\mu_e \sim 11.1\,{\rm MeV}\,{\left( \rho_{10}\, Y_e\right)}^{1/3}\sim 6\,{\rm MeV}$.

Low entropy, as we have at the outset of collapse, means that the baryonic component, the neutrons and protons, reside inside large nuclei for the most part. Free neutrons and protons, unbound in nuclei, and running helter-skelter, each with three translational degrees of freedom is a disordered, higher entropy state. By contrast, forcing those nucleons into big lumps (nuclei) which move around as collective units reduces the number of degrees of freedom accessible at the temperatures of relevance in collapse. This is the lower entropy regime that reigns in the \lq\lq in fall\rq\rq\ phase of stellar collapse.

The electron Fermi energies $\mu_e$ are prodigious during collapse, rising from $\sim 10\,{\rm MeV}$ to above $\sim 50\,{\rm MeV}$ later when the nuclei merge into bulk nuclear matter. 
%
This makes electron capture on protons favorable,
\[ e^- + p \rightarrow n +\nu_e, \]
\[ e^- + A(Z,N) \rightarrow A(Z-1,N+1)+\nu_e\]
In fact, only a tiny fraction of protons will be free particles, the rest
residing inside heavy nuclei, e.g., with mass number $A=Z+N$, the sum of the nuclear proton $Z$ and neutron $N$ numbers. This is significant, because the weak nuclear matrix element for electron capture is much larger for free protons than it is for protons inside nuclei. Moreover, in the initial stages of in-fall, the neutrinos escape the core. Those escaping $\nu_e$s produced by electron capture on free protons lower the entropy of the core. Not so for electron capture on nuclei, where nuclear selection rules favor capture into highly excited nuclear states. As a consequence, even though the neutrino produced escapes, electron capture on nuclei tends to increase the entropy, albeit modestly, with the temperature of the core rising from $T\sim 1\,{\rm MeV}$ to $\sim 2\,{\rm MeV}$. In either case, the escaping neutrinos carry off electron lepton number, and the entropy remains low.  

As the collapse proceeds, the mean nuclear mass number rises, and these nuclei become more neutron-rich as $Y_e$ falls. Eventually, when the density of the core reaches $\rho_{10} \sim 100$, the mean nuclear mass number $\langle A\rangle$ becomes so high that neutral current neutrino coherent scattering, whose cross section scales as $\sigma \propto \langle A^2 \rangle\, E_\nu^2$, causes the neutrino mean free path to fall below the core size, \lq\lq trapping\rq\rq\ the neutrinos. Thereafter the neutrinos quickly thermalize and approach chemical equilibrium (sometimes termed beta equilibrium), where the chemical potentials of electrons, neutrons and protons and $\nu_e$s are related by 
\[ \mu_e - \mu_{\nu_e} = {\hat{\mu}} + \delta{m_{\rm np}}, \]
where $\mu_{\nu_e}$ is the $\nu_e$ chemical potential and where the difference of neutron and proton {\it kinetic} chemical potentials is ${\hat{\mu}} = \mu_{\rm n}-\mu_{\rm p}$. Here $\delta{m_{\rm np}} \sim 1.293\,{\rm MeV}$ is the difference of neutron and proton rest masses. In NSE, $\hat{\mu}$ is, roughly, the difference between the neutron and proton Fermi levels {\it 
inside} the nuclei. These nuclei are very large and reside in highly excited states, with mean excitation energies $\sim a\,T^2$, where the nuclear level density parameter is $a\sim A/8\, {\rm MeV}^{-1}$. For example, mass number $A\sim 100$ nuclei, typical for the conditions near neutrino trapping, will have a mean excitation energy $\sim 30\,{\rm MeV}$ to $\sim 50\,{\rm MeV}$ above their ground states. NSE means that such excited states de-excite via photon and particle emission at the same rate as they are re-excited by the same processes, and both rates are very fast compared to dynamical times. However, the highly excited nuclei can de-excite by virtual $Z$-boson emission, producing neutrino-antineutrino pairs of all flavors. This process, along with plasmon decay and the Pauli blocking-hindered electron bremsstrahlung of virtual Z's and $\nu$-$\bar\nu$-pairs begins the build up of seas of mu and tau neutrinos.

Initially the whole core falls homologously, with in-fall speed proportional to radius. But as $Y_e$ and electron pressure fall because of electron capture, only an inner core can remain in homologous collapse. The outer boundary of this homologous core is where the in-fall speed equals the local sound speed. The mass of this causally connected inner core, $M_{HC} \sim 0.6 - 0.9\, {\rm M}_\odot$, is essentially the instantaneous Chandrasekhar mass, proportional to the local value of $\langle Y_e^2\rangle$.

When the collapsing homologous core reaches densities in excess of $\rho_{10} > {10}^{3}$ the large nuclei merge into what eventually becomes bulk nuclear matter. The collapse of the homologous core is halted abruptly when the nucleons touch at $\rho_{10} > {10}^{4}$. The outer core, comprising the remainder of the old initial Fe-core outside of the inner core, is in-falling at supersonic speed and slams into the halted inner core. A shock wave is formed at this boundary.
Put another way, since nuclear matter is rather incompressible ($\sim$ 200 MeV/f$^3$),
the nuclear equation of state is quite effective in halting the collapse:
maximum densities of 3-4 times nuclear are reached, e.g.,
perhaps $6 \times 10^{14}$ g/cm$^3$.  The innermost shell of matter
reaches this supernuclear density first, rebounds, sending a 
pressure wave out through the inner core.  This wave
travels faster than the infalling matter, as the inner 
core is characterized by a sound speed in excess of the in-fall
speed.  Subsequent shells follow.  The resulting pressure
waves collect at the edge of the inner iron core -- the radius at which
the infall velocity and sound speed are equal.  
As this point reaches nuclear density and comes to
rest, a shock wave breaks out and begins its traversal of the 
outer core. 

Initially the shock wave may carry an order of magnitude more energy
than is needed to eject the mantle of the star (less than 10$^{51}$
ergs).  But as the shock wave travels through the outer iron core,
it heats and \lq\lq melts\rq\rq\ the large nuclei that cross the shock front. 
The entropy-per-baryon is $s\sim 1$ ahead of the shock, 
but $s\sim 10$ behind it, in the shocked matter. 
This shifts the composition in NSE from heavy nuclei to free neutrons and protons 
and a few alpha particles. However, pulling nucleons out of those heavy nuclei costs 
$\sim$ 8 MeV/nucleon, or roughly $\sim {10}^{51}\,{\rm ergs}$ for every $0.1\,{\rm M}_\odot$ 
of material transiting the shock front.  Additional energy is lost by neutrino
emission, which increases after the melting.   These losses are comparable to 
the initial energy carried by the shock wave.  Most simplified
(e.g., one dimensional) numerical models fail to produce a successful ``prompt"
hydrodynamic explosion, for this reason.   

Much of the modeling in the past two decades has focused on the role
of neutrinos in reviving this shock wave, a process that becomes more 
effective in multi-dimensional models that account for convection and other 
hydrodynamic transport of matter.  In this delayed mechanism,
the shock wave stalls at a radius of 200-300 km, some tens of milliseconds
after core bounce.  But neutrinos diffusing out of the proto-NS
react frequently in the nucleon gas left in the wake of the shock wave,
depositing significant energy.  Over $\sim$ 0.5 seconds the increasing pressure
due to neutrino heating of this nucleon gas helps
push the shock outward.  This description is over-simplified -- a variety of
contributing effects are emerging from numerical simulations -- but there is
wide agreement that energy deposition by neutrinos is an essential ingredient
for successful explosions.

Regardless of explosion details, neutrinos dominate SN energetics.
The kinetic energy of the explosion and
SN's optical display account for less than 1\% of the available energy.
The remaining 99\% 
of the 3 $\times 10^{53}$ ergs released in the collapse is 
radiated in neutrinos of all flavors.  The timescale governing the leakage 
of trapped neutrinos out of the proto-NS
is about three seconds. 
The energy is roughly equi-partitioned
among the flavors (a consequence of reactions
among trapped neutrinos that equilibrate flavor).  The detailed
decoupling of the emitted neutrinos from the matter -- which occurs at a
density of about $10^{11}-10^{12}$ g/cm$^3$ --does depend on flavor.
This leads to differences in neutrino temperatures, with electron
neutrinos being somewhat cooler ($T \sim$ 3.5 MeV) than the 
heavy-flavor neutrinos ($T \sim$ 6 MeV). The steep density fall off near the surface of the proto-neutron star
helps create a ``neutrinosphere", where neutrinos decouple from matter, 
analogous to the familiar photosphere for optical emissions. This is only a loose 
analogy, however, and in practice neutrino decoupling does not occur at a sharp radius.

On February 23,
1987, a neutrino burst from a CCSN in the Large Magellanic Cloud was
observed in the proton-decay detectors Kamiokande and IMB \cite{snnus}.   The optical
counterpart reached an apparent magnitude of about 3, and could be
observed easily in the night sky with the naked eye.  This SN originated
160,000 light years from Earth.  Approximately 20 events were seen in the 
Kamiokande and IBM detectors, spread over approximately 10 seconds.  
Within the limited statistics possible with these first-generation detectors, the
number of events and the burst duration were consistent with standard 
estimates of the energy released by and cooling time of the SN.  

Today, due to massive detectors like Super-Kamiokande (and soon Hyper-Kamiokande and DUNE),
the neutrino ``light curve" produced by a galactic CCSN would be mapped out in
great detail, with $\sim 10^4$ events recorded.   At Kamioka depths the measurements would be
effectively free of background, so that the light curve could be followed for 30 seconds or more \cite{DUSELhandbook}.
This would provide valuable information on the total energy release, thereby constraining the mass and 
radius of the progenitor -- though see comments below on potential complications due to our uncertain
knowledge of the flavor of the emitted neutrinos.  Abnormal features in the time structure of the neutrino
burst could be signatures of a phase changes in the matter or of a delayed
collapse to a black hole, both phenomena that could be triggered by the evolving lepton number of the matter.



\subsection{Supernova Neutrino Physics: Microphysics and Modeling}
\label{sec:SNP}
CCSNe provide an extraordinary laboratory for studying properties of
neutrinos under conditions that cannot be replicated in the laboratory.  Conversely, the
enormous flux of neutrinos produced in these events can be utilized as a probe of
the dynamics and other underlying physics of SN cores.  Just as 
in the solar neutrino problem, they allow us to look inside the collapsing star in real time.
Some of the themes of CCSN neutrino physics include:
\begin{itemize} 
\item As $\sim$ 99\% of the collapse energy is radiated in neutrinos, one can in principle
deduce the the binding energy of the NS from neutrino flux measurements,
provided other parameters (such as the distance to the SN) are sufficiently well known.
\item Because of the trapping of neutrinos, the proto-NS core that forms is hot
and puffy, with a radius on the order of 50 km, but as neutrinos are radiated, this core cools and
contracts, evolving toward conditions that we can study in cold, isolated NSs.
It is apparent that by monitoring the neutrinos, we can track this evolution and consequently
deduce properties of the equation-of-state (EoS) of the matter under conditions of 
changing temperature and lepton number.   We know very little about the nuclear EoS
at the extremes of density and isospin that are relevant to SNe.   Thus data
from such events, together with new information on the EoS now available from observations
of isolated, cold two-solar-mass NSs and from NS mergers, can help us extend our
knowledge of the phase diagram of nuclear matter into regions not otherwise accessible.
\item Neutrinos from galactic CCSNe will not be obscured by intervening matter or
dust, unlike optical signals.  Thus a program monitoring SN neutrino bursts will, in time,
provide a reliable measure of the contemporary rate of galactic core collapse.
\item CCSN rates in nearby galaxies with active star formation regions are expected to be
considerably higher than in the Milky Way.  The detection of neutrino from such events 
would involve very small number of counts and would require very massive detectors --
but detection does not seem impossible.
\item With the advent of gravitational wave (GW) astronomy, both the neutrino burst and the gravitational
wave signature \cite{Abdikamalov} could provide prompt notice that a CCSN event is occurring, while restricting the region
of the sky where the event resides to a reasonably small solid angle.  Due to the time required
for the shock wave to reach the progenitor star's surface, the optical display may be delayed for
times of up to several days, depending on the progenitor type.  Thus the neutrino and GW signals
together with the field's growing capability in rapid scanning could allow us to see in real time the
shock breakout, potentially providing valuable new information on the progenitor.
\item There exists a so-far undetected diffuse background of SN neutrinos, produced
by all past CCSNe.  As discussed below, there is an effort underway at Super-K 
to detect these neutrinos, and in the future detectors at the
near-megaton scale should be able to see a few events from this source.  Detection of
these neutrinos would place a important constraint on the inventory of massive stars
undergoing core collapse, from the first epoch of star formation until now.
\item Supernovae are one of the most important engines for nucleosynthesis, controlling
much of the chemical enrichment of the galaxy.  As described in Sec. \ref{sec:nucleosynthesis},
neutrinos are directly and indirectly involved in this synthesis.  SN neutrino observations
that constrain the frequency, flux, and flavor of neutrinos would reduce uncertainties in
nucleosynthesis estimates.
\item As noted above, most of the energy released in core collapse is radiated as neutrinos over the
first several seconds, and neutrino emission at a lower level continues as the proto-NS
cools and radiates away its lepton number.  Direct observation of this cooling should be possible from
tens to perhaps 100 s, depending on distance to and other properties of the galactic CCSN.
However, while there are considerable uncertainties in
estimates, neutrino processes may continue to dominate NS cooling for times
$\sim 10^5$ years.  Thus understanding neutrino cooling in CCSNe is part of a larger program
that includes the cooling of recently created NSs.
\item  The neutrino burst could include other sharp features in time, marking interesting
astrophysics.  One such event is the passage of the shock wave through the edge of the outer core, the neutrino sphere, some $10\,{\rm ms}$ after core bounce. The melting of nuclei to nucleons with the passage of the shock wave through
the outer iron core is predicted to produce an electron capture-induced spike in the neutrino luminosity, lasting for
a few milliseconds, carrying an energy of $\sim 10^{51}$ ergs, and dominated by $\nu_e$s.  Continued accretion onto the NS surface could produce a
collapse to a black hole, and consequently a sudden termination in neutrino emission.
\item  Supernova cooling times place constraints on new physics associated with particles
that also couple weakly to matter.  For example, a light scalar called the axion could,
in principle, compete with neutrinos in cooling a supernovae.  The requirement that axion
emission not shorten the cooling time too much, which would be in conflict with SN1987A
data, constrains the mass and coupling of this hypothesized particle.
\item The ``neutrinosphere" for a CCSN -- the radius at which neutrinos last scatter, after which
they free-stream from the SN to Earth -- corresponds to a density $\sim 10^{12}$ g/cm$^3$.
Thus neutrinos with fixed distributions and flavor experience flavor-transformation potentials stemming from forward scattering on electrons and {\it other neutrinos} which far larger than 
those available in the Sun. However, the analogy with the Sun ends there, as neutrino-neutrino forward scattering renders the flavor conversion problem highly nonlinear. We discuss this below.
\end{itemize}
Several of the items above deserve further
comment, so additional discussion is included below.\\

\noindent
{\it Total Flux Determinations:} The determination of the total neutrino flux at Earth from a CCSN is not an easy job, as the responses
of most detectors are flavor dependent.  Thus in general one would need a strategy similar to that employed
by SNO for solar neutrinos -- several detection methods with distinct flavor sensitivities to unravel
the flavor content of solar neutrinos -- to reliably determine the total flux and its $\nu_e$, $\bar{\nu}_e$,
$\nu_\mu/\nu_\tau$, and $\bar{\nu}_\mu/\bar{\nu}_\tau$ 
components.   In some regards the field has regressed: when SNO was operating, we had a kton detector with a neutrino detection mode
(breakup of the deuteron) that was equally sensitive to all flavors.  Even then, as this channel provides
no spectral information, the determination of the total neutrino flux requires analysis.  Constraints from other detectors, such as
neutrino-electron scattering rates from a water detector, would be helpful in this effort.

There is a dedicated SN neutrino detector utilizing lead as a target, the HALO observatory \cite{HALO}, operating in SNOLab.  The signals are neutrons
produced by both neutral current and charge current breakup of lead.  Other neutral current reactions
have been discussed in connection with SN neutrinos.  In carbon-base scintillators the excitation of the
15.11 MeV 1$^+$1 state leads to a distinctive signal, a 15.11 MeV mono-energetic photon.   However, for a thermal Fermi-Dirac 
neutrino spectrum with a temperature of 5 MeV, the flux-averaged cross section is nearly a factor of 50 smaller than that
of the corresponding charged-current $\bar{\nu}_e$+p $\rightarrow$ n+e$^+$ cross section that dominates in water
detectors like Super-Kamiokande.   The resulting need for a large-volume detector is a significant drawback.  The high threshold
(reflected in the small cross section) could be helpful, however, in a flavor analysis that combines results from various detectors,
as this reaction samples the high-energy tail of the neutrino distribution, where heavy-flavor neutrinos might dominate.

Recently coherent neutrino scattering has been seen in the laboratory for the first time \cite{coherent},  in experiments using CsI and LiAr.
The neutrinos were produced in stopped pion decay, a reaction that produces a spectrum not too different from that expected
from a CCSN.  The cross section is large due to coherence, which scales approximately as N$^2$ where N is the neutron number of the
target nucleus.  The threshold is only limited by the ability to detect the recoil of the nucleus, the scattering signal.  There
have been numerous recent papers \cite{coherentpapers} on using coherent scattering to detect supernova neutrinos, including one by the
DarkSide/Argo Collaboration \cite{DarkSide} that concludes one could monitor CCSN out to the Small Magellanic Cloud ($\sim$ 200,000 light years from Earth)  by observing events 
in a 360-ton fiducial volume.  The fraction of the neutrino energy carried off by the nucleus is proportional to $\sim {E_\nu \over M_T}$,
where $M_T$ is the target mass: most of the events in a LiAr detector would have recoil energies $\lesssim$ 10 keV.\\

\noindent
{\it Nuclear Equation of State:}  The nuclear EoS relates the energy and pressure of a volume element to the temperature,
density, and composition of the matter, specifically its isospin.  The EoS determines the bulk properties of the nuclear matter, including
the saturation density at zero temperature, the binding energy per nucleon, the compressibility, and the symmetry energy and its slope, which
describe the energy cost of deviations from N=Z matter.  In a supernova the EoS influences the depth of the core bounce -- the maximum
central density that is reached during the core bounce -- the energy release, and properties of the neutron star, such as its mass and radius,
that may be the end state
of the collapse.   As central densities of $\sim 6 \rho_0$, where $\rho_0$ is the nuclear saturation density, could be reached in the core bounce, nuclear matter properties relevant to SNe 
are not simply related to nuclear properties measured in the laboratory.   The required extrapolations are quite uncertain because of the
evolving isospin (a cold neutron star core may have a 
proton fraction $Y_p \sim 0.1$) and because the production and role of hyperons is difficult to estimate.

In astrophysics the EoS links phenomena such as supernovae, cold neutron stars, and binary NS mergers.  The EoS is also
central issue in laboratory nuclear physics, relevant to phenomena ranging from the collective modes of ordinary nuclei to the properties of the fireball
produced in heavy ion collisions at the LHC and RHIC.   The uncertainties noted here impact a broad range of astrophysics and physics.

As a detailed discussion new astrophysical constraints on the nuclear EoS would take us beyond the scope of this chapter, we refer readers to a recent
summary \cite{2105.08688} of relevant data, including
\begin{itemize}
\item Precise timing measurements of the pulsars PSR J0348+0432 \cite{Antoniadis2013}, PSR J1614-2230 \cite{DeMorest2010,Fonseca2016}, and PSR J0740+6620 \cite{Cromartie2019} 
yield neutron star masses of  2.01$\pm$ 0.04, 1.928 $\pm$ 0.017, and 2.14$^{+0.10}_{-0.09}$ M$_\odot$,  establishing that the maximum NS mass is $\gtrsim$ 2M$_\odot$;
\item Continued observations of the double neutron star system PSR J0737-3039 \cite{Burgay2003,Lyne2004} could, over time, determine
the neutron star moments of inertia to good accuracy \cite{LattimerShutz2005};
\item X-ray observations by the NICER and XMM telescopes constraining the radius of PSR J0740+6620 \cite{NICER};
\item Gravitational wave observations of the NS mergers GW170817 and GW190425, and observation of their electromagnetic counterparts.
\end{itemize}

\noindent
{\it The Diffuse SN Neutrino Background:}  While SNe within our galaxy are rare events, occurring with a frequency $\sim$ 1/100y,
there is also a continuous background of such neutrinos, representing the net effect of the $\sim 10^{17}$ CCSNe that have occurred in the
cosmos since the first massive stars formed, some 300 My after the Big Bang.  Detection of these neutrinos would place an important constraint
on the cosmological history of massive star formation and on CCSN-associated nucleosynthesis.   From a cosmological model of star formation,
assumptions about the range of stellar masses that undergo core-collapse, and calculated red shifts, one can estimate the flux and spectrum
of the relic neutrinos.  While only limits have been established on this flux, the diffuse supernova neutrino background (DSNB) -- also called the
relic SN neutrino flux -- may be measurable.   The rate of interactions in Super-Kamiokande is predicted to be  $\sim$ 3/y \cite{Beacom,Lunardini}.

The key to detecting these neutrinos is their separation from background events, including those associated with solar and atmospheric
neutrinos.  In a water detector, the dominant detection channel is
\[ \bar{\nu}_e + p \rightarrow n + e^+ \]
The detection strategy is measurement the positron together with the produced neutron, as there a few background sources that can mimic
this coincidence.  Beacom and Vagins suggested a strategy for accomplishing this, the doping of Super-Kamiokande with Gd, which has
an enormous thermal neutron capture cross section (257,000 barns) \cite{Vagins}.  After a decade exploring whether a water detector could be doped with
Gd without degrading the detector's performance or its materials, Super-Kamiokande dissolved 13 tons of Gd$_2$(SO$_4$)$_3$-8H$_2$O into the tank
during summer 2020, which yields a Gd concentration by weight of 0.01\% and a neutron capture efficiency of 50\%.   Results from the
first run have not been announced.\\

\noindent
{\it Modeling and the Neutrino Flux:}  The realistic simulation of SN explosions has long been a grand challenge problem is numerical astrophysics,
requiring treatment of the hydrodynamics helping to drive ejection including shock wave production and propagation, transport of energy, entropy, and 
lepton number by neutrinos, and processes like accretion, convection, and turbulence that require modeling in 3D.   The most sophisticated explosion models 
generally limit their integrations to $\lesssim$ 5 s due to cost, which is generally 
long enough to determine whether or not a successful explosion occurs. This leave a considerable 
gap between the termination of the calculation and many late-time observables that can be used to constrain models -- nucleosynthetic output,
electromagnetic signals, and even the neutrino burst given that today's largest  detectors will be able to follow that burst for tens of seconds.   (However,  
a small number of groups have developed procedures for coupling explosions to hydrodynamic codes to continuing calculations to shock
breakout.  See Ref. \cite{Sandoval21} for a recent calculation as well as references to earlier work.)
Very light progenitors explode readily -- the mechanism is largely hydrodynamic.  In heavier progenitors the shock
wave typical stalls in the outer iron core, an accretion phase ensures, then the shock is revived perhaps a second later because of the energy deposited by the neutrinos in the wake of the
shock, a process that is facilitated by the turbulent convection that moves hot matter outward, replacing it with cold matter at small $r$ than can be more readily heated by the neutrinos.

There are several excellent recent reviews that together summarize the underlying physics of CCSN and the state-of-the-art in their numerically modeling, e,g., 
\cite{Burrows21,Glas19,OConnor18,Janka16,Janka12}.  Among the interesting
issues are mechanisms by which the neutrinos couple to the matter and deposit energy, characteristics of the remnants including the ranges
of masses of the NSs produced and the progenitor properties that determine whether NSs or BHs are formed, the composition and velocities of the ejecta, the nucleosynthetic
conditions created within SNe, pulsar kick velocities,
and gamma-ray burst associations

\begin{figure}
\begin{center}
\includegraphics[width=12cm]{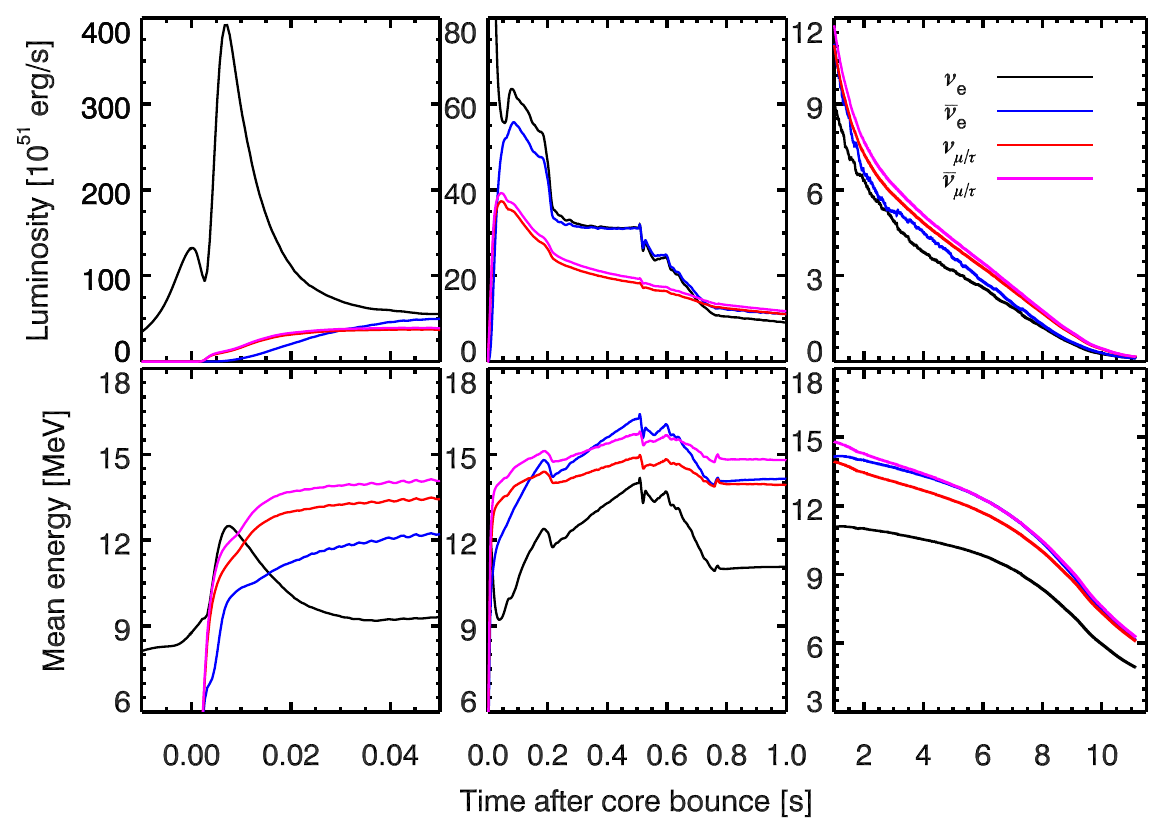}
\caption{Neutrino luminosities (upper panels) and mean energies (lower) are shown for the shock-breakout (left), accretion (middle),
and proto-NS cooling (right) phases of the CCSN explosion of a 27 M$_\odot$ progenitor.  Results for $\nu_e$s are shown in black,
$\bar{\nu}_e$ in blue, one species -- a $\nu_\mu$ or a $\nu_\tau$ -- in red, and one species -- a $\bar{\nu}_\mu$ or a $\bar{\nu}_\tau$ in
magenta.  See text for additional details.  Reprinted with permission from Ref. \cite{Janka1702}
\label{fig8:Janka}.}
\end{center}
\end{figure}

The key issue for us here is the neutrino ``light curve" produced by the models:  what can we learn from observing
neutrino luminosity as a function of time, the neutrino energetics, and the flavor decomposition of the burst?   Figure \ref{fig8:Janka},
from Janka \cite{Janka1702}, 
summarizes the time evolution of the relevant neutrino species in terms of both luminosity and  mean energy, for $\nu_e$s, $\bar{\nu}_e$s,
$\nu_\mu/\nu_\tau$'s, and $\bar{\nu}_\mu/\bar{\nu}_\tau$s.  The calculations were done for a 27 M$_\odot$ progenitor, and results are
presented in three panels corresponding to the shock-breakout phase (left), the accretion phase (middle), and the proto-NS cooling
phase (right).

The shock breakout is associated with the passage of the shock wave though the outer iron core a few ms after core bounce.
The iron is heated by shock and is essentially melted to free protons and neutrons:  this robs the shock wave of much of its energy,
while suddenly reducing the neutrino opacity of the material.  (The initial opacity of the material is determined by coherent neutrino
scattering off Fe, with a cross section $\sim$ N$^2$, where N is the neutron number.  The opacity of full dissociated matter is thus 
reduced by $\sim$ 1/N).  Passage of the shock wave generates rapid production of $\nu_e$s by electron capture on free protons,
e$^-$+p $\rightarrow$ n + $\nu_e$, and because of the reduced opacity, these neutrinos free-stream out of the star.  Similar, any
$\nu_e$s produced as a consequence of the dropping $Y_p$ and associated neutronization during infall, would also be freed at this
time.   Though the energy carried off by these shock breakout (or deleptonization) neutrinos is relatively modest, $\sim 2 \times 10^{51}$ erg and thus $\lesssim$ 1\% of
the total released in neutrinos, the time structure is abrupt ($\sim$ 10 ms width), the luminosity is high (up to $ 4 \times 10^{53}$ erg/s),
and the flavor distinctive, almost entire $\nu_e$s.  Thus as a neutrino source for oscillation studies, the shock breakout neutrinos are quite
interesting.

The accretion phase spans the time interval between the stalling of the shock wave in the outer iron core and its revival due to the
heating caused by charge-current interactions with the nucleons left in its wake, aided by convection and accretion.  These reactions generate 
comparable luminosities in $\nu_e$ and $\bar{\nu}_e$, though with a number excess in $\nu_e$s over $\bar{\nu}_e$s.  Significant fluxes
of heavy-flavor neutrinos are also produced, generated in $\nu-\bar{\nu}$ pairs by bremsstrahlung off nucleons, somewhat deeper in the core.  The higher mean energies
of the heavy-flavor and $\bar{\nu}_e$s, apparent in the lower middle panel, follow from their weaker coupling to the matter, and consequently 
their radiation from a ``neutrino sphere" that is somewhat deeper in the star and consequently slightly hotter:  The heavy-flavor
neutrinos lack charge-current reactions off nucleons, while there are fewer proton targets  for $\bar{\nu}_e + $p $\rightarrow$ n+e$^+$ than
neutron targets for $\nu_e + $n $\rightarrow$ p+e$^-$.  In the calculation illustrated in Fig. \ref{fig8:Janka} the shock wave begins to move
outward again after $\sim$ 0.5 s, with the neutrino production transitioning to the proto-NS stage.   Over this 0.5s the mean energies of all of
the species gradually increase, though the changes are larger for the $\nu_e$s and $\bar{\nu}_e$s.  As described in \cite{Janka1702} and
references therein, the pattern of neutrino fluxes during the accretion phase varies from star to star, showing sensitivity to both the
accretion rate and the compactness of the core of the progenitor star.  In addition, convection and hydrodynamic instabilities such as the
standing accretion-shock instability can cause large-scale variations of the accretion with time.  These variations in turn can alter the neutrino flux, 
producing quasi-periodic fluctuations and lepton-number asymmetries (with respect to the angle to the observer) that could be detectable, given detectors with sufficient
sensitivity and mass \cite{Lund2012,Tamborra2013}.

During the much longer cooling phase the total luminosity (summed over all six flavors) is $\sim$ few $\times 10^{52}$ erg/s.   There is little
variation in the luminosity per flavor.  The Kelvin-Helmholtz cooling produces a smooth decline in the luminosity and mean energy in each 
flavor.  The relatively small differences between the mean energies of neutrinos of differing flavors makes certain neutrino oscillation
tests more difficult:  a more pronounced hierarchy of temperatures would produce more distinctive signals of a flavor swap, e,.g., $\bar{\nu}_\mu \leftrightarrow \bar{\nu}_e$.
The associated spectra are reasonably represented by a Fermi-Dirac distribution with $\langle E \rangle \sim 3.1 T$, though with a somewhat
pinched high-energy tail, as higher energy neutrinos have larger cross sections for scattering, and have neutrino spheres at slightly larger $r$ than those of
lower energy neutrinos.   As we have noted elsewhere, underground detectors like Super- and Hyper-K and DUNE should be able to
monitor the Kelvin-Helmholtz cooling of the core for several tens of seconds.

\subsection{Neutrino Flavor Transformation}
Neutrino flavor transformation in the core collapse supernova environment can be far more complicated than in the solar case \cite{annrev10}. Unlike in the Sun, in CCSN a component of the \lq\lq medium\rq\rq\ through which a neutrino propagates consists of {\it other neutrinos}. This means that the potentials, for example from neutrino-neutrino forward scattering, that contribute to determining how neutrinos change flavor depend on the flavor of the neutrinos! Consequently, the neutrino flavor transformation problem can be fiercely nonlinear. Moreover, in the supernova environment, as we have discussed above, neutrinos may not always freely stream, but can scatter on any targets carrying weak charge, thereby changing direction and energy. Put in simplistic single particle parlance, a single neutrino propagating along might be a coherent superposition of flavor states, but a scattering event can be like a flavor measurement, by the rules of quantum mechanics \lq\lq collapsing\rq\rq\ the neutrino state into a particular flavor state, in essence transforming flavor. Of course, we actually have a neutrino flavor {\it field} in the supernova, with the neutrinos non-linearly coupled to each other. Neutrino flavor field evolution in compact objects is, in essence, a nonlinear many body physics problem, albeit one driven entirely by the weak interaction and involving spin-1/2 fermions with relativistic kinematics with nontrivial coupling to a matter background.  This is a daunting and unsolved problem, but one which is the subject of intense research activity. 

If all neutrino types had identical energy spectra and fluxes at all times and in all locations in the supernova environment, it would not matter if flavor labels were switched. But this happy circumstance does not apply. As discussed above, there can arise significant differences in the energy distributions and fluxes of the different neutrino flavors. The neutrino flavor problem must be addressed for three key reasons: (1) most of the gravitational binding energy in the collapse/explosion process resides in the neutrino seas; (2) the way neutrinos transport and deposit energy and entropy in this environment is neutrino flavor-dependent; and (3) the neutron-to-proton ratio, a key determinant of nucleosyhthesis as we will discuss, depends on the local lepton number content and associated interaction rates, specifically on the competition between neutron-destroying and -creating weak interaction processes, for example, $\nu_e+n\rightleftharpoons p +e^-$ and $\bar\nu_e+p\rightleftharpoons n +e^+$. Clearly, the stakes are high when it comes to neutrino flavor transformation's potential effects in CCSNe.

We can make some progress on understanding neutrino flavor evolution in the supernova environment by leveraging our above discussion of neutrino mass level crossings in the Sun, and generalizing this treatment to the supernova case. In fact, nearly as soon as the MSW mechanism was recognized and applied in solar neutrino studies, it was realized that a potentially richer pattern of level
crossings would occur in the mantle of a CCSN, providing an opportunity to probe a significantly wider range of possible $\delta m^2$s \cite{MSWSN}. First, let us ignore neutrino-neutrino scattering potentials, dangerous as that might be, and consider a simple coherent MSW treatment.

To make contact
with our previous discussion of solar neutrino oscillations, we start here with perhaps the simplest
opportunity, the early shock breakout neutrinos  \cite{thompson}, the so called neutronization neutrino burst, when the shock comes through the neutrino sphere. This burst occurs at about $10\,{\rm ms}$ post core bounce and has a duration of about $10\,{\rm ms}$, and during that interval the neutrino luminosity will be prodigious, perhaps $\sim {10}^{54}\,{\rm ergs}\,{\rm s}^{-1}$.  The prompt post core bounce appearance of these neutrinos makes them a good \lq\lq clock\rq\rq\ in a detected neutrino signal.  As a potential source for exploring oscillations, they have the attractive feature of
being produced nearly in a pure flavor, $\nu_e$s.  In addition, they originate in the outer iron core under conditions 
where they are not trapped, thus accounting for their sharp structure in time.

\begin{figure}
\begin{center}
\includegraphics[width=12cm]{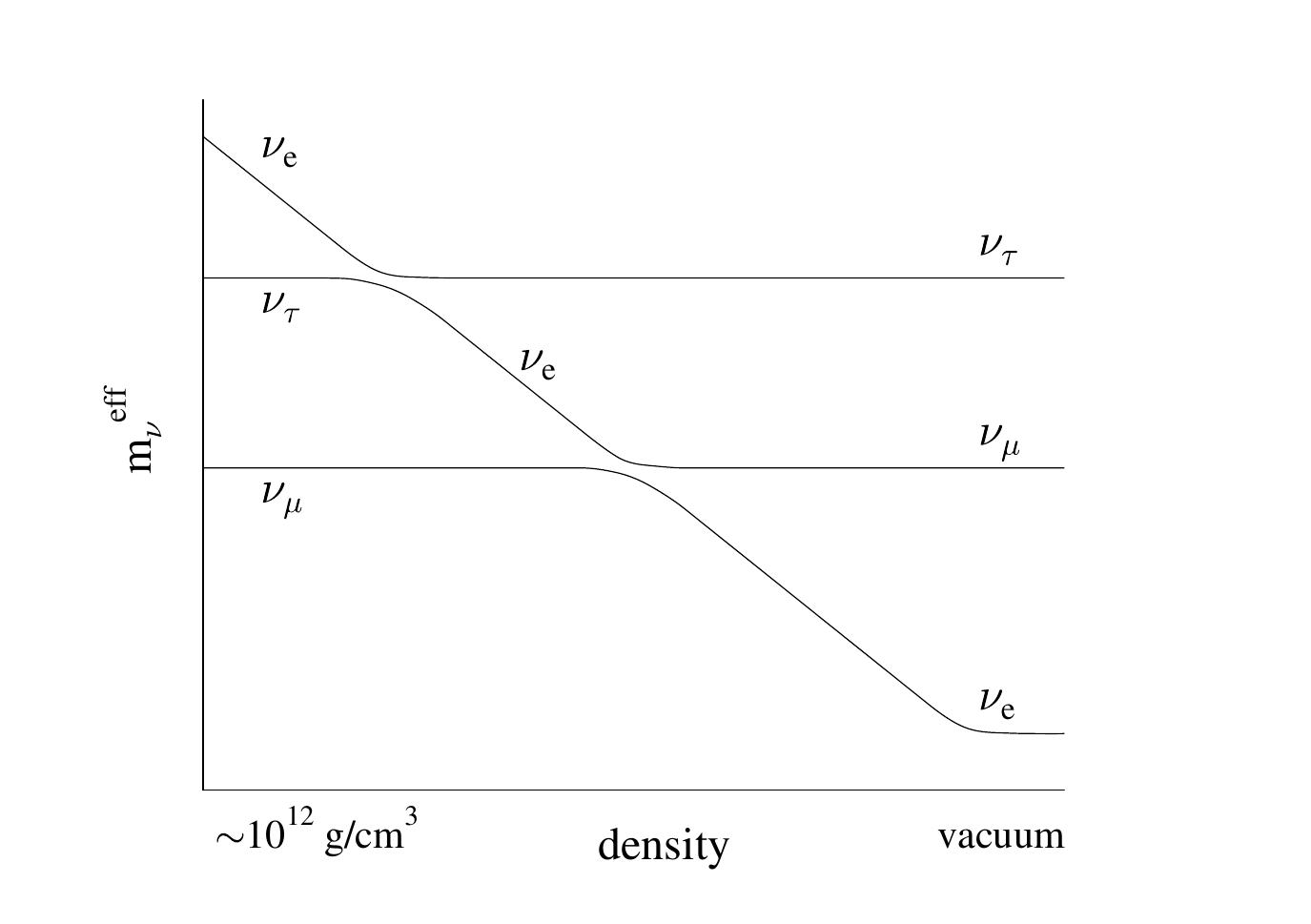}
\end{center}
\caption{A schematic illustration of the two level crossings that the $\nu_e$ would
experience, given the higher densities available in the mantle of a massive
star undergoing core collapse.}
\label{fig:threelevel}
\end{figure}

These neutrinos are produced at a density of $\sim 10^{11}$ g/cm$^3$, and consequently
experience a much larger MSW neutrino-electron forward scattering potential than do the solar neutrinos previously discussed.  This requires an
extension of our solar neutrino case, where we were free to limit our discussion to two flavors.  The
central density of the Sun, $\rho \sim 150$ g/cm$^3$, is sufficient to produce a level-crossing of the 2 and 1 local mass eigenstates
(at least for the more energetic half of the solar neutrino spectrum), where the mass difference is only $\delta m_{21}^2 \sim 7.4 \times 10^{-5}$ eV$^2$.
The crossing associated with the atmospheric mass difference, $\delta m_{31}^2 \sim 2.5 \times 10^{-3}$ eV$^2$, is not relevant,
as this crossing requires a density $\rho \sim 10^4$ g/cm$^3$.
But in a CCSN a density of $10^4$ g/cm$^3$ in found well out in the progenitor's mantle -- typically the carbon zone --
so that the three-flavor MSW level-crossing diagram of Fig.~\ref{fig:threelevel} is needed.  

The figure assumes that $\delta m_{31}^2 > 0$, one of two open possibilities.  If this is not
the case, no crossing would occur.  Thus there is at least the possibility that the shock breakout neutrinos
could be used to test whether the neutrino hierarchy is normal or inverted.  There are a number
of papers in the literature (e.g., \cite{Duan,Fullerbump}) discussing this possibility and the uncertainties that would need to
be under control, to reach a definite conclusion, including the consequences of neutrinos crossing the shock
front, as a sharp change in density can alter oscillations \cite{Fullerbump,shockdiffuse2010,turb2006,turb22013,Fried2020}, and contributions of neutrino self-interactions to
the MSW potential.


Given the obvious utility of this MSW approach, we are emboldened to take the next step and consider a general mean field treatment of supernova neutrino flavor transformation, but now one which includes the nonlinear potentials created by neutrino-neutrino scattering. In this approach we follow the flavor evolution of a given neutrino by assuming that it is experiencing an overall mean field potential arising from its interactions with all of the particles in the medium that carry weak charge. The flavor quantum kinetic equations (QKEs) \cite{QKE2014,SiglRaffelt93,Volpe2013,ZhangBurrows2013,Richers2019} that govern this evolution for a neutrino of momentum $p$ can be written as
\begin{equation}
    i {{ D {\hat{f}}(p, t)}\over{D \lambda }} =\left[ {\hat{H}}, {\hat{f}}(p, t)  \right] + i {\hat{C}}{\left[ {\hat{f}}(p, t) \right]}
    \label{QKE}
\end{equation}
where $D/D\lambda$ is a derivative along the world line of the neutrino (with Affine parameter $\lambda$, e.g., time $t$), ${\hat{H}}$ is a $3\times 3$ Hamiltonian operator arising from neutrino-electron and neutrino-neutrino {\it forward} scattering, and ${\hat{C}}$ is a $3\times 3$ collision operator encoding energy- and direction-changing neutrino scattering events and, in its off-diagonal components, encoding the flavor-decoherent scattering effects discussed above. Here ${\hat{f}}(p,t)$ is the $3\times 3$ neutrino density operator, which in the flavor basis would have matrix representation
\begin{equation}
\left[ {\hat{f}} \right]_{\rm flavor} = \begin{pmatrix} f_{\nu_e \nu_e} & f_{\nu_e \nu_\mu} & f_{\nu_e \nu_\tau}\\ f_{\nu_\mu \nu_e} & f_{\nu_\mu \nu_\mu} & f_{\nu_\mu \nu_\tau}\\ f_{\nu_\tau \nu_e} & f_{\nu_\tau \nu_\mu} & f_{\nu_\tau \nu_\tau}\end{pmatrix}.   
\end{equation}
The diagonal elements of this density matrix are the analogs of the usual neutrino occupation probability for momentum state $p$. The off-diagonal elements encode quantum coherence. If we zero these out and set the commutator $[ {\hat{H}}, {\hat{f}}]=0$ and consider only a flavor diagonal version of ${\hat{C}}$, then we recover the ordinary Boltzmann equation for neutrino occupation probabilities \--- solving that equation was the basis of the neutrino transport simulations for core collapse supernovae discussed above. On the other hand, if we set the collision operator ${\hat{C}}$ to zero, but retain the commutator, we recover a Schr\"odinger-like equation for the coherent evolution of neutrino flavor. For example, this equation, when solved for the matter potentials in the Sun, gives the result shown above in Figure 6. 

There is a similar set of equations governing the evolution of antineutrino flavor.  Strictly speaking, for Majorana neutrinos, neutrino and antineutrino flavor evolution is coupled by potentials which flip neutrino spin, tantamount to $\nu\rightleftharpoons\bar\nu$ inter-conversion. We could describe this with a $6\times 6$ generalization of the above operators. In practice spin flip conversion is small, as in the end the potentials that govern it are always proportional to the ratio of neutrino rest mass to neutrino energy, a quantity that is $< {10}^{-8}$ in typical compact object environments.

Let us first explore flavor evolution in the coherent limit, where we set ${\hat{C}}=0$. Given the small measured mass squared differences, and the generally high matter densities in CCSN and other compact object venues, the simple MSW-inspired treatment discussed above suggests than there should be scant neutrino flavor transformation during the shock re-heating accretion phase or in the later neutrino-driven wind epoch. Instead, when the coherent version of Equation 4 was first solved \cite{DFCQ2006,DFCQ2006Let} the result was shocking: large-scale neutrino flavor transformation, collective neutrino transformation, was ubiquitous. This outcome was completely unexpected, stemming from the nonlinear neutrino-neutrino forward scattering potential. Various modes of collective neutrino flavor oscillations were found. These include a bipolar mode, where systems with equal numbers of $\nu_e$s and $\bar\nu_e$s, experience nearly complete flavor conversion, and a synchronized mode. In CCSN neutrinos these modes can couple, producing complex collective behavior \cite{annrev10}. Aspects of these collective phenomena can be understood in a schematic way through analogies to the behavior of a pendulum (in flavor space) \cite{pen2006,splits2007,simple2007}.

The \lq\lq bulb model\rq\rq\ version of this problem incorporated a sharp neutrino sphere, coincident with the hot proto-neutron star surface, and roughly thermal, Fermi-Dirac black body neutrino emission from this surface. Neutrinos were taken to be freely streaming above the neutrino sphere, giving a neutrino emission geometry depicted in Figure 10.
\begin{figure}
\begin{center}
\includegraphics[width=12cm]{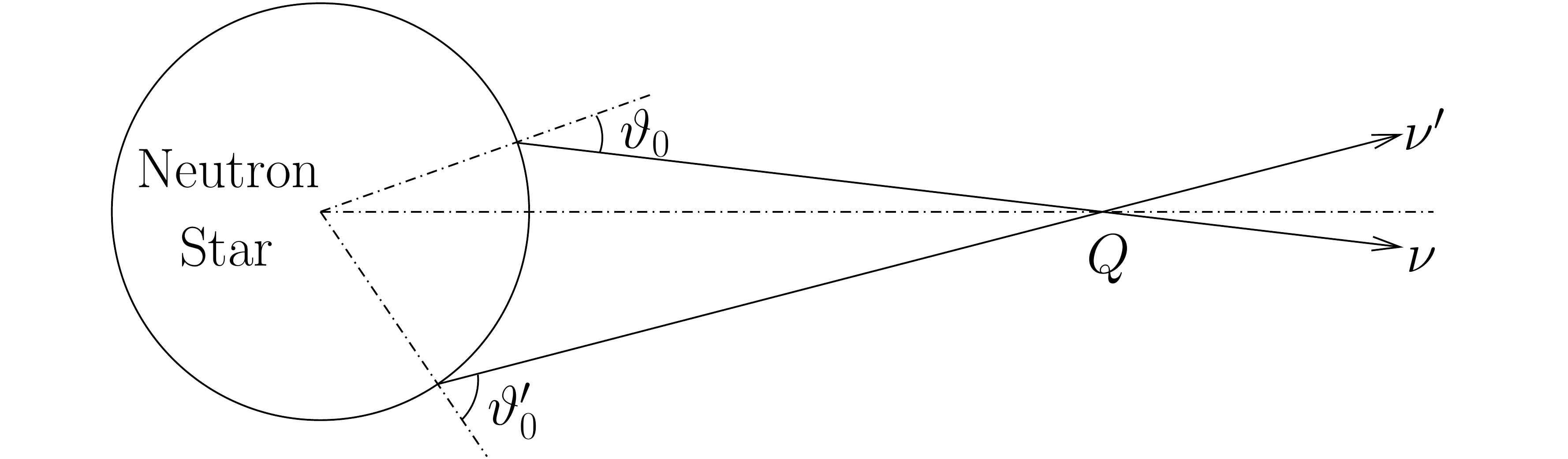}
\caption{Neutrino emission geometry in the bulb model. Angles relative to the normal at the emission point on the neutrino sphere (neutron star surface) are shown for two different neutrino trajectories. These trajectories intersect at location $Q$. Forward scattering of $\nu$ and $\nu^\prime$ at point $Q$ influences the subsequent flavor evolution of each neutrino. From Ref.~\cite{DFCQ2006Let}
(Copyright 2006 by the American Physical Society).}
\label{angles}
\end{center}
\end{figure}
From this figure it is obvious that neutrino-neutrino forward scattering couples the flavor evolution of intersecting neutrino trajectories. Successfully solving for neutrino flavor evolution in a typical case required some thousand neutrino angle bins and 100 or so energy bins, ultimately requiring the solution of $\sim {10}^7$ nonlinearly-coupled differential equations on a supercomputer. An example of the startling result is shown in Figure 11. 
\begin{figure*}
\begin{center}
$\begin{array}{@{}c@{\hspace{\myfigsep}}l@{}}
\includegraphics[scale=.34, keepaspectratio]{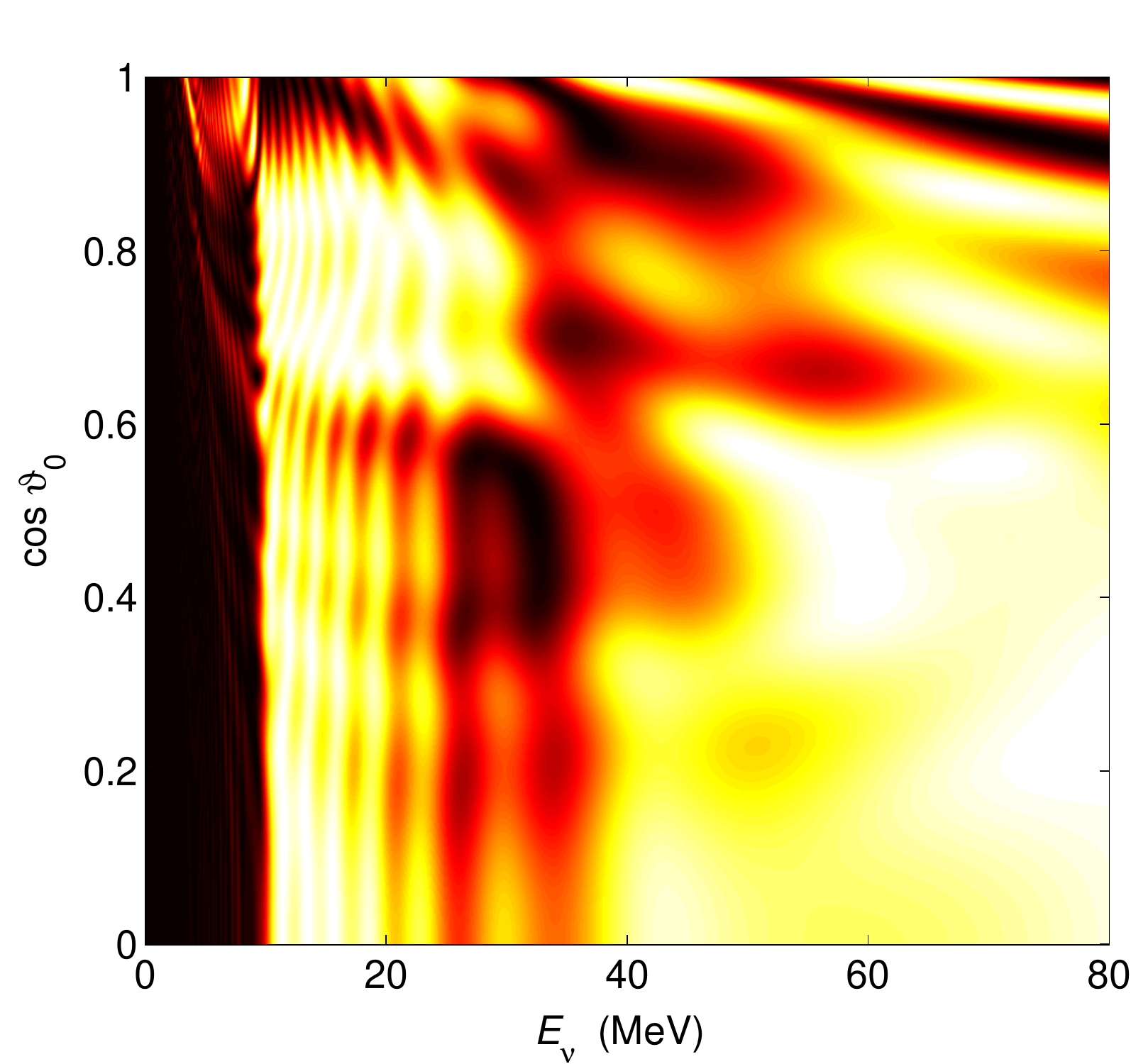} &
\includegraphics[scale=.34, keepaspectratio]{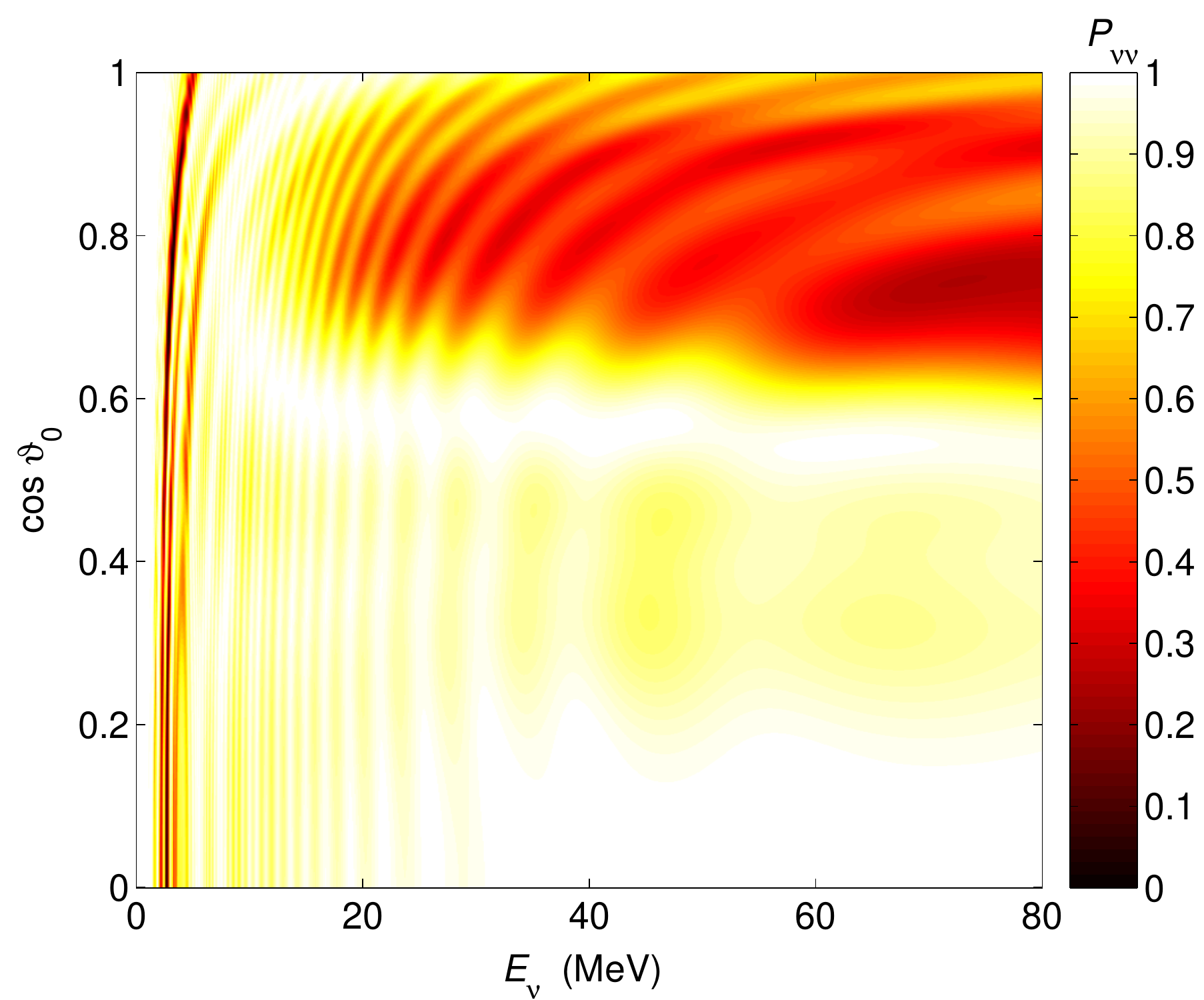} \\
\includegraphics[scale=.34, keepaspectratio]{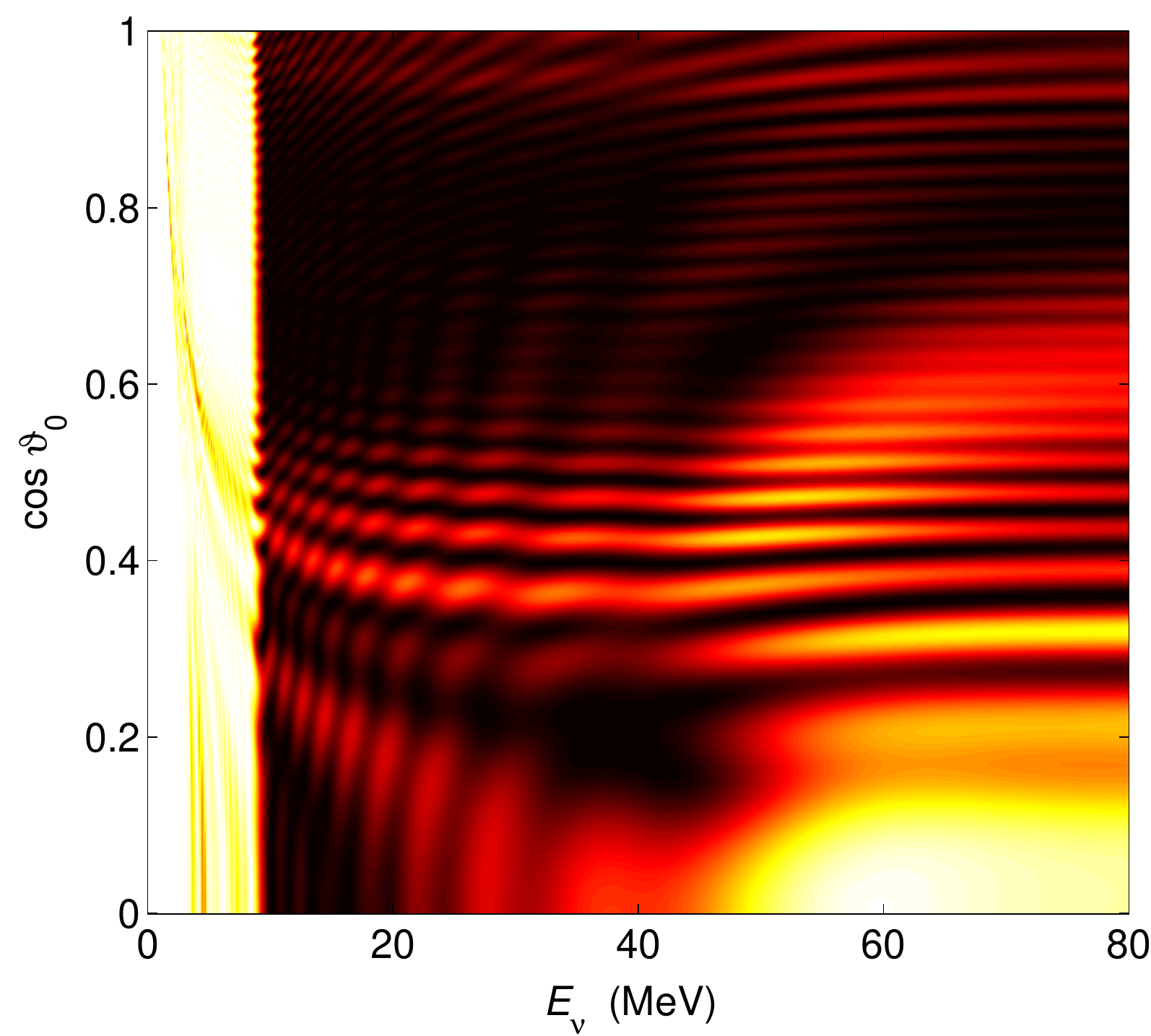} &
\includegraphics[scale=.34, keepaspectratio]{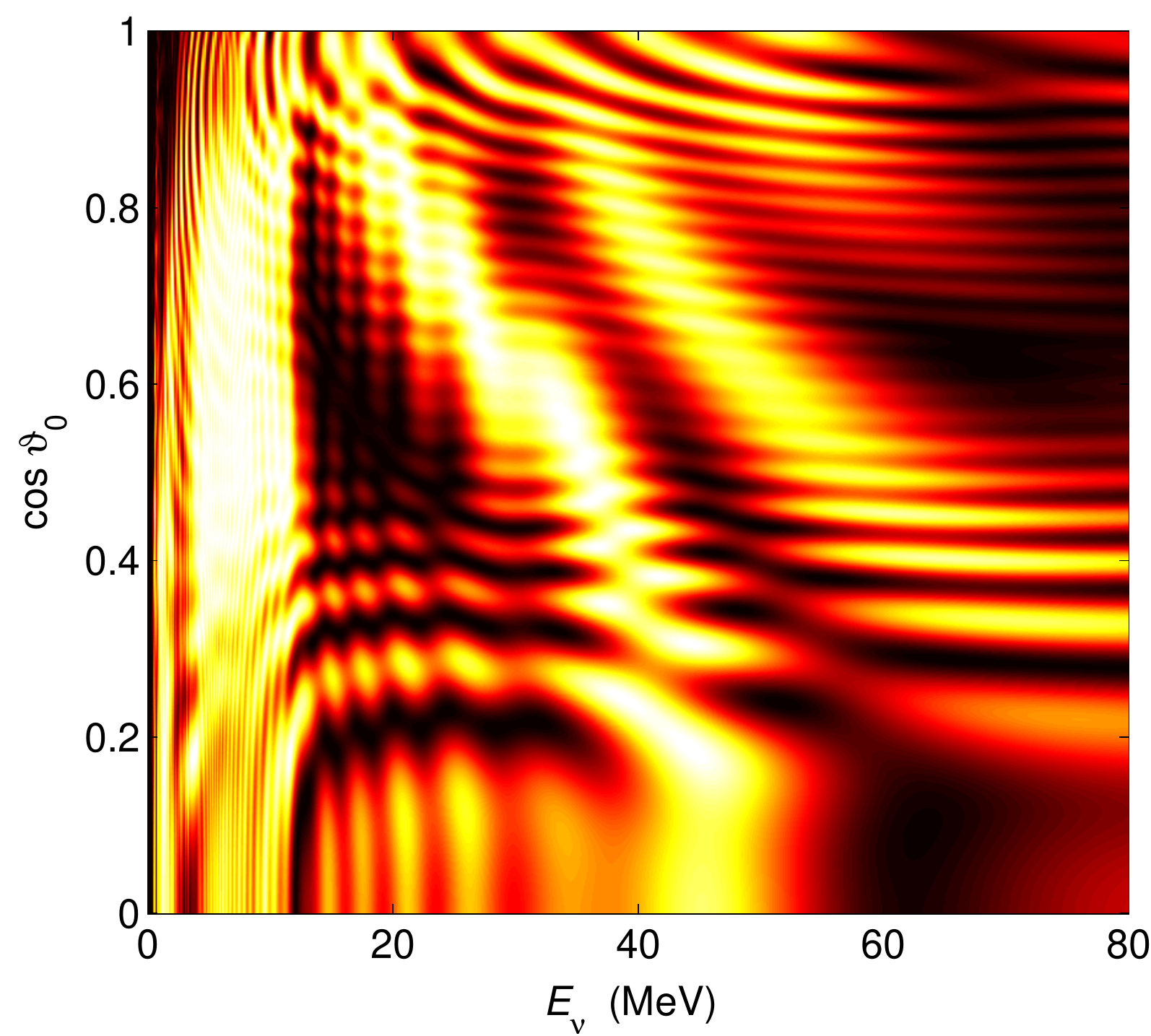}
\end{array}$
\caption{\label{fig:P-E-c}
Plots of survival probabilities $P_{\nu\nu}$ (color contour key at upper right) for neutrinos (left panels) and
antineutrinos (right panels) as functions of both neutrino energy $E_\nu$
and emission angle $\vartheta_0$ at radius $r=225$ km.
The upper panels employ a normal neutrino mass hierarchy,
and the lower panels employ an inverted neutrino mass hierarchy.
From Ref. \cite{DFCQ2006Let}
(Copyright 2006 by the American Physical Society).}
\end{center}
\end{figure*}

Perhaps the most prominent feature of these calculations is the spectral swap (or split): in the normal neutrino mass hierarchy nearly all of the neutrinos below a certain energy ($E_\nu \sim 10\,{\rm MeV}$ in Figure 11) are transformed out of their original flavor, regardless of emission angle; whereas the neutrinos above that energy are largely not transformed (see the panel on the upper right in Fig.~11). The swap has the opposite sense for neutrinos if we assume that nature has chosen the inverted mass hierarchy. The swap, or spectral split, phenomenon, arises from energy-like conserved quantities inherent in the coherent version of Equation 4 and the assumptions inherent in the bulb model.  

Though the neutrinos are isotropically distributed inside the hot proto-neutron star, well above it they are not. Figure 10 shows another important aspect of neutrino emission geometry.  In the bulb model, at a given location, the neutrinos contributing to the neutrino-neutrino forward scattering potential are contained in a cone. The further away from the neutron star this location is, the narrower the cone will be. This is significant because the neutrino-neutrino forward scattering contribution to the potential governing flavor transformation depends on the intersection angle $\phi$ of the forward-scattering neutrinos, going like $1-\cos\phi$. This implies that if the matter-potential tracks density and drops off with increasing radius roughly like $1/r^3$, then the corresponding neutrino-neutrino potential scales like $1/r^4$.

An assumption of the bulb model is that neutrinos do not suffer direction-changing scattering when they propagate above the neutrino sphere. Indeed, in the wind epoch ($> 3\,{\rm s}$ post core bounce) only of order one in a thousand neutrinos scatters this way. However, such scattered neutrinos can contribute substantially to the local potentials for flavor transformation. For example, the scattered neutrinos might have a larger value of $\phi$ at some location. Consequently, these scattered neutrinos can make up for their small numbers because their contribution to the neutrino-neutrino potential may not be as suppressed by the $1-\cos\phi$ factor as is the contribution from the un-scattered neutrinos in the bulb cone. This is the Halo Effect \cite{Cherry2012,Cherry2020}. It may be a significant determinant of neutrino flavor transformation. Because quantum flavor information can be scattered downward, toward the neutron star, it converts the initial value problem we discussed above into something more like a boundary value problem. Moreover, since neutrino neutral current coherent scattering scales like $E_\nu^2$ and like the square of nuclear mass, $A^2$, the Halo Effect couples nuclear composition into the neutrino flavor transformation problem, as well has adding new energy dependence.

At the very least, the Halo is a cautionary tale that begs a question. What other aspects of the anisoptropy of the neutrino field and the effects of neutrino scattering, not captured in the bulb model, could impact neutrino flavor transformation? In fact, ongoing research suggests that the symmetries built into the simplistic bulb model, may hide a plethora of neutrino flavor field instabilities, all augmented by the essential non-linearity of the problem. The bulb model symmetries are: Assumed spherical symmetry; Flavor oscillation patterns are constant in time; and, of course, the neglect of neutrino scattering and an assumption of a sharp decoupling surface. Linear stability analyses and numerical experiments suggest that spatial homogeneity of a slab distribution of neutrinos is unstable against formation of flavor domains \cite{AxialBreaking2013,Abbar2015}. Likewise, the neutrino field may be subject to temporal instabilities \cite{temporal2015}.

Relaxing the sharp decoupling surface assumption and taking into account neutrino scattering and absorption near the surface of the neutron star gives the different neutrino flavors correspondingly different \lq\lq neutrino spheres.\rq\rq\ With this effect, for example, the $\nu_e$ and $\bar\nu_e$ angular distributions could be different, with the latter species decoupling deeper in the neutron star surface region than the former. In turn, that can lead to net zero electron lepton number ($n_{\nu_e}-n_{\bar\nu_e}$) flux in certain locations and directions. These electron lepton number crossings set up the conditions for significant neutrino flavor transformation in the $\nu_e\rightleftharpoons\nu_{\mu,\tau}$ channel. This is termed fast flavor conversion and is potentially significant because it can occur deep in the supernova envelope, quite near the neutron star \cite{Sawyer2016,Basebu2017,MengRu2017}.

Solving the full QKEs in Equation 4, including the collision functional ${\hat{C}}({\hat{f}})$, is a daunting computational problem. QKE modeling and collision-induced decoherence in CCSN is the focus of intense recent research. Many open issues remain unresolved, including under what conditions collisions feed into or damp collective neutrino flavor transformation.

In the end, the QKEs are a mean field treatment. The efficacy of the mean field in calculating neutrino flavor evolution in compact object environments and the early universe seems likely, but may depend on the particular astrophysical conditions. At issue is whether multi-particle (i.e., multi-neutrino) quantum correlations are important and whether quantum entanglement of many body neutrino states can make a significant difference in neutrino flavor development. This issue has recently been explored in numerical experiments with small numbers of neutrinos \cite{Cervia2019}. The results suggest that entanglement entropy, an indicator of many body correlations, grows with increasing numbers of neutrinos. 

The behavior of flavor in the dense neutrino environments associated with compact objects like CCSN is a fascinating, if vexing problem at the heart of multi-messenger astrophysics.  It may yet hold many surprises for astrophysics and for fundamental neutrino physics.

\section{Neutrinos and Nucleosynthesis}
\label{sec:nucleosynthesis}
\begin{figure}[t]
\begin{center}
\includegraphics[width=12cm]{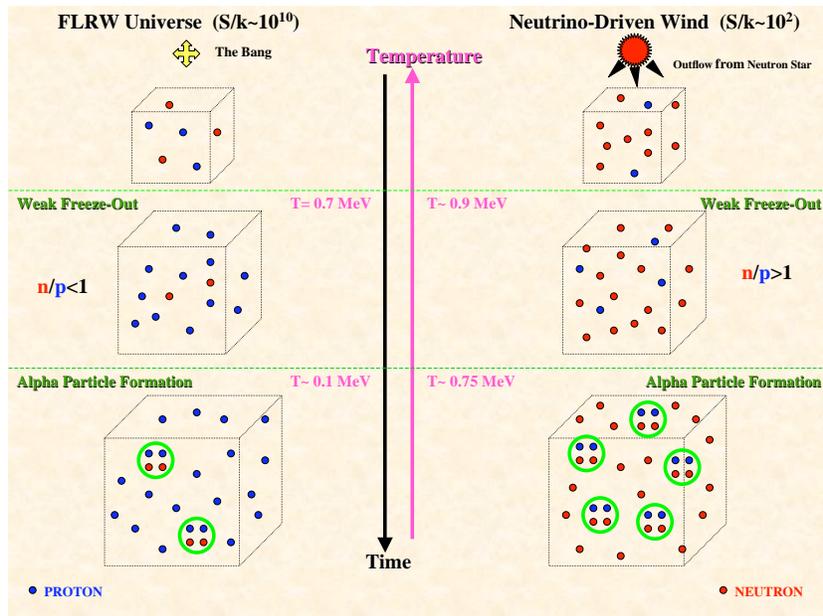}
\end{center}
\caption{ Cartoon showing the weak interaction, nuclear physics, and nucleosynthesis histories of a fluid element in (left) the expanding Friedman-LeMaitre-Robertson-Walker (FLRW) radiation-dominated early universe; and (right) the neutrino-heated outflow from a post core collapse- or binary neutron star merger-generated hot proto-neutron star. Each has a high (on a nuclear physics scale) entropy-per-baryon, $S/k_{\rm b}$, and each has a neutron-to-proton ratio determined by a competition among the charged-current lepton capture processes shown. Although these processes are in equilibrium at high temperature, they eventually slow down below the local expansion rate and freeze out. The early universe has an excess of protons (blue), while idealized neutrino-driven ejecta from a compact object could have an excess of neutrons (red). As described in the text, the high entropy dictates that alpha particles win the competition between binding and disorder that characterizes nucleosynthesis in a freeze out from nuclear statistical equilibrium. Coulomb barriers mean that the big bang nucleosynthesis scenario on the left makes only $^4{\rm He}$ and trace amounts of deuterium and $^7{\rm Li}/^7{\rm Be}$. Neutron capture in the scenario on the right could, in principle, produce heavy transuranic elements.}
\label{fig12:IsospinMirrors}
\end{figure}

Neutrinos and the weak interaction can play a key role in some cosmic sites where the synthesis of the elements, nucleosynthesis, takes place. At high enough temperature the sum of the rates of the nuclear reaction processes that build-up and, oppositely, tear-down nuclei are equal to each other and both are fast compared to the local matter expansion or decompression rate. This is the nuclear statistical equilibrium (NSE) condition discussed above. It generally obtains when the temperature is $T > 0.2\,{\rm MeV}$. As the matter expands and cools, eventually the nuclear reactions slow down and \lq\lq freeze out,\rq\rq\ leaving behind a pattern of nuclear abundances. This freeze-out from NSE is, very roughly, what happens in Big Bang Nucleosyntheis (BBN) in the early universe and in the neutrino-heated matter flowing out of a compact object venue like a CCSN or the ephemeral hot neutron star and disk left after a binary neutron star merger. Neutrinos can be the chief determinant of such nucleosynthesis because they are important in setting the entropy, the neutron-to-proton ratio, and the matter expansion rate. These are the three quantities that govern the nuclear abundances emerging from an NSE freeze-out scenario.  
In fact, the BBN and compact object outflow NSE freeze-out scenarios are similar and, in a sense, are nearly isospin mirrors of each other (one is neutron-rich, the other proton-rich), with the same sequence of weak interaction and nuclear physics events. These scenarios are depicted in Fig.~\ref{fig12:IsospinMirrors}.
We can flesh out these issues with a discussion of, first, BBN, and then follow that with a discussion of the r-process.

\subsection{Cosmological Neutrinos and BBN}
The advent of observations of the universe across the electromagnetic band, including especially cosmic microwave background (CMB) experiments, has produced a wealth of data and a clear picture for the large-scale evolution of the universe. 
Our best fit model for this data is one where the distribution of all matter and energy is, at any given snapshot in time, homogeneous and isotropic. This symmetry leads to two features that are key to understanding neutrinos and BBN: (1) A simple equation of motion, the Friedman equation for the scale factor $a$ appearing in the Friedman-LeMaitre-Roberstson-Walker (FLRW) metric; and (2) No preferred directions, meaning no spacelike heat flow, in turn implying that the entropy in a co-moving volume is constant so long as there is not a timelike heat source, for example, particles scattering or decaying out-of-equilibrium. Condition (1) gives the time dependence of the scale $a(t)$ given the mass-energy density $\rho(t)$, while condition (2) gives the relationship between scale factor and temperature $T$. 

The Friedman equation is tantamount to the conservation of total mechanical energy (expansion kinetic energy plus \lq\lq gravitational potential energy\rq\rq) on any co-moving 2-sphere region in the universe. For the radiation-dominated conditions in the early universe this is simply 
\[ H \equiv {1 \over a} {d a \over dt} = \sqrt{ 8 \pi G \rho \over 3}, \]
where $G=1/m_{\rm pl}^2$ is the gravitational constant, with the Planck mass $m_{\rm pl} \approx 1.22\times{10}^{22}\,{\rm MeV}$, $H$ is the expansion rate, the Hubble parameter, at FLRW time coordinate $t$, and $\rho$ is the total mass-energy density. Neutrinos, electrons, positrons, and photons, all with relativistic kinematics, make the largest and dominant contribution to the mass-energy density, with the (very few) baryons and dark matter making essentially negligible contributions at $T\sim 1\,{\rm MeV}$ where we are most concerned with neutrino physics. At that temperature, a figure of merit for the time-temperature relation is roughly $t\sim 1\,{\rm s}\, ({\rm MeV}/T)^2$. 

The takeaway message from the Friedman equation is that the expansion rate $H$ is governed by gravitation. But, gravitation is weak (the Planck mass is large) and, hence, the expansion rate in the early universe is comparatively {\it slow}. That, in turn, allows plenty of time for weakly interacting particles like neutrinos to come into thermal equilibrium. Consequently, at high enough temperature, neutrino number densities will be similar to those of photons, roughly $\sim T^3$. The expansion rate in radiation-dominated conditions ($\rho \sim T^4$) is $H \sim T^2/m_{\rm pl}$, whereas the neutrino scattering and absorption/emission rates are $\sim G_{\rm F}^2 T^5$. The weak interaction strength is governed by the Fermi constant, $G_{\rm F}\approx 1.166\times{10}^{-11}\,{\rm MeV}^{-2}$. Obviously, as the universe expands and the temperature drops, eventually the neutrino interaction rates will fall well below the expansion rate, and the neutrino component will \lq\lq freeze out.\rq\rq\ 

Thermal neutrino decoupling, where the neutrinos cease to scatter rapidly enough to exchange energy efficiently with the photon-electron/positron-baryon plasma, and chemical decoupling, where the lepton capture-induced (see Fig. 12) neutron-proton inter-conversion rates fall well below $H$, are protracted processes. These decoupling events, sometimes termed weak decoupling and weak freeze-out, respectively, actually proceed more or less concurrently over hundreds of Hubble times $H^{-1}$, between $T\sim 10\,{\rm MeV}$ and $T\sim 0.1\,{\rm MeV}$. In stark contrast, photon decoupling at $T\approx 0.2\,{\rm eV}$, which gives rise to the CMB, is driven by atomic bound state formation and, consequently, is abrupt, taking place in a small fraction of a Hubble time.
\begin{figure}
\begin{center}
\includegraphics[width=12cm]{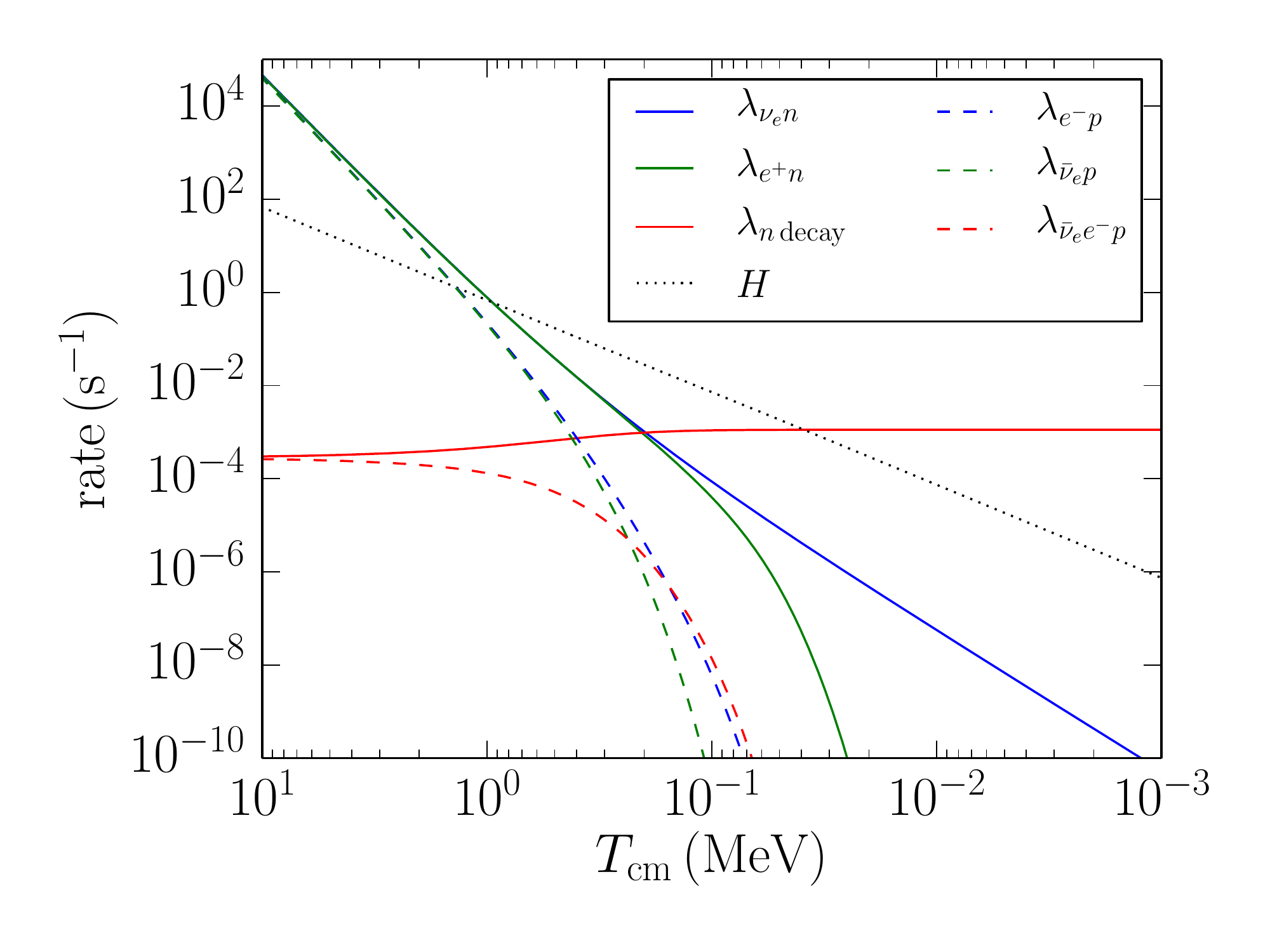}
\end{center}
\caption{ Rates for lepton capture and decay processes that inter-convert protons and neutrons as a function of co-moving temperature $T_{\rm cm}$ (a proxy for inverse scale factor, $1/a$). The Hubble expansion rate $H$ is also shown. Legend provides a key for the curves. Figure from Ref.~\cite{Surprising2016}
.}
\label{fig13:nprates}
\end{figure}

The rates of the weak interaction processes that inter-convert neutrons and protons are shown in Fig.~13 \cite{Surprising2016}. Neutrons are heavier than protons and, inevitably (for low lepton numbers) that fact means that the neutron-to-proton ratio will be $n/p<1$ when alpha particles form abruptly at $T\sim 0.1\,{\rm MeV}$. There is no threshold for the neutron-destroying processes of $\nu_e$ and $e^+$ capture, and the high entropy of the early universe guarantees that there will be plenty of positrons around even at temperatures $T\sim 0.1\,{\rm MeV}$. Neutron decay, $n\rightarrow p+e^-+\bar\nu_e$, also contributes to overall neutron destruction, though prior to alpha formation the rate of this process (red line in Fig. 13) is suppressed relative to \lq\lq free neutron decay\rq\rq\ by final state lepton Pauli blocking.

The nuclear physics of BBN is, in broad brush, simple: Nearly every nucleon that can be incorporated into an alpha particle {\it is} so incorporated at $T\approx 0.08\,{\rm MeV}$. Since $n/p \sim 1/7$ at this point, the alpha particle mass fraction will be roughly $25\%$. Only about one neutron in ${10}^5$ winds up in a deuteron, and one in $\sim {10}^{9}$ in a $^7{\rm Be}/^7{\rm Li}$ nucleus. In NSE at temperature $T$ and entropy-per-baryon $s$ (in units of $k_{\rm b}$), the abundance $Y_A$ of a nucleus with mass number $A$ and binding energy $Q$ is
\begin{equation}
Y_A \propto s^{1-A}\, \exp{(Q/T)}
\label{NSE}    
\end{equation}
The measured CMB acoustic peak amplitudes show that the baryon-to-photon ratio is $\eta \approx 6.1\times{10}^{-10}$, implying $s \approx 5.9\times{10}^9$ in the plasma of the early universe \cite{Steigman}. The Planck and seven-year WMAP results for $\eta$ are $(6.09 \pm 0.13) \times 10^{-10}$ \cite{Planck} and  $(6.19 \pm 0.15) \times 10^{-10}$ \cite{WMAP7}, respectively. The prodigious entropy means that heavier nuclei will not appear in appreciable numbers in NSE. High entropy, together with the significant Coulomb barriers in charged particle nuclear reactions, preclude significant assembly of nuclei heavier than $^4{\rm He}$. 

A more detailed evolution of the neutrino and baryonic components can be obtained by simultaneously and self-consistently following all relevant strong, electromagnetic and weak nuclear reactions together with Boltzmann neutrino transport treatment of neutrino scattering and decoupling. Results of such a calculation are shown in Fig.~14 \cite{BigPaper2016}.
\begin{figure}
\begin{center}
\includegraphics[width=12cm]{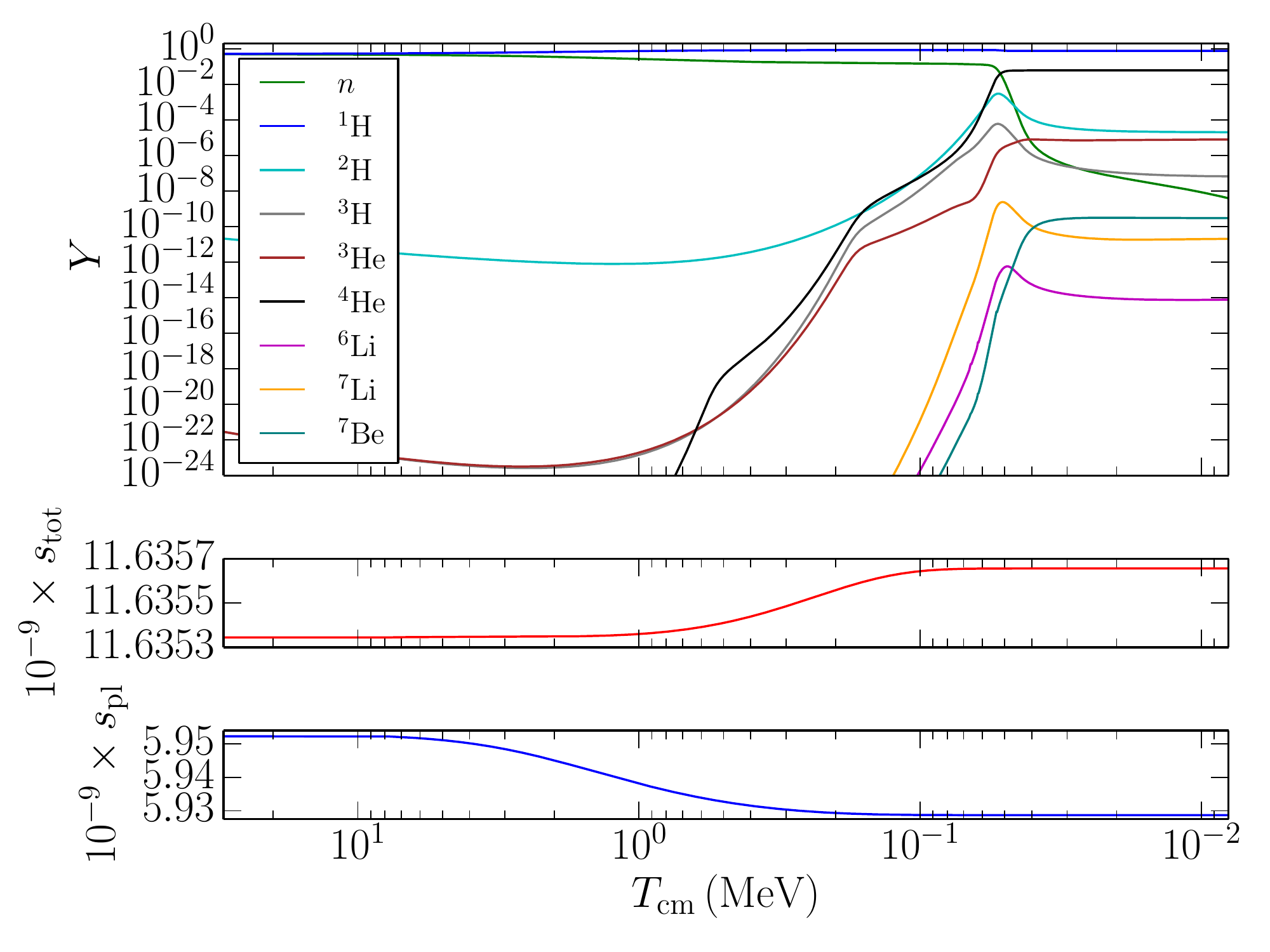}
\end{center}
\caption{BBN abundances $Y$ (see legend) for light nuclear species, total entropy per baryon (plasma plus the decoupling neutrino component) $s_{\rm tot}$, and entropy in the photon-electron/positron-baryon plasma $s_{\rm pl}$, are each shown as a function of co-moving temperature $T_{\rm cm}$ (a proxy for inverse scale factor, $1/a$) in the top, middle, and lower panels, respectively. Figure from Ref.~\cite{BigPaper2016}
.}
\label{fig14:abund}
\end{figure}
With the CMB-determined value of $\eta$ and entropy, the BBN abundance predictions for helium and deuterium are in good agreement with observationally-inferred values. In fact, the high redshift damped Lyman-alpha system hydrogen absorption line-derived deuterium abundance gave the first measure of $\eta$ \--- this was {\it confirmed} by the later CMB observations. Observations of old halo stars suggest a primordial $^7{\rm Li}$ abundance that is a factor of $3$ or $4$ below the BBN-predicted value for the $^7{\rm Be}/^7{\rm Li}$ yield (beryllium decays to lithium long after BBN). Resolution of this \lq\lq 
lithium problem\rq\rq\ may lie in the interpretation of observations and in stellar physics, or in beyond standard model (BSM) physics, but it is unlikely to be found in standard nuclear physics.

The entropy in a co-moving volume, $s\propto g\,a^3\,T^3$, will be a constant throughout the weak decoupling/BBN epoch if we stick to Standard Model physics and neglect out-of-equilibrium neutrino scattering. The implications of this simple scenario: The entropy carried by $e^\pm$-pairs at high temperature is transferred to the photons, but not to the decoupled neutrinos, as the temperature drops and the pairs go away. The statistical weight in relativistic particles is $g_{\rm DEC}=2+7/8(2+2)= 11/2$ at temperatures high enough that $e^\pm$-pairs are copiously produced. If neutrinos decouple in those conditions, and never scatter again, they will free fall through spacetime and their 3-momenta will simply redshift like $p\propto 1/a$ \--- they will retain a Fermi-Dirac black body-shaped energy distribution but with \lq\lq temperature\rq\rq\ at scale factor $a$ related to the decoupling temperature and decoupling scale factor values by $T_\nu = T_\nu^{\rm DEC}\,(a^{\rm DEC}/a)$. Of course, the photon (plasma) temperature is the same as the neutrino temperature prior to neutrino decoupling. At low temperature, after the $e^\pm$-pairs have gone away, photons carry all of the entropy so $g_{\rm low}=2$. Co-moving entropy conservation then dictates that the ratio of the neutrino \lq\lq temperature\rq\rq\ to the plasma (photon) temperature is $(g_{\rm low}/g_{\rm DEC})^{1/3} = [2/(11/2)]^{1/3} \approx 0.714$. This is the origin of the relic neutrino background, the C$\nu$B, shown in Fig.~1. 

Accounting for the timelike entropy source from out-of-equilibrium neutrino-$e^\pm$ scattering makes a very small alteration in this picture, as shown in Fig.~14. The photon-electron/positron-baryon plasma is a little hotter than the neutrinos, so that about one part in $300$ of the entropy in this plasma, $s_{\rm pl}$, is transferred to the decoupling neutrino component. Of course, this transfer is accompanied by an even smaller overall increase (a few parts in ${10}^5$) in the {\it total} entropy.

Though as yet the C$\nu$B has not been detected {\it directly}, the overall success of BBN implies that it is there, at least at the BBN epoch. Moreover, its presence is readily observable in the analysis of CMB data. The expansion rate at the CMB decoupling epoch, $T_\gamma \approx 0.2\,{\rm eV}$, depends, in part, on the contribution to the overall energy density from {\it all} extant particles with relativistic kinematics. Excluding the photon contribution, the energy density contribution from {\it all} other relativistic components is commonly expressed as $(7 \pi^2/120)(4/11)^{4/3}\, T_\gamma^4\, N_{\rm eff}$. If the only relativistic particles are from the neutrino background, and these have perfect zero chemical potential Fermi-Dirac black body energy spectra, we would expect $N_{\rm eff}=3$. The out-of-equilibrium neutrino scattering discussed above gives rise to small distortions in the neutrino energy spectra. Taking account of these gives $N_{\rm eff}\approx 3.046$. This Standard Model value is well within the current CMB bounds and uncertainties, as are the BBN-predicted helium and deuterium abundances \cite{IT,WMAP7}. The advent of 30-m class telescopes may offer sub-one percent precision on the primordial deuterium abundance. Likewise, Stage-4 CMB experiments promise comparable constraints on $N_{\rm eff}$ and primordial helium \cite{ScienceBook}. CMB observations, coupled with large scale structure limits, also allow constraints on $\sum m_\nu$, sometimes termed the sum of the light neutrino masses. In fact, $\sum m_\nu$ really measures the neutrino collision-less damping scale, essentially how far neutrinos stream and their effectiveness at damping the growth of structure. Stage-4 CMB experiments will have $15\,{\rm meV}$ $1\,\sigma$ sensitivity to this quantity, which is significant given that with perfect black body energy spectra and the normal neutrino mass hierarchy, and a very small lightest neutrino mass eigenvalue, we expect $\sum m_\nu \approx 57\,{\rm meV}$. The synergistic possibilities for CMB experiments, long baseline neutrino oscillation experiments (e.g., to pin down the neutrino mass hierarchy), next generation neutrino rest mass probes (e.g., tritium endpoint and neutrino-less double beta decay experiments), and 30-m class telescopes are tantalizing. Taken together, all of these capabilities and their anticipated data, plus the simple nuclear physics of BBN, comprise a sensitive set of tools for probing fundamental neutrino sector physics and beyond standard model physics \cite{Field,Pospelov}.

\subsection{The r-process}
Some of the pathways for the synthesis of the elements heavier than the iron peak, i.e., nuclear mass numbers $A>100$, may also involve neutrinos and the weak interaction. 
Material outflow from a hot, neutrino-emitting, neutron-rich source like the proto-neutron star remnant of a CCSN or the transient supra-massive neutron star and disk remnant of a binary neutron star merger  could follow a weak interaction, thermodynamic, and nuclear physics history broadly similar to that in BBN. Such a nucleosynthesis scenario, a near (isospin) mirror of BBN, is depicted
on the right in Fig.~\ref{fig12:IsospinMirrors}. Notably, quite unlike BBN, this scenario is neutron-rich, so that when NSE freeze out occurs and the alpha particles form, neutrons are isolated. Some of these neutrons and alphas may assemble heavier \lq\lq seed\rq\rq\ nuclei like iron. However, the high entropy conditions obviously dictate (see Eq.~6) that there not be very many of these seeds. With no Coulomb barriers to hinder them, neutrons can readily capture on the seed nuclei. Instead of the dead end of BBN at alpha particles and tiny amounts of trace light elements, this scenario could make uranium! This is one way the rapid-neutron-capture process, or r-process, could occur. There are many models and proposed sites for the r-process. Each can meet the key necessary condition for the r-process:  providing in excess of 100 neutrons per seed nucleus. 

The primary determinants of the neutron-to-seed nucleus ratio are: (1) the neutron-to-proton ratio $n/p$, or $Y_e = 1/(1+n/p)$; (2) the entropy-per-baryon $s$ (e.g., $\sim 100\,k_{\rm b}$ in the CCSN $\nu$-driven wind); and (3) the matter expansion rate. Various astrophysical environments have been invoked as sites of r-process nucleosynthesis, and each of these sites \lq\lq turns\rq\rq\ these three \lq\lq knobs\rq\rq\ to produce the requisite neutron-to-seed ratio \cite{qian}. For example, there are schemes that utilize: low entropy and extreme neutron excess; high entropy, fast expansion rates and modest neutron excess; or high energy particle-induced spallation to \lq\lq mine\rq\rq\ neutrons from existing nuclei. The site or sites of the r-process have been debated since the seminal 1957 BBFH paper that brokered the marriage of nuclear and particle physics with astronomy \cite{BBFH}.
 
However, until recently there was no direct observational confirmation for any of the proposed r-process production sites. Gravitational wave astronomy and multi-messenger astrophysics (MMA) have changed that. To see how this development highlights neutrino physics as a vital part of MMA, some background on neutron capture nucleosynthesis and the r-process is in order.

Beyond the iron peak, nuclear Coulomb barriers become so high
that charged particle reactions become ineffective, leaving
neutron capture as the only viable mechanism for producing
the heavy nuclei.
If the neutron abundance is modest,
this capture occurs in such a way that each newly synthesized
nucleus has the opportunity to $\beta$ decay, if it is energetically
favorable to do so. In this limit weak equilibrium is maintained among
the nuclei, so that synthesis is along the nuclear valley of stability.
This is called the s- or slow-process.  However a
plot of the s-process in the (N,Z) plane reveals that this
path misses many stable, neutron-rich nuclei.  This suggests that another mechanism is also at
work.  Furthermore, the abundance peaks found in nature 
near masses A $\sim$ 130 and A $\sim$ 190, which mark the closed
neutron shells where neutron capture rates and $\beta$ decay
rates are slower, each split into two subpeaks.  One set of subpeaks
corresponds to the closed-neutron-shell numbers N $\sim$ 82
and N $\sim$ 126, and is clearly associated with the s-process.
The other set is shifted to smaller N $\sim$ 76 and $\sim$ 116,
respectively, suggestive of a much more dynamic 
neutron capture environment. 
  
This second process is the r- or rapid-process. Clearly, it requires a neutron fluence
so large that neutron capture is fast compared to $\beta$ decay.  In this case,
nuclei rapidly absorb neutrons until they approach the neutron drip line.  That is,
equilibrium is maintained by $(n,\gamma) \leftrightarrow
(\gamma,n)$, not by weak interactions.  Consequently the nuclei participating in the r-process are very
different from ordinary nuclei -- very neutron rich nuclei that would decay immediately
in the low-temperature environment of Earth.   The rate of nucleosynthesis is
controlled by the rate of $\beta$ decay: a new neutron can be captured only
after $\beta$ decay, $n \rightarrow p + e^- +\bar{\nu}_e$,
opens up a hole in the neutron Fermi sea.  Consequently one expects abundance
peaks near the closed neutron shells at N $\sim$ 82 and 126, as $\beta$ decay 
is slow and mass will pile up at these ``waiting points."  By a series of rapid neutron captures 
and slower $\beta$ decays, synthesis can proceed all the way to the transuranics. 
Typical r-process conditions include neutron
densities $\rho(n) \sim 10^{18}-10^{22}$/cm$^3$, temperatures $\sim$
10$^9$ K, times $\sim$ 1 second, and ratios of free neutrons to heavy seed 
nuclei of $\gtrsim$ 100.
  
The path of the r-process is typically displaced by just $\sim$ (2-3) MeV from
the neutron drip line (where no more bound neutron levels exist).
After the r-process neutron exposure ends,
the nuclei decay back to the valley of stability by $\beta$
decay.  This involves conversion of neutrons into protons,
shifting the r-process peaks from the parent-nucleus values of N $\sim$ 82 and 126
to lower values and explaining the double-peak structure of the 
r-process/s-process closed-shell abundance peaks.

One possible neutrino role in the r-process is in producing the required
explosive, neutron-rich environments.   One such site is the supernova neutrino-driven wind -- the
last ejecta blown off the proto-neutron star. This is the inspiration for the scenario on the right side of Fig.~\ref{fig12:IsospinMirrors}.
This material is hot, high entropy, dominated by radiation much like in BBN, and
contains neutrons
and protons, often with an excess of neutrons.  Freeze out from NSE produces mostly alpha particles, locking up most all pf the protons, and leaving a sea of free neutrons.
Then the $\alpha$s interact through 3-body reactions like 
\[ \alpha + \alpha + n \rightarrow {^{9}{\rm Be}} + \gamma  \]
producing a gateway for the formation of heavier seed nuclei.  A fast expansion rate, as well as high entropy,
both hinder the formation of these seed nuclei, thereby giving a higher neutron-to-seed nucleus ratio for a given reservoir of neutrons. 

The neutrinos can be crucial in this picture for the r-process. The same $\nu_e$ and $\bar\nu_e$ capture reactions (shown in Fog.~\ref{fig12:IsospinMirrors}) whose competition helps sets the $n/p$ ratio, also heat the out-flowing matter in the envelope above the neutron star to high entropy, $s/k_{\rm b} \sim 100$. Additionally, if we want the neutrinos to supply the energy to lift a baryon out of the gravitational potential well of the neutron star, then the neutrino flux has to be high enough that each nucleon interacts $\sim 10$ times via these neutrino capture processes. This is because the gravitational binding energy of a nucleon near the neutron star is $\sim 100\,{\rm MeV}$, $\sim 10\%$ of its rest mass, and each neutrino has an energy $\sim 10\,{\rm MeV}$. That is a prodigious neutrino flux, albeit one that is expected as discussed above. However, once the alphas form, $\nu_e+n\rightarrow p+e^-$ can destroy neutrons, while the counterbalancing reaction $\bar\nu_e+p\rightarrow n+e^+$ is effectively energetically disabled because the protons are sheltered inside alpha particles. This is the so-called $\alpha$-Effect \cite{FullerMeyer1995}, the key stumbling block to an r-process sited in neutrino-heated ejecta.

There have been many ideas on how to get around the $\alpha$-Effect, and other assorted problems associated with the wind entropy and ejection speed \cite{Roberts}. One way is to have some other energy source, perhaps based on high magnetic fields and multi-dimensional hydrodynamics, effect the ejection of material at high enough speed that the neutrino capture reactions are ineffective in resetting the very low $Y_e$ (because of electron degeneracy) material near the neutron star surface. Another suggestion is to invoke medium-enhanced active-sterile neutrino oscillations \cite{McLaughlin1999}. 

However, if it could be made to work, there are some very nice aspects of this site: the amount of
matter ejected is $\sim 10^{-6}$ solar masses,
a production per event that if integrated over the lifetime of the
galaxy gives the required total abundance of r-process metals,
assuming typical supernova rates.  

For this reason, considerable effort has been invested in looking for other
viable r-process sites.   Possibilities include a so-called ``cold" neutrino-driven r-process
operating in the ${}^4$He mantles of early, metal-poor supernova progenitors \cite{Banerjee}
and binary neutron star (NS) mergers, which could be the dominant r-process
site once our galaxy has evolved to the point that these events become common \cite{Argast}.
The most important recent result concerning the r-process comes from observations
of the NS merger event GW170817, which produced some r-process ejecta and likely left a balck hole (BH) remnant.

Nearly 50 years ago Lattimer and Schramm \cite{LSr} and Symbalisty and Schramm \cite{SSr} estimated the amount of neutron-rich material
that might be ejected in the tidal tails of NS-BH and NS-NS mergers, concluding that the
mass was sufficient to make these events potentially important r-process sites.  Because the matter is very
neutron rich, the conditions for creating an r-process are more robust than in the case of CCSN,
where small changes in dynamics can influence whether or not the ejected material is neutron-rich enough or evades the $\alpha$-Effect problem.   As the frequency of
such events is typically estimated to 3$\pm$1 orders of magnitude rarer than CCSNe, mergers
can be a major source of the r-process if the production is $\gtrsim 10^{-3} M_\odot$ per event.
Numerical simulations of such events confirmed  long ago that such productions are reasonable
outcomes \cite{Davies1994,Ruffert1997,Freiburghaus1999}.  Examples of more recent work
can be found in \cite{Just2015,Nedora2020,Chen2021}.

The NS-NS merger event GW170817 was observed in gravitational waves by the LIGO and
Virgo collaborations.   After the event, a long-lived transient with distinctive features was observed -- a kilonova \cite{kilonova} -- 
in multiple electromagnetic wavelengths.  During the merger matter is tidally stripped from the stars,
flung out as tails.  Additional ejection occurs over longer times, a second or so after the merger,
as matter held up in the accretion disk around the merged NSs is blown away by winds.  This material
is ejected at lower velocities and is exposed to intense neutrino irradiation.  Consequently, as in the neutrino-driven winds of CCSNe,
associated charged-current reactions can reduce the neutron mass fraction. Nevertheless, some of the jecta may be at high enough speed to evade these problems \cite{kilonova2}.

While the hydrodynamic expansion and nuclear reactions  may continue for several seconds, the longer term
evolution of the ejecta is governed by the energy released into the plasma by the decay of the newly synthesized
parents of r-process nuclei, as they decay.  This energy is emitted thermally (once the material becomes
translucent) over a period of days, with the luminosity determined approximately by the heating rate.  The light
curve depends on the material's opacity, which can be dramatically impacted by the presence of lanthanides 
and other heavy r-process nuclei.  

An analysis of the observed light curve \cite{kilonova2} led to the conclusion that the ejecta consisted on two
components, a rapidly expanding early component responsible for emission in optical wavelengths,
and a longer, slower component that radiated in the infra-red over a period of about two weeks.  The former
is associated with material containing dominantly light r-process nuclei and few lanthanides, with the resulting
low opacity producing an optical spectrum that faded in days.  The latter component was rich in lanthanides and
other heavy r-process nuclei, with the resulting high opacity accounting for the length and redness of the emission.
The details of the light curve for the kilonova associated with GW170817 match the predictions of this model,
which had been developed before the event \cite{kilonova}. Whether this event made the actinides, like uranium, remains an open question.

The estimated ejecta yield from the GW170817 kilonova, combined with reasonable estimates of NS-NS merger
rates, suggest that the mechanism could account for the bulk of the r-process material we observe in our galaxy --
though uncertainties on this estimate are large.

But is there an r-process, or in fact are there several?   The NS-NS merger rate is uncertain, requiring an estimate of
both the NS production rate as well as the coalescence timescale.  But an exploration of likely parameters typically yields a
chemical evolution pattern for r-process material that begins a steep rise when the iron content of our galaxy 
is about $10^{-2}$ solar \cite{Argast}.   Yet there are remarkable data from metal-poor halo stars, whose surfaces
were enriched by some nearby r-process event, showing that the r-process operated at metallicities $[\mathrm{Fe/H}] \sim 10^{-4}$ \cite{Sneden}.
This has caused some to conclude that an additional mechanism or mechanisms must have operated at early times.
A very recent analysis based on correlations between r-process elements and Fe argued for at least three such mechanisms \cite{Farouqi}.

\subsection{The Neutrino Process}
One of the more amusing roles for neutrinos in nucleosynthesis is found in
the neutrino process, the direct synthesis of new elements through neutrino reactions \cite{woosley}.
Core-collapse supernovae provide the enormous neutrino fluences
necessary for such synthesis to be significant.  They also eject newly 
synthesized material into the interstellar medium, where it can be incorporated
into a new generation of stars.

Among the elements that might be made primarily or partially in the $\nu$-process,
the synthesis of ${}^{19}$F is one of the more interesting examples \cite{woosley}.
The only stable isotope of fluorine,
${}^{19}$F has an abundance
\[ {^{19}\mathrm{F} \over ^{20}\mathrm{Ne}} \sim {1 \over 3100}. \]
Ne is one of the hydrostatic burning products in massive stars, produced in
great abundance and ejected in core-collapse supernovae.  Thus a mechanism
that converts $\sim$ 0.03\% of the ${}^{20}$Ne in the star's mantle into ${}^{19}$F could
account for the entire observed abundance of the latter.

The Ne zone in a supernova progenitor star
is typically located at a radius of $\sim$ 20,000 km.  A simple calculation
that combines the neutrino fluence through the Ne zone
with the cross section for inelastic neutrino scattering off ${}^{20}$Ne shows that
approximately 0.3\% of the ${}^{20}$Ne nuclei would interact with the
neutrinos produced in the core collapse.  Almost all of these
reactions result in the production of ${}^{19}$F, e.g., 
\begin{eqnarray}
 {}^{20}\mathrm{Ne}(\nu,\nu')^{20}\mathrm{Ne}^* &\rightarrow& {}^{19}\mathrm{Ne} + n 
\rightarrow ^{19}\mathrm{F} + e^+ + \nu_e + n \nonumber \\
  {}^{20}\mathrm{Ne}(\nu,\nu')^{20}\mathrm{Ne}^* &\rightarrow& {}^{19}\mathrm{F}
+ p, \nonumber
\end{eqnarray}
with the first reaction occurring half as frequently as the 
second.  Thus one would expect the abundance
ratio to be ${}^{19}\mathrm{F}/ {}^{20}\mathrm{Ne} \sim 1/ 300$, corresponding to an order
of magnitude more ${}^{19}$F than found in nature.  

This example shows that stars are rather complicated factories for nucleosynthesis.
Implicit in the reactions above are mechanisms that also destroy ${}^{19}$F.  
For example, about 70\% of the neutrons coproduced with ${}^{19}$F in the first reaction
immediately recapture on ${}^{19}$F, destroying the product of interest.  
Similarly, many of the coproduced protons destroy ${}^{19}$F via ${}^{19}$F(p,$\alpha)^{16}$O --
unless the star is rich in ${}^{23}$Na, which readily consumes protons via
${}^{23}$Na(p,$\alpha)^{20}$Ne.  Finally, some of the ${}^{19}$F produced in the
neon shell is destroyed when the shock wave passes through that zone:
the shock wave can heat the inner portion of the Ne zone above $1.7 \times 10^9$K, the
temperature at which ${}^{19}$F can be destroyed by ${}^{19}$F($\gamma,\alpha)^{15}$N.

If all of this physics is treated carefully in a nuclear network code, one finds that the
desired ${}^{19}$F/${}^{20}$Ne $\sim$ 1/3100 is achieved for a heavy-flavor neutrino
temperature of about 6 MeV.  This is quite consistent with the temperatures that come
from supernova models.

The neutrino process produces interesting abundances of several relatively
rare, odd-A nuclei including ${}^7$Li, ${}^{11}$B, 
${}^{138}$La, ${}^{180}$Ta, and ${}^{15}$N.  Charged-current neutrino reactions
on free protons can produce neutrons that, through $(n,p)$ and $(n,\gamma)$
reactions, lead to the nucleosynthesis of the so-called ``p-process" nuclei from
A=92 to 126.  The production of such nuclei has been a long-standing puzzle in nuclear astrophysics.

\section{Neutrino Cooling and Red Giants}
Several neutrino cooling scenarios have already been discussed,
including cooling of the proto-neutron star produce in core collapse 
and cooling connected with the expansion of the early universe.
Red giant cooling provides an additional example of the use of 
astrophysical arguments to constrain
fundamental properties of neutrinos.

\subsection{Red Giants and Helium Ignition}
In a solar-like star, when the hydrogen in the central core has been
exhausted, an interesting evolution ensues:
\begin{itemize}
\item  With no further means of producing energy, the core
slowly contracts, thereby increasing in temperature as gravity
does work on the core.
\item Matter outside the core is still hydrogen rich, and
can generate energy through hydrogen burning.  Thus a hydrogen-burning
shell forms, generating the gas pressure supporting the outside layers of the
star.  As the ${}^4$He-rich core contracts, the matter outside the
core is also pulled deeper into the gravitational potential.
Furthermore, the H-burning shell continually adds more mass to the core.
This means the burning in the shell must intensify to generate
the additional gas pressure to fight gravity.  The shell also
thickens, as more hydrogen is above the
burning temperature.
\item The resulting increasing gas pressure causes the outer
envelope of the star to expand by a large factor, up to a 
factor of 50.  The increase in radius more than compensates for
the increased internal energy generation, so that a cooler
surface results.  The star reddens.  Stars of this type are
called red supergiants.
\item This evolution is relatively rapid, perhaps a few
hundred million years: the dense core requires large energy
production.  The helium core is supported
by its degeneracy pressure, and is characterized by densities
$\sim 10^6$ g/cm$^3$.  This stage ends when the
core reaches densities and temperatures that allow helium burning
through the reaction
\[ \alpha + \alpha + \alpha \rightarrow ^{12}C + \gamma . \]
As this reaction is quite temperature dependent,
the conditions for ignition are very sharply defined.
This has the consequence that the core mass at the helium flash point
is well determined.
\item  The onset of helium burning produces a new source of
support for the core.  The energy released elevates the temperature
and the core expands: He burning, not electron degeneracy, now 
supports the core.  The burning shell and envelope move
outward, higher in the gravitational potential.  Thus shell
hydrogen burning slows (the shell cools) because less gas pressure
is needed to satisfy hydrostatic equilibrium.  All of this
means the evolution of the star has now slowed: the red giant
moves along the ``horizontal branch," as interior temperatures
increase slowly, much as in the main sequence.
\end{itemize}

The 3$\alpha$ process involves some fascinating nuclear physics
that will not be recounted here:  the existence of certain nuclear resonances
was predicted based on the astrophysical requirements for this process.
The resulting He-burning rate exhibits a sharp temperature dependence
$\sim$ T$^{40}$ in the range relevant to red giant cores.
This dependence is the reason the He flash
is delicately dependent on conditions in the core.

\subsection{Neutrino Magnetic Moments and Helium Ignition}
Prior to the helium flash, the degenerate He core radiates
energy largely by neutrino pair emission.  The process is
the decay of a plasmon --- which one can think of as a photon
``dressed" by electron-hole excitations, thereby given the photon 
an effective mass of $\sim$ 10 keV.  The plasmon couples to
an electron particle-hole pair that
then decays via $Z_\mathrm{o} \rightarrow \nu \bar{\nu}$. 

If this cooling is somehow enhanced, the degenerate helium core 
would not ignite at the normal
time, but instead continue to grow.    When the core does finally
ignite, the larger core will alter
the star's subsequent evolution.

One possible mechanism for enhanced cooling is a neutrino
magnetic moment.  Then the plasmon could directly couple to
a neutrino pair.  The strength of this coupling would 
depend on the size of the magnetic moment.

A delay in the time of He ignition has several observable
consequences, including changing the ratio of red giant to
horizontal branch stars.  Thus, using the standard theory of
red giant evolution, investigators have attempted to determine
what size of magnetic moment would produce unacceptable 
changes in the astronomy.  The resulting limit \cite{raffelt} on diagonal or
transition neutrino magnetic moments,
\[ \mu_{ij} \lesssim 3 \times 10^{-12} \mathrm{~electron~Bohr~magnetons} ,\]
is about an order of magnitude more stringent than the best limits
so far obtained from reactor neutrino experiments \cite{GEMMA}.

This example is just one of a number of such constraints that
can be extracted from similar stellar cooling arguments.
The arguments above, for example, can be repeated for 
neutrino electric dipole moments, or for
axion emission from red giants.   As noted previously, the
arguments can be extended to supernovae: anomalous
cooling processes that shorten the cooling time in a way that is
inconsistent with SN1987A observations are ruled out.  
For example, large Dirac neutrino masses are in conflict with SN1987A observations:
the mass term would allow neutrinos to scatter into sterile right-handed states,
which would then immediately escape, carrying off energy.




\section{Acknowledgements}
This article is based in part on ``Neutrino Astrophysics" appearing in the Wiley Online Library.
The work was supported in part by the National Science Foundation under  the
cooperative agreements 2017-228 and 2020-275 supporting the Network for Neutrinos,
Nuclear Astrophysics, and Symmetries;  U.S. Department of Energy
under contracts DE-SC00046548 (UC Berkeley) and
DE-AC02-98CH10886 (LBNL); the Heising Simons Foundation; and NSF Grant No. PHY-1914242 at UCSD.


\begin{thebibliography}{000}

\bibitem{hh00} W. C. Haxton and B. R. Holstein, Am. J. Phys. {\bf 68} (2000) 15 and
{\bf 72} (2004) 18.

\bibitem{Chadwickbeta} J. Chadwick, Verhandlungen der Deutschen Physikalischen Gesellschaft {\bf 16} (1914) 383.

\bibitem{Wooster} C. D. Ellis and W. A. Wooster, Proc. Roy. Soc. {\bf 117} (1927) 109.

\bibitem{Pauli} W. Pauli,  https://www.symmetrymagazine.org/article/march-2007/neutrino-invention (1930).

\bibitem{Chadwickneutron} J. Chadwick, Proc. Roy. Soc. {\bf A136} (1932) 692.

\bibitem{Fermi} E. Fermi, Il Nuovo Cimento {\bf 9} (1934) 1.

\bibitem{GamowTeller} G. Gamow and E. Teller, Phys. Rev. {\bf 49} (1936) 895.

\bibitem{CowanReines} C. L. Cowan Jr., F. Reines, F. B. Harrison, H. W. Kruse, and A. D. McGuire, Science {\bf 124} (1956) 103.

\bibitem{KATRIN} M. Aker et al., Phys. Rev. Lett. {\bf 123} (2019) 221802.

\bibitem{Hannestad} S. R. Choudhury and Steen Hannestad, JCAP {\bf 07} (2020) 037.

\bibitem{bahcallbook} J. N. Bahcall, {\em Neutrino Astrophysics}, (Cambridge University,
Cambridge, 1989).

\bibitem{lin} W.C. Haxton and W. Lin, Phys. Lett. {\bf B486} (2000) 263.

\bibitem{edoardo1} E. Vitagliano, J. Redondo, and G. Raffelt, JCAP {\bf 12}, 010 (2017).

\bibitem{edoardo2} E. Vitagliano, I. Tamborra, and G. Raffelt, Rev. Mod. Phys. {\bf 92}, 45006 (2020).

\bibitem{suwa} K. Abe et al. (Hyper-Kamiokande Collaboration), submitted to Ap. J. (arXiv:2101.05269).

\bibitem{SNhistory} J. N. Bahcall and R. Davis, Jr., in {\it Essays in Nuclear Astrophysics}, eds. C. A. Barnes, D. D. Clayton, and D. Schramm
(Cambridge University Press, 1982) pp. 243-285.

\bibitem{BetheC1938} H. A. Bethe and L. Chritchfield, Phys. Rev. {\bf 54} (1938) 248.

\bibitem{Bethe1939} H. A. Bethe, Phys. Rev. {\bf 55} (1939) 434.

\bibitem{Crane48} H. R. Crane, Rev. Mod. Phys. {\bf 20} (1948) 278.

\bibitem{davis} R. Davis Jr., D. S. Harmer, and K. C. Hoffman, Phys. Rev. Lett. {\bf 20} (1966) 1205.

\bibitem{HRS} W. C. Haxton, R. G. H. Robertson, and A. M. Serenelli, Ann. Rev. Astron. Astrophys. {\bf 51} (2013) 21.

\bibitem{LUNA} C. Broggini, D. Bemmerer, A.Caciolli, and D. Trezzi, Prog. Part. Nucl. Phys. {\bf 98C} (2018) 55.

\bibitem{SHP}  A. M. Serenelli, W. C. Haxton, and C. Pe\~{n}a-Garay, Ap. J. {\bf 743} (2011) 743.

\bibitem{SFII}  E. G. Adelberger et al., Rev. Mod. Phys. {\bf 83} (2011) 195.

\bibitem{faint} G. Feulner, Rev. Geophysics {\bf 50} (2012) RG2006.

\bibitem{Dilke} F. Dilke and D. Gough, Nature {\bf 240} (1972) 262.

\bibitem{SKgen}  Y. Nakajima, talk presented at Neutrino 2020 (Fermilab); Super-Kamiokande Coillaboration, Phys. Rev. {\bf D94} (2016) 052010; Phys. Rev. {\bf D83} (2011) 052010;  Phys. Rev. {\bf D78} (2008) 032002; and Phys. Rev. Phys. Rev. {\bf D73} (2006) 112001.

\bibitem{SNOgen} SNO Collaboration, Q. R. Ahmad {\it et al.}, Phys. Rev. Lett. {\bf 89} (2002) 011301;
SNO Collaboration, B. Aharmim {\it et al.}, Phys. Rev. {\bf D72} (2005) 052010 and
{\bf C81} (2010) 055504.

\bibitem{Borgen} Borexino Collaboration, G. Bellini {\it et al.}, Phys. Rev. Lett. {\bf 107} (2011) 141302
and {\bf 108} (2012) 051302; Phys. Rev. {\bf D82} (2010) 033006; Phys. Lett. {\bf B707} (2012) 22.

\bibitem{Maltoni} M. Maltoni and A. Yu. Smirnov, Eur. J. Phys. {\bf A52} (2016) 87.

\bibitem{MS} S. P. Mikheyev and A. Smirnov, Sov. J. Nucl. Phys. {\bf 42} (1985) 913.

\bibitem{Wolf} L. Wolfenstein, Phys. Rev. {\bf D17} (1979) 2369.

\bibitem{MNS} Z. Maki, M. Nakagawa, and S. Sakata, Prog. Theor. Phys. {\bf 28} (1962) 870.

\bibitem{Pontecorvo67} B. Pontecorvo, Zh. Eksp. Teor. Fiz. {\bf 53} (1967) 1717.

\bibitem{KamLAND} KamLAND Collaboration, A. Gando {\it et al.}, Phys. Rev. {\bf D83} (2011) 052002;
KamLAND Collaboration, S. Abe {\it et al.}, Phys. Rev. {\bf C84} (2011) 035804.

\bibitem{atmos} T. Kajita, Ann. Rev. Nucl. Part. Sci. {\bf 64} (2014) 343.

\bibitem{PDG} P. A. Zyla et al. (Particle Data Group), Prog. Theor. Exp. Phys. {\bf 2020} (2020) 083C01.

\bibitem{Abe16c} K. Abe et al. (Super-Kamiokande Collaboration), Phys. Rev. {\bf D94} (2016) 052010.

\bibitem{Gando13} A. Gando et al. (KamLand Collaboration), Phys. Rev. {\bf D88} (2013) 033001.

\bibitem{vissani} F. Vissani, in {\it Solar Neutrinos}, ed. Mikko  Meyer (World Scientific, Singapore, 2019) pp. 121-141.

\bibitem{vescovi} D. Vescovi, C. Mascaretti, F. Vissani, L. Piersanti, and O. Straniero, J. Phys. G {\bf 48} (2021) 015201.

\bibitem{Asplund0} M. Asplund, N. Grevesse, and A. J. Sauval, in ASP Conf. Ser. 336, ed. T. G. Barnes III and F.N. Bash (ASP, San Francisco, 2005) pp 25.

\bibitem{Asplund} M. Asplund, A. M. Amarsi, and N. Grevesse, arXiv:2105.01661 (to be published in A\&A)

\bibitem{Opacity} J. Colgan et al., Ap. J. {\bf 817} (2016) 116; G. Mondet, C. Blancard, P. Cosse, and G. Faussurier, Ap. J. Suppl. {\bf 220} (2015) 2.

\bibitem{Bergemann} M. Bergemann et al., to be published in MNRAS (2021)

\bibitem{HaxtonSerenelli} W. C. Haxton and A. M. Serenelli, Ap. J. {\bf 687} (2008) 678.

\bibitem{Guzik} J. A. Guzik and K. Mussack, Ap. J. {\bf 713} (2010) 1108; J. A. Guzik, in ESA Special Publication, Vol. 624;
M. Castro, S. Vauclair, O. Richard, A\&A {\bf 463} (2007) 755.

\bibitem{McClure} M. K. McClure, A\&A {\bf 632} (2019) A32; M. K. McClure, C. Dominik, and M. Kama, A\&A {\bf 642} (2020) L15.

\bibitem{Booth} R. A. Booth and J. D. Ilee, MNRAS {\bf 487} (2019) 3998.

\bibitem{BBAL} H. A. Bethe, G. E. Brown, J. Applegate, and J. Lattimer, Nuclear Physics {\bf A324}, 487 (1979).

\bibitem{VartBur20} A. Burrows and D. Vartanyan, Nature {\bf 589}, 29 (2021).

\bibitem{models} H.-Th. Janka, K. Langanke, A. Marek, G. Martinez-Pinedo, and B. Muller, Phys. Rep. {\bf 442} (2007) 38; A. Mezzacappa, Ann. Rev. Nucl. Part. Sci. {\bf 55} (2005) 467;
K. Kotake, K. Sato, and K. Takahashi, Rep. Prog. Phys. {\bf 69} (2006) 971;
S. Woosley and J. S. Bloom, Ann. Rev. Astron. Astrophys. {\bf 44} (2006) 507.

\bibitem{snnus} K. Hirata {\it et al.}, Phys. Rev. Lett. {\bf 58} (1987) 1490; R. M. Bionta {\it et al.},
Phys. Rev. Lett. {\bf 58} (1987) 1494.

\bibitem{DUSELhandbook} Homestake Collaboration, arXiv:nucl-ex/0308018; S. W. Li, L. F. Roberts, and J. F. Beacom, Phys. Rev. {\bf D103} (2021) 023016.

\bibitem{Abdikamalov} E. Abdikamalov, G. Pagliaroli, and D. Radice, in {\it Handbook of Gravitational Wave Astronomy}, ed. C. Bambi, S. Katsanevas, and K. Kokkotas
(Springer, Singapore, 2021).

\bibitem{HALO} HALO Collaboration, Nucl. Part. Phys. Proc. {\bf 265-266} (2015) 233.

\bibitem{coherent} D. Akimov et al. (COHERENT Collaboration), Phys. Rev. Lett. {\bf 126} (2021) 012002; 
D. Akimov et al. (COHERENT Collaboration), Science {\bf 357} (2017) 1123.

\bibitem{coherentpapers} A. L. Foguel, E. S. Fraga, C. Bonifazi, Astrop. Phys. {\bf 127} (2021) 102534; N. F. Bell, J. B. Dent, J. L. Newstead, S. Sabharwal, T. J. Weiler, Phys. Rev. D101 (2021) 015012;
T. Kozynets, S. Fallows, C. B. Krauss, Astrop. Phys. {\bf 105} (2019) 25; XMASS Collaboration, Astrop. Phys. {\bf 89} (2017) 51.

\bibitem{DarkSide} P. Agnes et al., arXiv:2011.07819v2

\bibitem{2105.08688} P. T. H. Pang, I. Tews, M. W. Coughlin, M. Bulla, C. Van Den Broeck, and T. Dietrich, arXiv:2105.08688; F. Ozel and P. Freire, Ann. Rev. Astron. Astrophys. {\bf 54} (2016) 401.

\bibitem{Antoniadis2013} J. Antoniadis et al., Science {\bf 340} (2013) 6131.

\bibitem{DeMorest2010} P. DeMorest, T. Pennucci, S. Ransom, M. Roberts, and J. Hessels, Nature {\bf 467} (2010) 1081.

\bibitem{Fonseca2016} E. Fonseca et al., Ap. J. {\bf 832} (2016) 167.

\bibitem{Cromartie2019} H. T. Cromartie et al., Nature Astron. {\bf 4} (2019) 72.

\bibitem{Burgay2003} M. Burgay et al., Nature {\bf 426} (2003) 531.

\bibitem{Lyne2004} A. G. Lyne et al., Science {\bf 303} (2004) 1153.

\bibitem{LattimerShutz2005} J. M. Lattimer and B. F. Schutz, Ap. J. {\bf 629} (2005) 979.

\bibitem{NICER} M. C. Miller et al., arXiv:2105.06979.

\bibitem{Beacom} J. F. Beacom, Ann. Rev. Nucl. Part. Sci. {\bf 60} (2010) 439.

\bibitem{Lunardini} Cecilia Lunardini, Astropart. {\bf 26} (2006) 190; Cecilia Lunardini and O. L. G. Peres, JCAP {\bf 08} (2008) 033.

\bibitem{Vagins} J. F. Beacom and M. R. Vagins, Phys. Rev. Lett. {\bf 93} (2004) 171101

\bibitem{Sandoval21} M. A. Sandoval, W. R. Hix, O. E. B. Messer, E. J. Lentz, and A. Harris, arXiv:2106.01389.

\bibitem{Burrows21}  A. Burrows and D. Vartanyan, Nature {\bf 589} (2021) 29.

\bibitem{Glas19} R. Glas, H.-T. Janka, T. Melson, G. Stockinger, and O. Just, Ap. J. {\bf 881} (2019) 36;
H. Andresen, R. Glas, and H.-T. Janka, MNRAS {\bf 503} (2021) 3552.

\bibitem{OConnor18} E. O'Connor and S. Couch, Ap. J. {\bf 865} (2018) 81.

\bibitem{Janka16} H.-T. Janka, T. Melson, and A. Summa, Ann. Rev. Nucl. Part. Sci. {\bf 66} (2016) 341.

\bibitem{Janka12} H.-T. Janka, Ann. Rev. Nucl. Part. Sci. {\bf 62} (2012) 407.

\bibitem{Janka1702} H.-T. Janka, in  {\it Handbook of Supernovae}, ed. A. Alsabti and P. Murdin (Springer, 2017) p. 1575.

\bibitem{Lund2012} T. Lund, A. Wongwathanarat, H.-T. Janka, E. Mueller, and G. Raffelt, Phys. Rev. {\bf D86} (2012) 105031.

\bibitem{Tamborra2013} I. Tamborra, F. Hanke, B. Mueller, H.-T. Janka, and G. G. Raffelt, Phys. Rev. Lett.{\bf 111} (2013) 121104;
 I. Tamborra, F. Hanke, H.-T. Janka, B. Mueller, G. G. Raffelt, and A. Marek, Ap. J. {\bf 792} (2014) 96.
 
 \bibitem{annrev10} H. Duan, G. M. Fuller, and Y.-Z. Qian, Annual Review of Nuclear and Particle Science, {\bf 60}, 569 (2010).
 
 \bibitem{MSWSN} G. M. Fuller, R. W. Mayle, J. R. Wilson, and D. N. Schramm, Ap. J. {\bf 322} (1987) 795;
D. Notzold and G. Raffelt, Nucl. Phys. {\bf B307} (1988) 924.
 
 \bibitem{thompson} T. A. Thompson, A. Burrows, and P. A. Pinto, Ap. J. {\bf 592} (2003) 434.

\bibitem{Duan} H. Duan, G. M. Fuller, J. Carlson, Y.-Z. Qian, Phys. Rev. Lett. {\bf 100} (2008) 021101.
 
 \bibitem{Fullerbump} R. C. Shirato and G. M. Fuller, arXiv:astro-ph/0205390.
 
 \bibitem{shockdiffuse2010} S. Galais, J. Kneller, C. Volpe, and J. Gava, Phys. Rev. D {\bf 81}, 053002 (2010).
 
 \bibitem{turb2006} A. Friedland and A. Gruzinov, arXiv:astro-ph/0607244

\bibitem{turb22013} J. Kneller and A. Mauney, Phys. Rev. D {\bf 88}, 025004 (2013).

 \bibitem{Fried2020} A. Friedland and P. Mukhopadhyay, arXiv:2009.10059
 
 \bibitem{QKE2014} A. Vlasenko, G. M. Fuller, V. Cirigliano, Phys. Rev. D {\bf 89}, 105004 (2014).
 
 \bibitem{SiglRaffelt93} G. Sigl and G. Raffelt, Nuclear Physics B {\bf 406}, 423 (1993).
 
 \bibitem{Volpe2013} C. Volpe, D. V\"a\"an\"anen, C. Espinoza, Phys. Rev. D {\bf 87}, 113010 (2013).
 
 \bibitem{ZhangBurrows2013} Y. Zhang and A. Burrows, Phys. Rev. D {\bf 88}, 105009 (2013).
 
 \bibitem{Richers2019} S. Richers, G. McLaughlin, J. Kneller, A. Vlasenko, Phys. Rev. D {\bf 99}, 123014 (2019).
 
 \bibitem{DFCQ2006} H. Duan, G. M. Fuller, J. Carlson, and Y.-Z. Qian, Physical Review D {\bf 74}, 105014 (2006).
 
 \bibitem{DFCQ2006Let} H. Duan, G. M. Fuller, J. Carlson, Y.-Z. Qian, Physical Review Letters {\bf 97}, 241101 (2006).
 
 \bibitem{pen2006} S. Hannestad, G. Raffelt, G. Sigl, and Y. Y. Wong, Phys. Rev. D {\bf 74}, 105010 (2006).
 
 \bibitem{splits2007} G. Raffelt and A. Yu. Smirnov, Phys. Rev. D {\bf 76}, 125008 (2007).
 
 \bibitem{simple2007} H. Duan, G. M. Fuller, and Y.-Z. Qian, Phys. Rev. D {\bf 76}, 085013 (2007).
 
 \bibitem{Cherry2012} J. Cherry, J. Carlson, A. Friedland, G. M. Fuller, and A. Vlasenko, Phys. Rev. Lett. {\bf 108}, 261104 (2012).
 
 \bibitem{Cherry2020} J. Cherry, G. M. Fuller, H. Shunsaku, K. Kotake, T. Takiwaki, and T. Fisher, Phys. Rev. D~{\bf 102}, 023022 (2020).
 
 \bibitem{AxialBreaking2013} G. Raffelt, S. Sarikas, D. Seixas, Phys. Rev. Lett. {\bf 111}, 091101 (2013).
 
 \bibitem{Abbar2015} S. Abbar, H. Duan, S. Shalgar, Phys. Rev. D {\bf 92}, 065019 (2015). 
 
 \bibitem{temporal2015} B. Dasgupta and A. Mirizzi, Phys. Rev. D {\bf 92}, 125030 (2015).
 
 \bibitem{Sawyer2016} R.~F.~Sawyer, Phys. Rev. Lett. {\bf 116}, 081101 (2016).
 
 \bibitem{Basebu2017} B. Dasgupta, A. Mirizzi, and M. Sen, Journal of Cosmology and Astroparticle Physics, {\bf 2017}, issue 2, 019, (2017).
 
 \bibitem{MengRu2017} M.-R. Wu and I. Tamborra, Phys. Rev. D {\bf 95}, 103007 (2017).
 
 \bibitem{Cervia2019} M.~J. Cervia, A. Patwardhan, A.~B.~Balantekin, S. Coppersmith, C.~Johnson, Phys. Rev. D {\bf 100}, 083001 (2019).
 
 \bibitem{Surprising2016} E. Grohs and G. M. Fuller, Nuclear Physics B, {\bf 911}, 955 (2016).

\bibitem{Steigman} G. Steigman, Ann. Rev. Nucl. and Part. Sci. {\bf 57} (2007) 463.

\bibitem{Planck} R. Adam et al. (Planck Collaboration), A\&A {\bf 594} (2016) A10.

\bibitem{WMAP7}  WMAP Collaboration, E. Komatsu {\it et al.}, Astrophys. J. Suppl. {\bf 192} (2011) 18.

 \bibitem{BigPaper2016} E. Grohs, G. M. Fuller, C. T. Kishimoto, M. W. Paris, and A. Vlasenko, Phys. Rev. D {\bf 93}, 083522 (2016).

\bibitem{IT} Y. I. Izotov and T. X. Thuan, Astrophys. J. {\bf 710} (2010) L67; E. Aver, K. A. Olive,
and E. D. Skillman, J. Cosmol. Astropart. Phys. {\bf 03} (2011) 043.

\bibitem{ScienceBook} CMB-S4 Science Book, First Edition, CMB S-4 community, K. N. Abazajian {\it et al.} (2016). arXiv:1610.02743.

\bibitem{Field} B. D. Fields, Ann. Rev. Nucl. Part. Science {\bf 61} (2011) 47.

\bibitem{Pospelov} M. Pospelov and J. Pradler, Ann. Rev. Nucl. Part. Science {\bf 60} (2010) 539.

\bibitem{qian} Y.-Z. Qian, Prog. Part. Nucl. Phys. {\bf 50} (2003) 153.

\bibitem{BBFH} E. M. Burbidge, G. R. Burbidge, W. A. Fowler, and F. Hoyle, Rev. Mod. Phys. {\bf 29} (1957) 547.

\bibitem{FullerMeyer1995} G.~M.~Fuller and B.~Meyer, Astrophysical Journal, {\bf 453}, 792 (1995).

\bibitem{Roberts} L. F. Roberts, S. E. Woosley, and R. D. Hoffman, Ap. J. {\bf 722} (2010) 1.

\bibitem{McLaughlin1999} G.~C.~McLaughlin, J.~M.~Fetter, A.~B.~Balantekin, and G.~M.~Fuller, Phys. Rev. C {\bf 59}, 2873 (1999).

\bibitem{Banerjee} P. Banerjee, W. C. Haxton, and Y.-Z. Qian, Phys. Rev. Lett. {\bf  106},  201104 (2011).

\bibitem{Argast} D. Argast, M. Samland, F.-K. Thielemann, and Y.-Z. Qian,
Astron. Astrophys. {\bf 416}, 997 (2004).

\bibitem{LSr} J. M. Lattimer and D. N. Schramm, Ap. J. {\bf 192} (1974) L145.

\bibitem{SSr} E. Symbalisty and D., N. Schramm, Ap. J. {\bf 22} (1982) 143; D. Eichler, M. Livio, T. Piran, and D. N. Schramm, Nature {\bf 340} (1989) 126.

\bibitem{Davies1994} M. D. Davies, W. Benz, T. Piran, F. K. Thielemann, Ap. J. {\bf 431} (1994) 742.

\bibitem{Ruffert1997} M. Ruffert, H.-T. Janka, K. Takahashi, and G. Schaefer, A\&A {\bf 319} (1997) 122.

\bibitem{Freiburghaus1999} C. Freiburghaus, S. Rosswog, and F. R. Thielemann, Ap. J. {\bf 525} (1999) L121.

\bibitem{Just2015} O. Just, A. Bauswein, R. Andevol Pulpillo, S. Goriely, and H.-T. Janka, MNRAS {\bf 448} 541.

\bibitem{Nedora2020} V. Nedora et al., Ap. J. {\bf 906} (2020) 98.

\bibitem{Chen2021} H.-Y. Chen, S. Vitale, and F. Foucart, arXiv:2107.02714.

\bibitem{kilonova} L. Li and B Paczynski, Ap. J. {\bf 507} (1998) L59;  B. D. Metzger et al., 
MNRAS {\bf 406} 2650; L. F. Robertson, D. Kasen, W. H. Lee, and E. Ramirez-Ruiz, Ap. J. {\bf 736} (2011) L21.

\bibitem{kilonova2} D. Kasen, B. Metzger, J. Barnes, E. Quataert, and E. Ramirez-Ruiz, Nature {\bf 551} (2017) 80.

\bibitem{Sneden} C. Sneden, J. J. Cowan, and R. Gallino, Ann. Rev. Astron. Astrophys. {\bf 46} (2008) 241.

\bibitem{Farouqi} K. Farouqi, F.-K. Thielemann, S. Rosswog, and K.-L. Kratz, arXiv:2107.03486.

\bibitem{woosley} S. E. Woosley, D. H. Hartmann, R. D. Hoffman, and W. C. Haxton, 
Ap. J. {\bf 356} (1990) 272.

\bibitem{raffelt} G. G. Raffelt, Phys. Rev. Lett. {\bf 64} (1990) 2856.

\bibitem{GEMMA} A. G. Beda {\it et al.}, Ad. High Energy Phys. {\bf 2012} (2012) 350150.


\end{thebibliography}
\end{document}